\documentclass[final,3p,times]{elsarticle}
\usepackage{amsmath,hyperref}
\usepackage{lineno}

\usepackage{epstopdf}
\journal{Chemical Engineering Journal}

\usepackage{lineno,hyperref}
\usepackage{graphicx}
\usepackage{dcolumn}
\usepackage{bm}
\usepackage{epsfig}
\usepackage{booktabs}
\usepackage{subfigure}
\usepackage{graphics}
\usepackage{graphicx}
\usepackage{amssymb}
\usepackage{amsmath}
\usepackage{array}
\usepackage{color}
\usepackage{booktabs}
\usepackage{multirow}
\usepackage{caption}
\usepackage{chngpage}
\usepackage{subfigure}
\usepackage{mathrsfs,gensymb,float}

\biboptions{numbers,sort&compress}
\modulolinenumbers[1]

\hypersetup{hidelinks}
\begin{document}

\captionsetup[figure]{labelfont={bf},name={Fig.},labelsep=period}


\begin{frontmatter}
	
\title{Droplet formation dynamics in a T-junction via pulsed electrohydrodynamics}
\author[1,2]{He Yan}
\author[3]{Kang Luo}
\author[1]{Bin Gong}
\author[2]{Lei Wang\corref{mycorrespondingauthor}}
\cortext[mycorrespondingauthor]{Corresponding author}
\ead{leiwang@cug.edu.cn}

\address[1]{School of Earth Resources, China University of Geosciences, Wuhan 430074, China}
\address[2]{School of Mathematics and Physics, China University of Geosciences, Wuhan 430074, China}
\address[3]{School of Energy Science and Engineering, Harbin Institute of Technology, Harbin 150001, China}	

	
\begin{abstract}
Numerical simulations based on the phase-field lattice Boltzmann method are employed to investigate droplet formation dynamics in a T-junction microchannel under pulsed electrohydrodynamics. The applied electric field modulates the interfacial morphology, thereby altering the hydrodynamic stress distribution acting on the liquid-liquid interface. Three configurations are examined: no electric field, top-biased electrode, and bottom-biased electrode. The results show that the top-biased electrode promotes rapid transverse migration of the dispersed phase towards the upper wall, thereby accelerating the transition to the channel-blocking stage. By contrast, under the bottom-biased electrode, the dispersed phase first adheres to the lower wall and subsequently deforms upward under the electric force. As a result, the droplet formation dynamics are strongly affected by the spatial distribution of the electric field. The droplet generation process is further divided into a filling stage and a squeezing stage based on theoretical analysis. During the filling stage, the electric force modifies the filling duration but has little influence on the filling flux. During the squeezing stage, the necking rate and breakup time are governed by the pinning of the dispersed-phase contact line on the channel wall. Compared with constant electric forcing, pulsed electric forcing introduces an additional temporal degree of freedom, enabling stage-dependent modulation of interfacial evolution and breakup dynamics. Finally, droplet sequence diagrams and regime maps are presented for different pulse parameters. The results demonstrate that pulsed electrohydrodynamics in a T-junction enables active and programmable control of consecutive droplet generation.

\end{abstract}

\begin{keyword}
droplets \sep pulsed electrohydrodynamics \sep microfluidics
\end{keyword}
	
\end{frontmatter}

\section{Introduction}
Droplet-based microfluidic devices are systems that use microfluidic technology to generate and manipulate discrete fluid volumes within immiscible phases \cite{Schneider,Whitesides_nat2006}. These systems provide precise control over small-volume fluids and droplet characteristics, allowing the generated droplets to function as independent chemical reactors \cite{Song_ACIEE2003,Korczyk_NC2019}. By encapsulating samples in droplet form, sample dilution caused by Taylor diffusion can be mitigated, thereby 1enhancing mixing performance \cite{Bontoux2006,Tice2003}. These features facilitate high-throughput experiments and provide a more efficient reaction environment. Additionally, it prevents direct interaction between the sample and the channel walls, thereby eliminating surface adsorption and cross-contamination. Consequently, droplet-based microfluidics has found broad applications in fields such as biotechnology, chemical reactions, and advanced material manufacturing \cite{Zhang_NC2024,Elvira_NC2013,Shang_Labonchip2016}.

The core of droplet microfluidics lies in controlling the frequency and size of droplet generation. Depending on the actuation mechanism, these devices can be categorised into 'active' and 'passive' types \cite{Shi2021}. Passive control relies on modifying the channel geometry or the flow properties to regulate droplet generation without external energy input. Microfluidic devices based on this method have been extensively studied and classified according to their geometric structure, including co-flow, T-junction, and flow-focusing \cite{Umbanhowar_langmuir2000,Thorsen_PRL2001,Haringa_PRF2019,Anna_APL2003}. Under passive conditions, droplet formation and breakup are governed by the competition among interfacial tension, viscous shear, and local pressure buildup. However, the stabilising effects of surface tension and fluid viscosity on the liquid--liquid interface make it difficult for passive methods to stably generate droplets of smaller size. Although reducing the droplet size is possible to some extent by adjusting the flow-rate ratio, this strategy often requires a large excess of the continuous phase and may therefore lead to excessive dilution of the product.  To overcome these challenges, researchers have developed external actuation methods that enhance droplet control, such as magnetic fields, temperature fields, and acoustic fields \cite{SHYAM_JFM2022,Miralles_LabonChip2015,JUNG_Labonchip2016}. These external fields locally modify the force balance acting on the fluid interface, providing additional means for droplet flow control. Although these fields can alter droplet sizes, each has inherent drawbacks. For instance, magnetic fields require the introduction of magnetic particles, which may lead to system contamination. Thermal control is often limited by the slow thermal diffusion of the medium, precluding an instantaneous response. Similarly, acoustic control faces limitations regarding response time and complex standing wave setups \cite{Huang_Labchip2017}.

Electrical control methods have long been applied to the transport of single-phase fluids or particles \cite{Park2016}, yet their application in micro-scale multiphase droplet manipulation emerged later. Compared with other active methods, electric field-based droplet manipulation offers the distinct advantage of rapid switching response. Moreover, the electric field method avoids the introduction of magnetic particles, thereby preventing contamination of the fluid. Since Taylor \cite{Taylor1966} first proposed the existence of vortex flow inside droplets under the influence of electric fields, droplet deformation, migration, and breakup under electric fields have become important topics in electrohydrodynamics. For weakly conducting two-phase liquids, the mismatch in permittivity and conductivity across the interface leads to charge redistribution and the generation of Maxwell electric stress at the liquid--liquid interface \cite{Yariv_JFM2016,Papageorgiou_ARFM2019,Vlahovska_PRF2016,Vlahovska2019}. The normal component of this electric stress competes with surface tension and alters the interface curvature, giving rise to different deformation modes such as elongation or flattening along the electric-field direction. Meanwhile, the tangential component restructures the flow near the interface and induces circulatory motion both inside and outside the droplet, thereby further affecting interfacial migration. On this basis, Link et al.\cite{Link_ACIEE2006} first demonstrated droplet generation and manipulation in a flow-focusing microfluidic device using a DC electric field. Subsequent experiments \cite{Yeh_MN2012} and numerical simulations \cite{Jiang_JMF2021} further showed that electric fields can effectively control droplet size and generation frequency. Beyond flow-focusing configurations, experiments have also shown that the application of electric fields in T-junctions can not only reduce the size of generated droplets, but also enable active programming of droplet sequences by controlling the size of individual droplets through variations in field strength and actuation time. In the experiments of Shojaeian and Hardt \cite{Shojaeian_APL2018}, the electric field was shown to accelerate droplet breakup and reduce the resulting droplet size by inducing interfacial electric stress and pinning the droplet to the channel wall. Subsequent studies further indicated that the application of electric forces may also lead to transitions in droplet flow patterns \cite{Agarwal_PRF2020,Zhu_CEJ2024}.

Although existing studies have demonstrated that electric fields can effectively regulate droplet generation in T-junctions, most of them are still restricted to constant electric fields. In contrast to constant electric fields, which continuously modify the spatial distribution of interfacial stresses, time-varying electric fields introduce an additional time scale into droplet dynamics. As a result, the interfacial response depends not only on the field strength, but also on the temporal characteristics of the applied forcing. Under a pulsed DC electric field, the electric stress switches periodically between on and off states, leading to alternating stretching and relaxation of the interface \cite{Wang_IJHMT2016,Huang_AICHE2020}. More generally, studies under temporally varying electric forcing have shown that droplet deformation depends sensitively on the time scale of the imposed field relative to the intrinsic interfacial response time \cite{Sengupta_PRF2020}. Existing studies have further shown that, in T-junctions subjected to AC electric fields, the droplet size is jointly determined by the applied voltage, the excitation frequency, and the electrical response of the dispersed phase, indicating that time-varying electric forcing can modify droplet behaviour through the coupling between interfacial electric stress and fluid response time \cite{Teo_MN2020}. Although Shojaeian and Hardt \cite{Shojaeian_APL2018} demonstrated the programming of droplet sequences using rectangular voltage pulses, this type of pulsed control still acts as an overall on--off modulation of a complete droplet formation event. It has not yet clarified how the interfacial evolution, breakup location, and subsequent droplet sequence are altered when the pulsed electric field is applied only during specific stages within a single droplet formation cycle, such as filling, squeezing, or pinch-off. Motivated by this gap, the present study investigates the stage-dependent action of a pulsed electric field within an individual droplet formation cycle in a T-junction.

\section{Formulation}
\label{sec:method}

The geometric configuration of the T-junction microchannel employed in this study is illustrated in figure 1. The dispersed phase and continuous phase are injected into the main and side channels with flow rates $Q_{d}$ and $Q_{c}$, respectively, where the subscripts '$c$' and '$d$' denote the continuous and dispersed phases. Consistent with previous studies, the electrodes are positioned downstream near the junction of the side channel. We assume an initial state with zero electric potential and free charge distribution within the fluid. The top and bottom electrodes are connected to an external power source, while the remaining channel walls are electrically insulated. The width of the main and side channels is set to $w=100~\mu\text{m}$, a dimension widely reported in prior numerical \cite{Liu_POF2009,Liu_POF2022} and experimental \cite{Graaf_langmuir2006} investigations. In the present simulation, the channel width corresponds to 30 lattice units, resulting in a total computational domain of $601 \times 91$ grid points. This resolution was selected to accurately capture interfacial topological deformations and necking dynamics while minimizing computational cost; furthermore, the domain length is sufficient to ensure that the outlet boundary conditions do not perturb the internal flow field.

The physicochemical properties of the fluids are chosen to match typical experimental systems \cite{Ha2000,Singh_JFM2020}. The continuous and dispersed phases share identical densities, $\rho_{c} = \rho_{d} = 1000~\text{kg m}^{-3}$, while the dynamic viscosities are given by $\mu_{c} = 4\mu_{d} = 4~\text{mPa s}$. The interfacial tension is fixed at $\gamma = 16~\text{mN m}^{-1}$. In the context of EHD, the system is characterized by permittivity ($\hat S = \epsilon_{c}/\epsilon_{d}$) and conductivity ($\hat R = \sigma_{c}/\sigma_{d}$) ratios of $5$. Here, $\hat R = \hat S = 5$ is chosen to ensure a sufficient Maxwell stress at the interface while avoiding excessively strong local electric-field effects that may arise under extreme conductivity or permittivity ratios \cite{Singh_JFM2020}. The system dynamics are governed by the capillary number ($Ca$) and the electric capillary number ($Ca_{E}$), defined as
\begin{equation}
	Ca = \frac{\mu_c u_c}{\gamma}, \quad Ca_E = \frac{\epsilon_c E^2 w}{\gamma}
\end{equation}
where $u_c$ denotes the characteristic velocity of the continuous phase and $E$ is the electric field strength.
\begin{figure}
	\centerline{\includegraphics[width=0.6\linewidth]{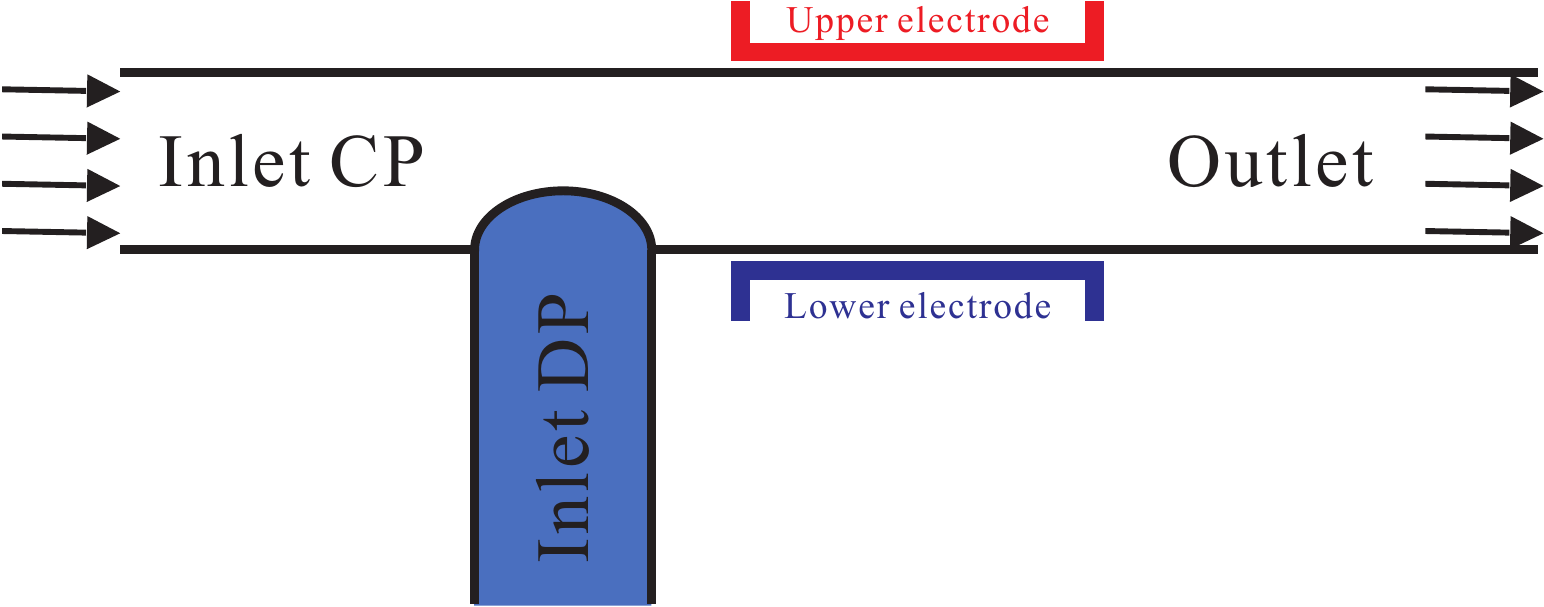}} 
	\caption{Schematic illustration of droplet generation in a T-junction with electrodes positioned downstream near the junction. CP and DP denote the continuous- and dispersed-phase inlets, respectively.}
	\label{fig1}
\end{figure}

\subsection{Governing equations}
The phase-field lattice Boltzmann method is used to simulate the dynamic evolution of the droplet interface in the T-junction microchannel, which is widely applied in the field of multiphase fluid dynamics \cite{Badalassi_jcp2003,Chiu_jcp2011}. In this model, the phase interface is described by a rapidly changing but continuous phase field variable, also known as the order parameter ($\phi$). In this work, the second-order conserved Allen-Cahn equation is used to track the multiphase interface. According to prior work \cite{SunJCP2007}, the interface advection equation for the order parameter can be written as
\begin{equation}
     \frac{{\partial \phi }}{{\partial t}} + \left( {{u_n} + {\bf{u}}} \right) \cdot \nabla \phi  = 0	
\end{equation}
where ${\bf{u}}$ is the fluid velocity, ${u_n}$ and ${\bf{n}}$ are the normal interface speed and unit normal vector, which can be defined as
\begin{equation}
	{u_n} = - {M_\phi }\kappa ,{\bf{n}} = \frac{{\nabla \phi }}{{\left| {\nabla \phi } \right|}},
\end{equation}
in which ${M_\phi}$ is the mobility and $\kappa$ is the interface curvature, given by
\begin{equation}
	\kappa = \nabla \cdot n = \nabla \cdot \left( {\frac{{\nabla \phi }}{{|\nabla \phi |}}} \right) = \frac{1}{{|\nabla \phi |}}\left[ {{\nabla ^2}\phi - \left( {\nabla \cdot \nabla \phi } \right)/|\nabla \phi |} \right].
\end{equation}
We use ${\phi _d}=1$ and ${\phi _c}=-1$ to represent the dispersed and continuous phases, respectively. Therefore, ${\phi_{0}=0}$ is used to represent the interface between the continuous and dispersed phases, with the assumption that it has a finite thickness $W$. Specifically, the order profile at the equilibrium state can be written as 
\begin{equation}
	\phi (x) = 0.5\tanh \left( {\frac{{2\left\| {{\bf{x}} - {{\bf{x}}_0}} \right\|}}{W}} \right),
\end{equation}
where ${\left\| {{\bf{x}} - {{\bf{x}}_0}} \right\|}$ represents the spatial distance from the interface ${\bf{x}_0}$. Based on the phase-field formulation proposed by Folch \textit{et al.}~\cite{FolchPRE1999}, the local Allen-Cahn equation can be expressed as \cite{Chiu_jcp2011}
\begin{equation}\label{eq2.1}
	{\partial _t}\phi  + \nabla  \cdot \left( {\phi {\bf{u}}} \right) = \nabla  \cdot {M_\phi }\left( {\nabla \phi  - \sqrt {\frac{{2\beta }}{\kappa }} \left( {{\phi _c} - \phi } \right)\left( {\phi  - {\phi _d}} \right)\frac{{\nabla \phi }}{{\left| {\nabla \phi } \right|}}} \right),
\end{equation}
in which $\beta$ and $\kappa$ are physical quantities related to the surface tension($\gamma $) and the interface thickness
\begin{equation}
	W = \frac{1}{{\left| {{\phi _c} - {\phi _d}} \right|}}\sqrt {\frac{{8\kappa }}{\beta }} ,\gamma  = \frac{{{{\left| {{\phi _c} - {\phi _d}} \right|}^3}}}{6}\sqrt {2\kappa \beta }.
\end{equation}
In addition to the aforementioned interface capturing equation, we also need to introduce the governing equations for the two-phase flow. Assuming that the two fluids are immiscible and incompressible, the flow field is governed by the Navier-Stokes equations \cite{Unverdi_JCP1992}.
\begin{equation}\label{eq2.4}
	\nabla  \cdot {\bf{u}} = 0,
\end{equation}
\begin{equation}\label{eq2.5}
	\rho \left( {\frac{{\partial {\bf{u}}}}{{\partial t}} + {\bf{u}} \cdot \nabla {\bf{u}}} \right) =  - \nabla p + \nabla  \cdot \left[ {\mu \left( {\nabla {\bf{u}} + \nabla {{\bf{u}}^{\bf{{\rm T}}}}} \right)} \right] + {{\bf{F}}_s} + {{\bf{F}}_e},
\end{equation}
where ${{\bf{F}}_s}$ and ${{\bf{F}}_e}$ represent the surface tension force and the electric force, respectively. They can be expressed as
\begin{equation}
	{{\bf{F}}_s} = {\mu _\phi }\nabla \phi ,
\end{equation}
\begin{equation}\label{eq2.11}
	{{\bf{F}}_e} = {\rho _e}{\bf{E}} - \frac{1}{2}{\bf{E}} \cdot {\bf{E}}\nabla \varepsilon ,
\end{equation}
where ${\mu _\phi } = 4\beta \left( {\phi  - {\phi _c}} \right)\left( {\phi  - {\phi _d}} \right)\left( {\phi  - {\phi _0}} \right) - \kappa {\nabla ^2}\phi$ is the chemical potential, ${\rho _e} = \nabla  \cdot (\varepsilon {\bf{E}})$ represents the charge density and $\varepsilon$ is the permittivity. In the external electric field force ${{\bf{F}}_e}$, the first term represents the Coulomb force, which applies in the direction of the electric field, and the second term is the polarization stress, which acts along the normal direction of the interface.

In this model, the Poisson equation is employed to describe the existing electric field. Consistent with prior research \cite{Liu_pof2019, Singh_PRFluids2019}, we adopt the leaky dielectric model, and the macroscopic equation for the electric field is given by \cite{Hua_pof2008}
\begin{equation}\label{eq2.3}
	\nabla  \cdot \left( {\sigma {\bf{{\rm E}}}} \right) = 0.
\end{equation}
Since it is assumed that the electric field strength ${\bf{{\rm E}}}$ is irrotational ($\nabla  \times {\bf{E}} = 0$) , the electric field can be expressed as the gradient of the electric potential ($\psi$), i.e., ${\bf{{\rm E}}} =  - \nabla \psi $.

\subsection{Lattice Boltzmann method}
To solve the macroscopic governing equations involved above using the lattice Boltzmann method, three distribution functions: the particle distribution function ${f_i}$, the order parameter distribution function ${g_i}$, and the electric potential distribution function ${h_i}$, which describe the evolution of the flow field, phase field, and electric field, respectively. The evolution equations for ${f_i}$, ${g_i}$, and ${h_i}$ are \cite{Wang_PRE2003,Liu_pof2019}
\begin{equation}\label{eq8}
	{f_i}\left( {{\bf{x}} + {{\bf{c}}_i}\Delta t,t + \Delta t} \right) = {f_i}\left( {{\bf{x}},t} \right) - \frac{1}{{{\tau _f}}}\left[ {{f_i}\left( {{\bf{x}},t} \right) - f_i^{eq}\left( {{\bf{x}},t} \right)} \right] + \Delta t\left( {1 - \frac{1}{{2{\tau _f}}}} \right){{\bf{F}}_i}\left( {{\bf{x}},t} \right),
\end{equation}
\begin{equation}\label{eq9}
	{g_i}\left( {{\bf{x}} + {{\bf{c}}_i}\Delta t,t + \Delta t} \right) = {g_i}\left( {{\bf{x}},t} \right) - \frac{1}{{{\tau _\phi }}}\left[ {{g_i}\left( {{\bf{x}},t} \right) - g_i^{eq}\left( {{\bf{x}},t} \right)} \right] + \Delta t\left( {1 - \frac{1}{{2{\tau _\phi }}}} \right){{\bf{G}}_i}\left( {{\bf{x}},t} \right),
\end{equation}
\begin{equation}\label{eq10}
	{h_i}\left( {{\bf{x}} + {{\bf{c}}_i}\Delta t,t + \Delta t} \right) = {h_i}\left( {{\bf{x}},t} \right) - \frac{1}{{{\tau _h}}}\left[ {{h_i}\left( {{\bf{x}},t} \right) - h_i^{eq}\left( {{\bf{x}},t} \right)} \right],
\end{equation}
where ${{\tau _f}}$, ${{\tau _\phi}}$ and ${{\tau _h}}$ are the relaxation parameters, ${f_i^{eq}}$,${g_i^{eq}}$ and ${h_i^{eq}}$ are the equilibrium distribution functions of ${f_i}$, ${g_i}$ and ${h_i}$, they are be defined as \cite{Wang_PRE2003,Liu_pof2019}
\begin{equation}
	f_i^{eq} = 
	\begin{cases} 
		\frac{p}{c_s^2}(\omega_0 - 1) + \rho s_i(\mathbf{u}), & i = 0, \\
		\frac{p}{c_s^2} \omega_i + \rho s_i(\mathbf{u}), & i = 1-8,
	\end{cases}
\end{equation}
\begin{equation}
	g_i^{eq} = {\omega _i}\phi \left( {1 + \frac{{{{\bf{c}}_i} \cdot {\bf{u}}}}{{c_s^2}}} \right),
\end{equation}
\begin{equation}
	h_i^{eq} = 
	\begin{cases} 
		(\omega_0 - 1)\psi, & i = 0, \\
		\omega_i \psi, & i = 1-8,
	\end{cases}
\end{equation}
where the ${s_i}\left( {\bf{u}} \right)$ is given by
\begin{equation}
	s_i(\mathbf{u}) = \omega_i \left[ \frac{\mathbf{c}_i \cdot \mathbf{u}}{c_s^2} + \frac{(\mathbf{c}_i \cdot \mathbf{u})^2}{2c_s^4} - \frac{\mathbf{u} \cdot \mathbf{u}}{2c_s^2} \right],
\end{equation}
For the D2Q9 model, the weight factors ${\omega _i}$ and the discrete velocities ${{{\bf{c}}_i}}$ are given by
\begin{equation}
	\omega_i =
	\begin{cases}
		{4 \mathord{\left/
				{\vphantom {4 9}} \right.
				\kern-\nulldelimiterspace} 9},  & i=0,\\[4pt]
		{1 \mathord{\left/
				{\vphantom {4 9}} \right.
				\kern-\nulldelimiterspace} 9},  & i=1-4,\\[4pt]
		{1 \mathord{\left/
				{\vphantom {4 9}} \right.
				\kern-\nulldelimiterspace} 36}, & i=5-8.
	\end{cases}
\end{equation}
\begin{equation}
	{{\bf{c}}_i} = \left\{ {\begin{array}{*{20}{l}}
			{c{{(0,0)}^{\rm{T}}},}&{i = 0,}\\
			{c{{(\cos [(i - 1)\pi /2],{\mkern 1mu} \sin [(i - 1)\pi /2])}^{\rm{T}}},}&{i = 1-4,}\\
			{\sqrt 2 {\mkern 1mu} c{{(\cos [(2i - 1)\pi /4],{\mkern 1mu} \sin [(2i - 1)\pi /4])}^{\rm{T}}},}&{i = 5-8,}
	\end{array}} \right.
\end{equation}
where $c = {{\Delta x} \mathord{\left/
		{\vphantom {{\Delta x} {\Delta t,}}} \right.
		\kern-\nulldelimiterspace} {\Delta t,}}$ is the lattice speed, ${c_s}$ is the sound speed of lattice fluid which can be computed as ${c_s} = {c \mathord{\left/
		{\vphantom {c {\sqrt 3 }}} \right.
		\kern-\nulldelimiterspace} {\sqrt 3 }}$ and ${\Delta x}$ is the lattice spacing. The ${{\bf{F}}_i}$ is the force distribution function and ${{\bf{G}}_i}$ is the source term, they are expressed as \cite{Wang_PRE2003,Liu_pof2019}
\begin{equation}
	{{\bf{F}}_i} = {\omega _i}\left[ {{\mkern 1mu} {\bf{u}} \cdot \nabla \rho  + \frac{{{{\bf{c}}_i} \cdot \left( {{{\bf{F}}_s} + {{\bf{F}}_e}} \right)}}{{c_s^2}} + \frac{{{\bf{u}}\nabla \rho :\left( {{{\bf{c}}_i}{{\bf{c}}_i} - c_s^2{\bf{I}}} \right)}}{{c_s^2}}{\mkern 1mu} } \right],
\end{equation}
\begin{equation}
	{{\bf{G}}_i} = {\omega _i}\frac{{{{\bf{c}}_i} \cdot \left[ {{\partial _t}\left( {\phi {\bf{u}}} \right) + c_s^2\sqrt {\frac{{2\beta }}{\kappa }} \left( {{\phi _c} - \phi } \right)\left( {\phi  - {\phi _d}} \right)\frac{{\nabla \phi }}{{\left| {\nabla \phi } \right|}}} \right]}}{{c_s^2}},
\end{equation}
In the present model, the macroscopic variables, such as the velocity, pressure, the order parameter and the electric potential are given by \cite{Wang_PRE2003,Liu_pof2019}
\begin{equation}
	\rho {\mkern 1mu} {\bf{u}}\; = \;\sum\limits_i {\mkern 1mu}  {{\bf{c}}_i}{\mkern 1mu} {f_i}\; + \;0.5{\mkern 1mu} \Delta t{\mkern 1mu} \left( {{{\bf{F}}_s} + {{\bf{F}}_e}} \right),
\end{equation}
\begin{equation}
	p\; = \;\frac{{c_s^2}}{{{\mkern 1mu} 1 - {\omega _0}{\mkern 1mu} }}{\mkern 1mu} \left[ {\sum\limits_{i \ne 0} {{f_i}} \; + \;\frac{{\Delta t}}{2}{\mkern 1mu} {\bf{u}} \cdot \nabla \rho \; + \;\rho {\mkern 1mu} {s_0}({\bf{u}})} \right],
\end{equation}
\begin{equation}
	\phi  = \sum\limits_i {\mkern 1mu}  {\mkern 1mu} {g_i},
\end{equation}
\begin{equation}
	\psi  = \sum\limits_i {\mkern 1mu}  {\mkern 1mu} {h_i}.
\end{equation}
Through the Chapman-Enskog analysis on the evolution \eqref{eq8},  \eqref{eq9} and \eqref{eq10} indicate that the three lattice Boltzmann evolution equations can recover the Navier-Stokes equation, Allen-Cahn equation and Poisson equation. The relationships between the macroscopic variable (dynamic viscosity $\mu$ , mobility ${M_\phi}$ and electrical conductivity $\sigma$) and the relaxation time are derived as \cite{Wang_PRE2003,Liu_pof2019}
\begin{equation}
	\mu  = \rho c_s^2\left( {{\tau _f} - \frac{1}{2}} \right)\Delta t,{M_\phi } = c_s^2\left( {{\tau _\phi } - \frac{1}{2}} \right)\Delta t,\sigma  = c_s^2\left( {{\tau _h} - \frac{1}{2}} \right)\Delta t.
\end{equation}
In this model, the fluid is considered as a mixture of two-phase fluids. Therefore, dynamic physical parameters such as density ($\rho$), viscosity ($\mu$), electrical conductivity ($\sigma$), and permittivity ($\varepsilon $) can be defined by the order parameter as follows
\begin{equation}
	\mu  = \frac{1}{2}\left( {{\mu _d} + {\mu _c}} \right) + \frac{1}{2}\phi \left( {{\mu _d} - {\mu _c}} \right),\rho  = \frac{1}{2}\left( {{\rho _d} + {\rho _c}} \right) + \frac{1}{2}\phi \left( {{\rho _d} - {\rho _c}} \right),
\end{equation}
\begin{equation}
	\sigma  = \frac{1}{2}\left( {{\sigma _d} + {\sigma _c}} \right) + \frac{1}{2}\phi \left( {{\sigma _d} - {\sigma _c}} \right),\varepsilon  = \frac{1}{2}\left( {{\varepsilon _d} + {\varepsilon _c}} \right) + \frac{1}{2}\phi \left( {{\varepsilon _d} - {\varepsilon _c}} \right).
\end{equation}

\subsection{Boundary conditions}
In the present study, the boundary conditions consist of the no-slip boundary, the wetting boundary, and the inlet-outlet boundary \cite{Graaf_langmuir2006,Chen_JFM2017}. For the solid walls, the no-slip boundary condition is applied, which is realized through the halfway bounce-back scheme. The boundary conditions considered in the present research include the no-slip boundary, the wetting boundary, and the inlet-outlet boundary. The no-slip boundary condition is applied to the solid walls, implemented using the halfway bounce-back scheme. For both the flow field and the phase field, the boundary conditions are realized by \cite{Ladd_jfm1994}
\begin{equation}
	\begin{array}{l}
		{f_{\bar i}}\left( {{{\bf{x}}_{b}},t + \Delta t} \right) = f_i^ + \left( {{{\bf{x}}_{b}},t} \right),\\
		{g_{\bar i}}\left( {{{\bf{x}}_{b}},t + \Delta t} \right) = g_i^ + \left( {{{\bf{x}}_{b}},t} \right),
	\end{array}
\end{equation}
in which ${{\bf{x}}_b}$ represents the boundary node. While for the electric field, a Dirichlet boundary condition is applied at the electrodes \cite{Zhang_PRE2012}
\begin{equation}
	{h_{\bar i}}\left( {{{\bf{x}}_{b}},t + \Delta t} \right) = h_i^ + \left( {{{\bf{x}}_{b}},t} \right) + 2{\omega _i}\psi 
\end{equation}
where $f_i^ + $, $g_i^ + $ and $h_i^ + $ are the posicollision distribution functions, and ${{\bf{c}}_{\bar i}}$ is the opposite velocity of ${{{\bf{c}}_i}}$(${{\bf{c}}_{\bar i}} =  - {{\bf{c}}_i}$). The wetting boundary condition is prescribed according to the method proposed by van der Graaf \textit{et al.} \cite{Graaf_langmuir2006}. This approach assumes that the wall is a mixture of two fluids, with the order parameter ${\phi_{wall}}$ representing this mixture. By combining the interfacial tension between the fluid and the solid wall with Young’s equation, an expression for the contact angle can be derived
\begin{equation}
	\cos \left( \theta  \right) = \frac{1}{2}{\phi _{wall}}\left( {3 - \phi _{wall}^2} \right).
\end{equation}
The inlet boundary condition follows the format proposed by Ladd, which has shown good performance in many previous studies. For the flow field and phase field, the distribution functions of the unknowns can be expressed \cite{Ladd_jfm1994}
\begin{equation}
	\begin{array}{l}
		{f_{\bar i}}\left( {{{\bf{x}}_{in}},t + \Delta t} \right) = f_i^ + \left( {{{\bf{x}}_{in}},t} \right) - 2{\omega _i}\rho \left( {{{\bf{x}}_{in}},t + \Delta t} \right)\frac{{{{\bf{c}}_i} \cdot {{\bf{u}}_{in}}}}{{c_s^2}},\\
		{g_{\bar i}}\left( {{{\bf{x}}_{in}},t + \Delta t} \right) = f_i^ + \left( {{{\bf{x}}_{in}},t} \right) - 2{\omega _i}\phi \left( {{{\bf{x}}_{in}},t + \Delta t} \right)\frac{{{{\bf{c}}_i} \cdot {{\bf{u}}_{in}}}}{{c_s^2}},
	\end{array}
\end{equation}
in which ${{{\bf{u}}_{in}}}$ is the velocity at the inlet boundary. The convective outflow boundary conditions are used at the outlet, which have good results in dealing with open boundaries of multiphase flows. For the unknown distribution function at the outlet it can be given by the following equation \cite{Lou_PRE2013}
\begin{equation}
	\begin{array}{l}
		{f_i}\left( {N,j,t + \Delta t} \right) = \frac{{{f_i}\left( {N,j,t} \right) + \lambda {f_i}\left( {N - 1,j,t + \Delta t} \right)}}{{1 + \lambda }},\\
		{g_i}\left( {N,j,t + \Delta t} \right) = \frac{{{g_i}\left( {N,j,t} \right) + \lambda {g_i}\left( {N - 1,j,t + \Delta t} \right)}}{{1 + \lambda }},
	\end{array}
\end{equation}
where $\lambda  = {{U\left( {t + \Delta t} \right)\Delta t} \mathord{\left/
		{\vphantom {{U\left( {t + \Delta t} \right)\Delta t} {\Delta x}}} \right.
		\kern-\nulldelimiterspace} {\Delta x}}$, which is known after the streaming step at the (N-1)th layer. The $U$ is a typical velocity normal to the outlet boundary, in this work,
\begin{equation}
	U = \frac{1}{{M + 1}}\sum\limits_j {u\left( {N - 1,j,t} \right)} ,
\end{equation}
in which $M+1$ is the number of nodes in the ($N-1$)th layer in the $y$ direction.

\section{Validation}
\label{sec:validation}

In this section, two numerical examples are used to validate the present phase-field lattice Boltzmann model. For the first benchmark case, we simulate the process of droplet formation in a T-junction microchannel and compare our numerical data with piror experimental results \cite{Graaf_langmuir2006}. Simulations were run in a system containing $400 \times 120$ lattice cells (i.e., one lattice cell corresponds to $2.5$ $\mu m$). Initially, the dispersed phase flow rate ${Q_d}$ is maintained at $0.2~\text{ml h}^{-1}$, and the continuous phase flow rate ${Q_c}$ is increased from $0.2$ to $4.0~\text{ml h}^{-1}$. The corresponding physical values of the dispersed and continuous phases are same as experiment.

\begin{figure}[H]
	\centering
	\subfigure[]{\includegraphics[height=6.5cm]{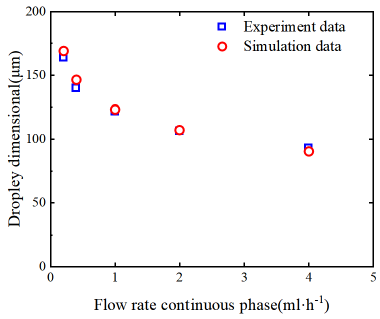}}
	\subfigure[]{\includegraphics[height=6.5cm]{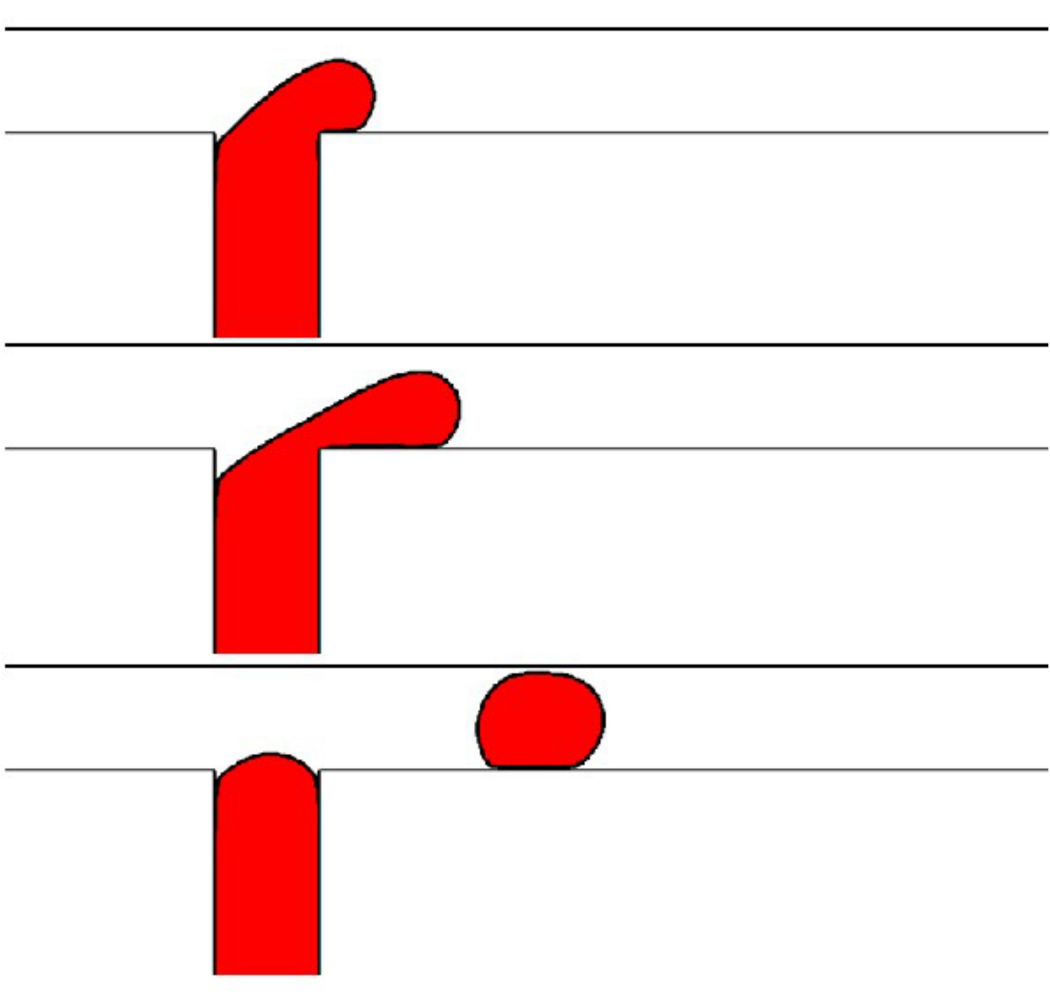}}
	\caption{(a) Comparison of droplet diameter variation with the flow rate ratio between the simulation (red circles) and experiment (blue squares), with the dispersed-phase flow rate fixed at $Q_d = 0.2 \text{ ml h}^{-1}$. (b) Time-evolution snapshots of the droplet formation process at $Q_c = 2.0 \text{ ml h}^{-1}$.}
	\label{fig:fig2}
\end{figure}

Figure \ref{fig:fig2}(a) illustrates the dependence of the droplet diameter on the continuous phase flow rate, $Q_c$, where the equivalent diameter is evaluated using (\ref{eq3.1}). Operating within a low capillary number regime ($Ca \sim 10^{-2}$), the droplet formation is governed by geometric confinement, marking a transition from the squeezing to the dripping regime. The numerical simulations show excellent agreement with experimental observations, accurately capturing the reduction in droplet size with increasing flow rate, with a maximum deviation of less than $4\% $. To elucidate the breakup dynamics, figure \ref{fig:fig2}(b) depicts the temporal evolution of the interfacial morphology during a single generation cycle at $Q_c = 2.0~\text{ml h}^{-1}$. Initially, the dispersed phase intrudes into the main channel, effectively blocking the cross-section while leaving only a narrow gap for the continuous phase to bypass. Subsequently, driven by the combined effects of upstream pressure build-up and viscous shear stress, the dispersed phase neck progressively thins and eventually pinches off to release the droplet.
\begin{equation}\label{eq3.1}
	D = {\left( {\frac{{4A}}{\pi }} \right)^{{1 \mathord{\left/
					{\vphantom {1 2}} \right.
					\kern-\nulldelimiterspace} 2}}}.
\end{equation} 

\begin{figure}[H]
	\centering
	\subfigure[]{\includegraphics[width=0.3\textwidth]{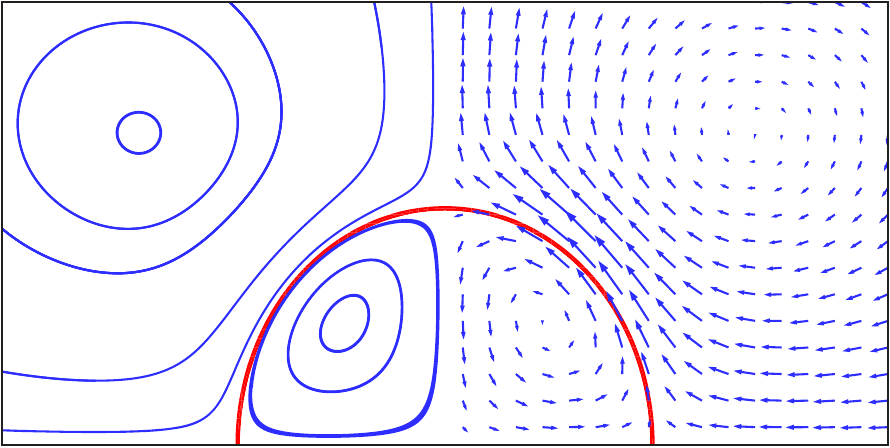}}
	\hspace{0.5cm}
	\subfigure[]{\includegraphics[width=0.3\textwidth]{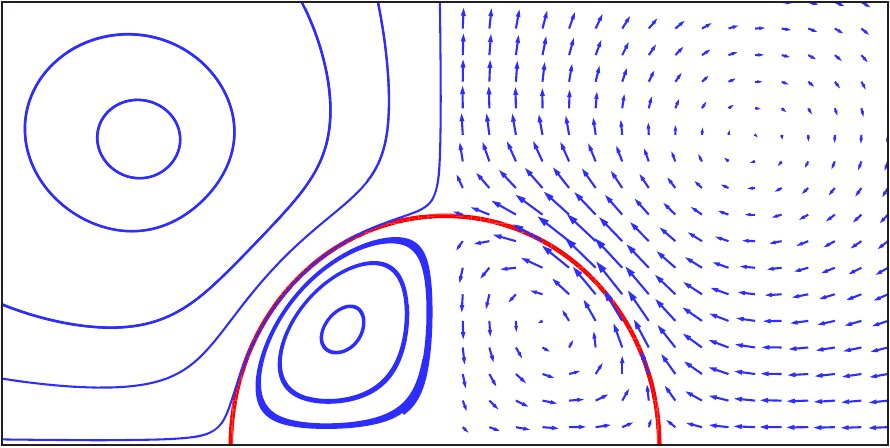}}
	\hspace{0.5cm}
	\subfigure[]{\includegraphics[width=0.3\textwidth]{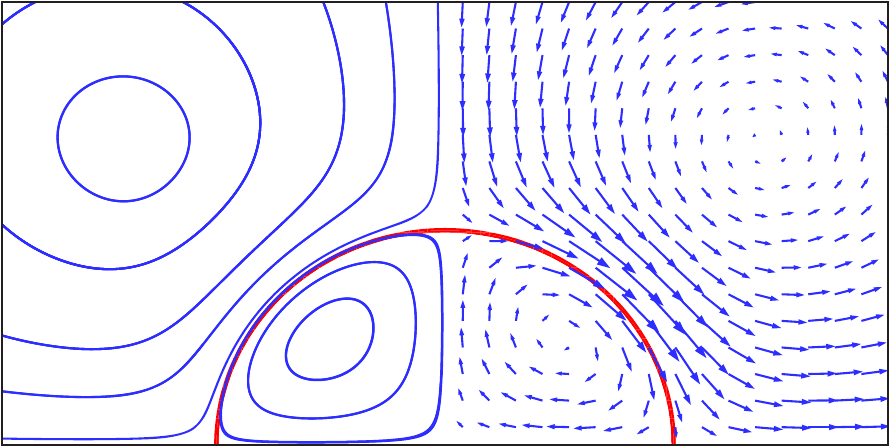}}
	\caption{Stream lines (left half), velocity vectors (right half) and the profile of the droplet (red lines) for three representative steady-state droplets: (a) prolate deformation obtained at $\left( {\hat R,\hat S} \right) = \left( {5,0.5} \right)$; (b) slight prolate deformation at $\left( {\hat R,\hat S} \right) = \left( {5,5} \right)$; (c) oblate deformation at $\left( {\hat R,\hat S} \right) = \left( {5,15} \right)$.}
	\label{fig:fig3}
\end{figure}

The above test does not involve the coupled electric field and two-phase flow model. To validate the electric field model adopted here, we simulate droplet deformation in an electric field. Initially, a droplet with radius $R$ is placed stationary in a cavity with a computational domain of $8R \times 8R$, where the electric potentials at the top and bottom walls are set to ${\varphi _H}$ and ${\varphi _L}$ (${\varphi _H} > {\varphi _L}$), respectively. In this setup, the droplet can deform into either a prolate shape (aligned with the electric field) or an oblate shape (perpendicular to the electric field). The deformation pattern is determined by the combined effects of conductivity and permittivity ratios. To show that the current model is capable of handling such a problem, our numerical results of simulation are compared with the theoretical solution and the available numerical data. For consistency, the parameters of the continuous phase and dispersed phase are the same as those in previous work. Figure \ref{fig:fig3} shows the streamlines and velocity vectors under different permittivity ratios ${\hat S}$ and conductivity ratios ${\hat R}$, where the permittivity ratio ${\hat R}$ is fixed at $5.0$. The trends in droplet deformation and the directions of the vortices inside and outside the droplet, under different conductivity ratios, are in good agreement with the theoretical predictions. 

Furthermore, to quantify the results, we present the deformation factor ${\hat D}$ of the droplet under various parameters in Table \ref{tab:tb1}. This factor is defined as $\hat D = {{\left( {\hat L - \hat H} \right)} \mathord{\left/
		{\vphantom {{\left( {\hat L - \hat H} \right)} {\left( {\hat L + \hat H} \right)}}} \right.
		\kern-\nulldelimiterspace} {\left( {\hat L + \hat H} \right)}}$, where ${\hat L}$ and ${\hat H}$ represent the lengths of the droplet parallel and perpendicular to the electric field, respectively. In Feng’s work \cite{Feng_PRSA1999}, the value of ${\hat D}$ for the deformation of a 2D droplet in a steady electric field can be predicted 
\begin{equation}\label{eq32}
	\hat D = \frac{{C{a_E}}}{{3{{\left( {1 + 2\hat R} \right)}^2}}}\left[ {{{\hat R}^2} + \hat R + 1 - 3\hat S} \right].
\end{equation}
From the Table \ref{tab:tb1}, it is evident that our current numerical results align closely with both existing theoretical and numerical benchmark data.
\begin{table}[H]
	\centering
	\caption{Quantitative comparison of the results from the present work with theoretical solutions and existing data.}
	\label{tab:tb1}
	\setlength{\tabcolsep}{5mm}{
		\begin{tabular}{lcccccc}
			\hline
			\hline
			\multirow{2}*{$\hat{R}$} & \multirow{2}*{$\hat{S}$} & \multirow{2}*{$Ca_{E}$} & \multicolumn{4}{c}{Deformation factor ($\hat{D}$)} \\
			\cline{4-7}
			{} & {} & {} & Present & Eq.(\ref{eq32}) & Ref. \cite{Luo_pof2020} & Ref. \cite{Xiong_JSC2025} \\
			\hline
			5     & 5    & 0.2 &  0.03347  &  0.02960  &  0.03150  &  0.03503 \\
			5     & 60   & 0.2 & -0.27384  & -0.27590  & -0.27510  & -0.26304 \\
			1     & 2    & 0.2 & -0.04956  & -0.05000  & -0.05390  & -0.04102 \\
			1.75  & 3.5  & 0.1 & -0.02126  & -0.02070  & -0.02000  & -0.01968 \\
			3.25  & 3.5  & 0.1 &  0.00826  &  0.00800  &  0.00900  &  0.00848 \\
			4.75  & 3.5  & 0.1 &  0.02017  &  0.01800  &  0.02090  &  0.02236 \\
			\hline
			\hline
		\end{tabular}
	}
\end{table}

\section{Results and Discussion}
\label{sec:results}
\subsection{Droplet formation under constant electric field}
\label{sec:4.1}

\begin{figure}[H]
	\centering
	
	\includegraphics[width=0.5\textwidth]{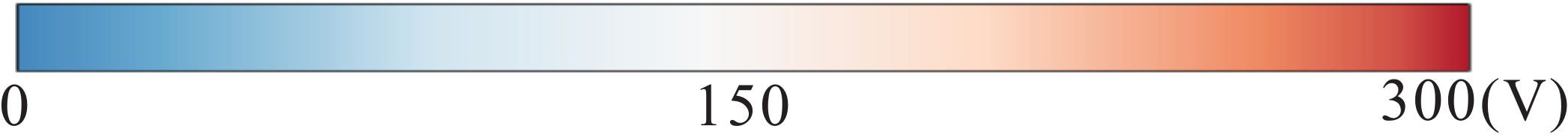}
	
	\vspace{0.2cm}
	
	\includegraphics[width=0.3\textwidth]{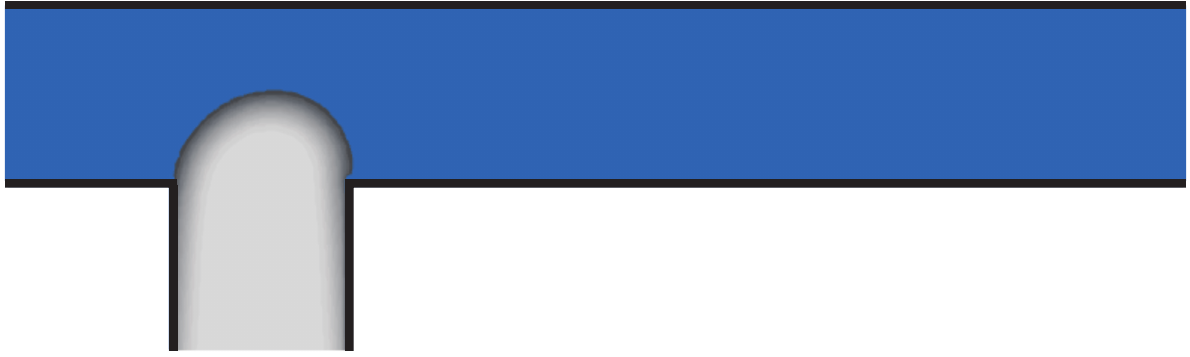}
	\put(-150, 35){\scriptsize \textit{(a)}}
	\put(-70, -5){\scriptsize \textit{(i)}}
	\quad
	\includegraphics[width=0.3\textwidth]{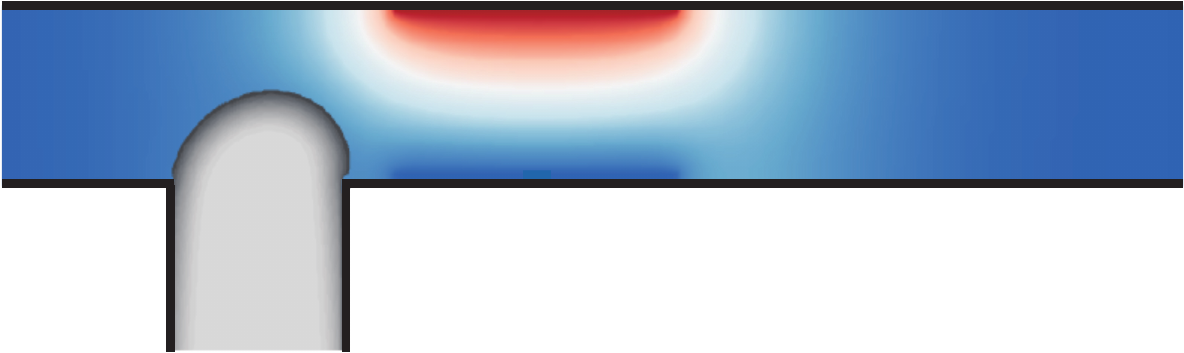}
	\put(-150, 35){\scriptsize \textit{(b)}}
	\put(-70, -5){\scriptsize \textit{(v)}}
	\quad
	\includegraphics[width=0.3\textwidth]{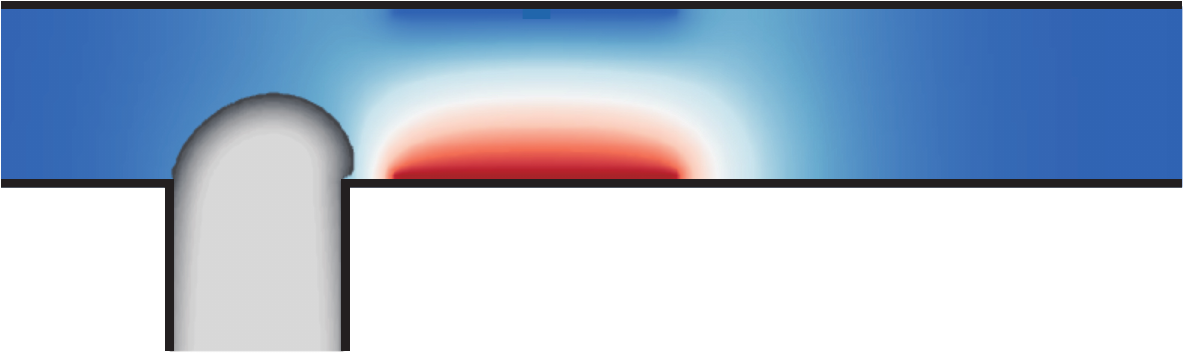}
	\put(-150, 35){\scriptsize \textit{(c)}}
	\put(-70, -5){\scriptsize \textit{(ix)}}
	
	\vspace{0.2cm}
	
	\includegraphics[width=0.3\textwidth]{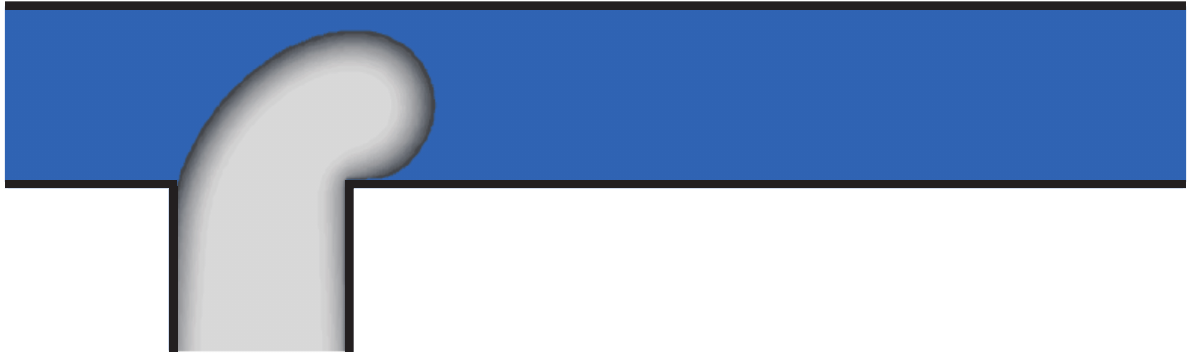}
	\put(-70, -5){\scriptsize \textit{(ii)}}
	\quad
	\includegraphics[width=0.3\textwidth]{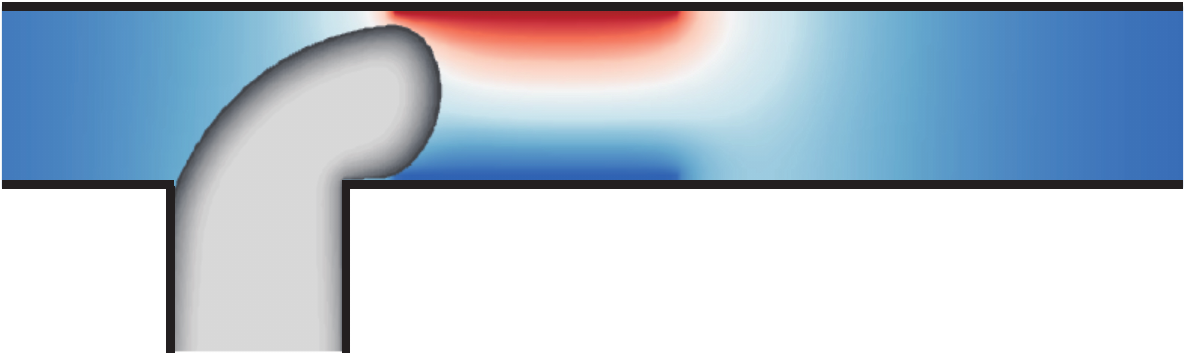}
	\put(-70, -5){\scriptsize \textit{(vi)}}
	\quad
	\includegraphics[width=0.3\textwidth]{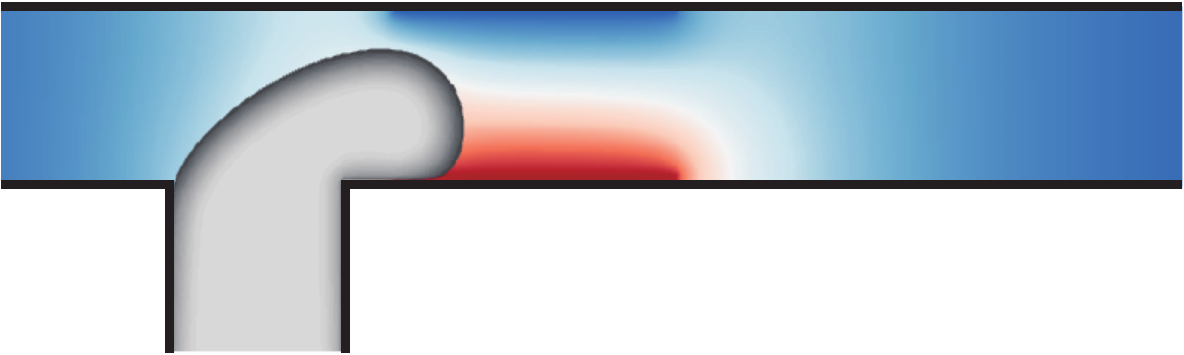}
	\put(-70, -5){\scriptsize \textit{(x)}}
	
	\vspace{0.2cm}
	
	\includegraphics[width=0.3\textwidth]{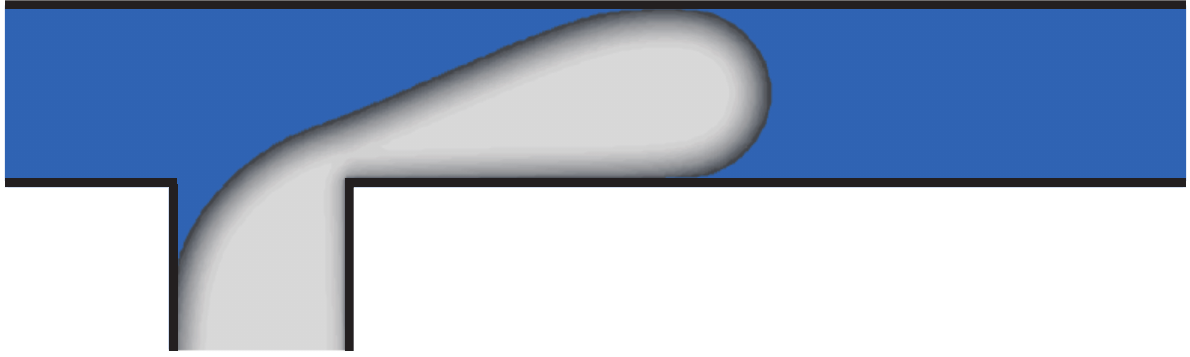}
	\put(-70, -5){\scriptsize \textit{(iii)}}
	\quad
	\includegraphics[width=0.3\textwidth]{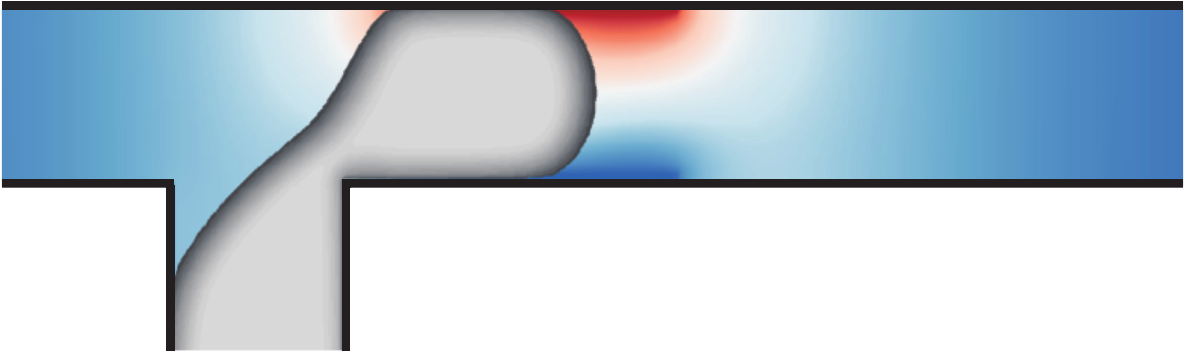}
	\put(-70, -5){\scriptsize \textit{(vii)}}
	\quad
	\includegraphics[width=0.3\textwidth]{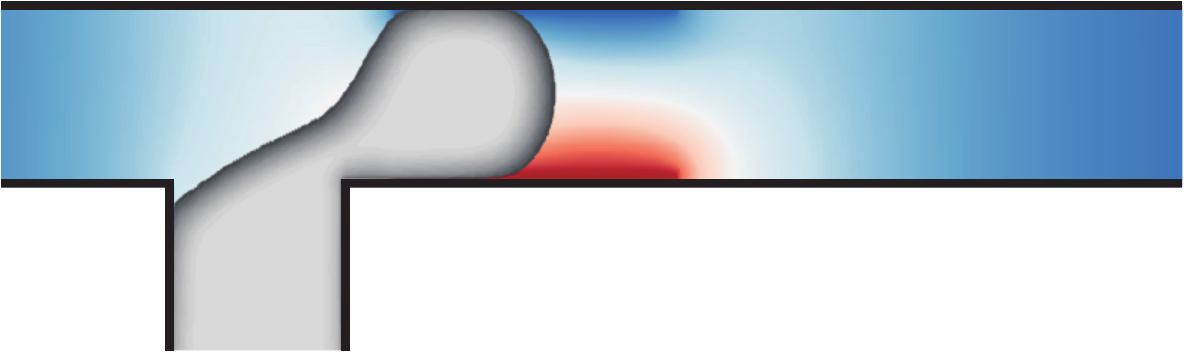}
	\put(-70, -5){\scriptsize \textit{(xi)}}
	
	\vspace{0.2cm}
	
	\includegraphics[width=0.3\textwidth]{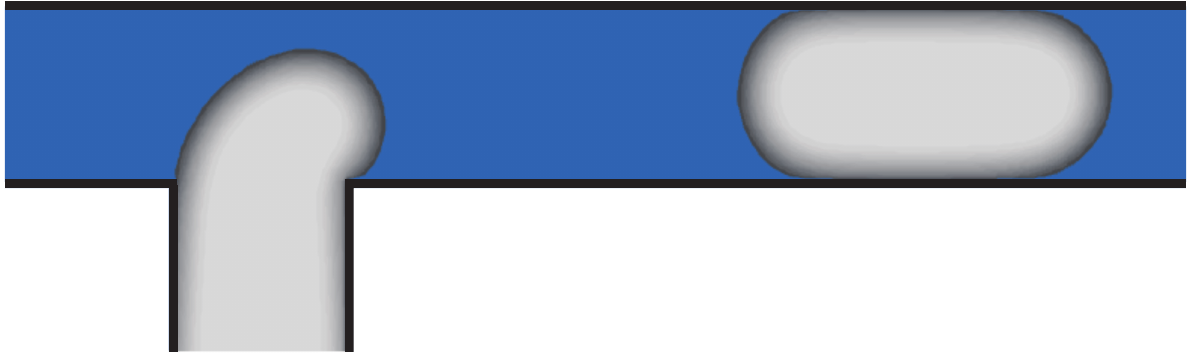}
	\put(-70, -5){\scriptsize \textit{(iv)}}
	\quad
	\includegraphics[width=0.3\textwidth]{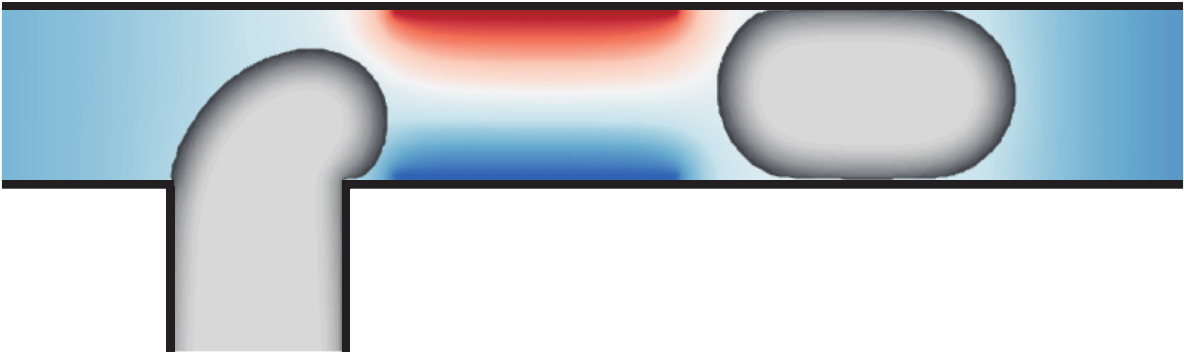}
	\put(-70, -5){\scriptsize \textit{(viii)}}
	\quad
	\includegraphics[width=0.3\textwidth]{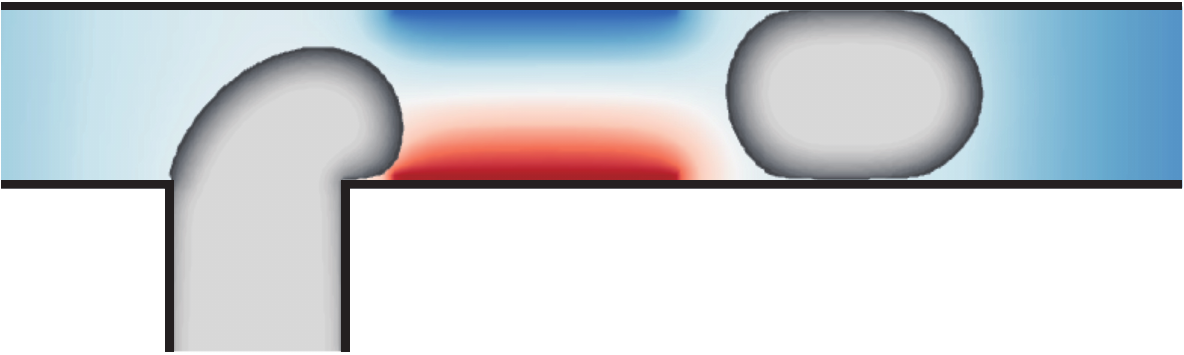}
	\put(-70, -5){\scriptsize \textit{(xii)}}
	
	\caption{Snapshots of droplet formation in a T-junction under three different electric field conditions: (a) no electric field, (b) top-biased CEF, and (c) bottom-biased CEF.}
	\label{fig4}
\end{figure}

We first compare the droplet formation process under three conditions, namely no electric field, top-biased CEF, and bottom-biased CEF, at a fixed flow-rate ratio of ${Q_r}={Q_d}/{Q_c}=0.4$ (figure \ref{fig4}). As widely reported, droplet formation in a T-junction can be divided into two successive stages, namely filling and squeezing \cite{Garstecki_Labonchip2006}. In the absence of an electric field, the dispersed phase progressively penetrates into the main channel while being advected downstream by the hydrodynamic shear imposed by the continuous phase. Once the dispersed phase spans the channel width, the system enters the squeezing stage. The resulting blockage causes a rapid pressure build-up upstream of the droplet, which compresses the liquid–liquid interface toward the T-junction. A neck is then formed and progressively thinned until rupture occurs, ultimately leading to droplet breakup. When an electric field is applied, the electric stress acting on the interface induces a transverse migration of the dispersed phase toward the electrode with the higher potential, thereby markedly modifying the interfacial topology, as shown in figure \ref{fig4}(vi) and (x). Under top-biased CEF, the dispersed phase is pulled upward, which accelerates channel blockage. By contrast, under bottom-biased CEF, the dispersed phase tends to adhere to the lower wall, delaying the vertical expansion required to block the channel. These results indicate that the spatial distribution of the electric field strongly affects the interfacial evolution during the filling stage. During the squeezing stage, the dispersed phase continues to deform under the combined action of hydrodynamic and electric stresses \cite{Singh_JFM2020}. Nevertheless, in both electric-field configurations, the dispersed phase eventually blocks the channel completely (figure \ref{fig4}(vii) and (xi)), after which the neck undergoes further thinning and finally ruptures to form a droplet.

\begin{figure}
	\centerline{\includegraphics[width=0.6\textwidth]{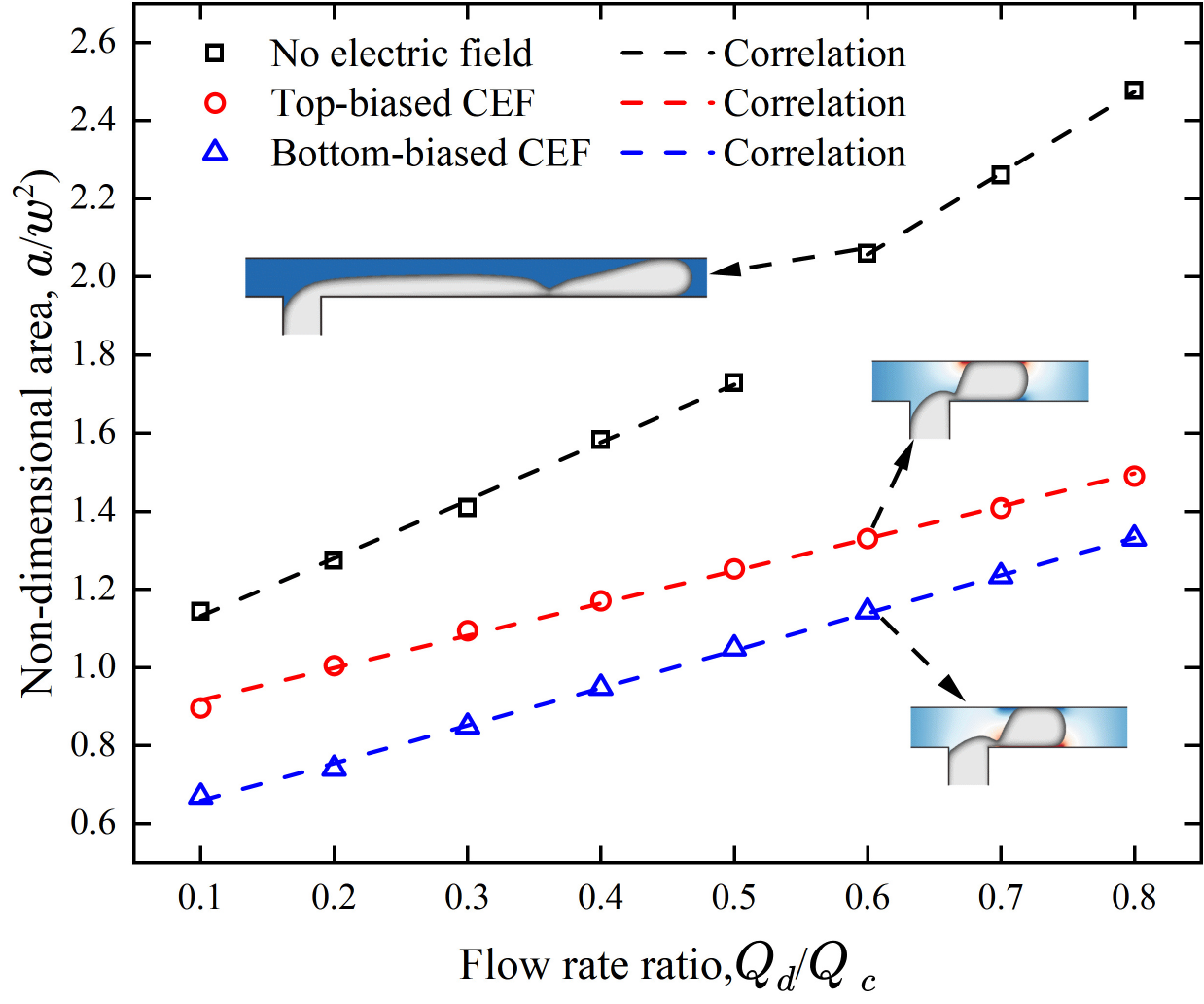}} 
	\caption{The effect of the flow rate ratio on droplet size under different electric field conditions. LBM simulation results (hollow symbols) and linear fitting (dashed lines).} 
	\label{fig5} 
\end{figure}

To quantitatively assess the effect of the electric field on droplet formation, figure \ref{fig5} compares the droplet area for ${Q_r}$ ranging from $0.1$ to $0.8$. Consistent with previous numerical and experimental studies \cite{Singh_JFM2020,Demench_JFM2008,Garstecki_Labonchip2006}, the droplet area increases approximately linearly with the flow-rate ratio in the squeezing regime. However, under the no-electric-field condition at ${Q_r}=0.6$, the dispersed phase ruptures downstream of the junction, causing the droplet area to deviate from the linear trend characteristic of squeezing breakup. In contrast, for both top-biased and bottom-biased CEF cases, droplet breakup consistently occurs at the channel junction, and the resulting droplet area remains approximately linear with ${Q_r}$. In addition, the droplets generated under bottom-biased CEF are smaller than those formed under top-biased CEF.

\subsection{Droplet formation under pulsed electric field}
\label{sec:4.2}
\subsubsection{Impact of PEF duration}
\label{sec:4.2.1}

\begin{figure}[H]
	\centering
	
	\includegraphics[width=0.5\textwidth]{fig4z-eps-converted-to.pdf}
	
	\vspace{0.2cm}
	
	\includegraphics[width=0.3\textwidth]{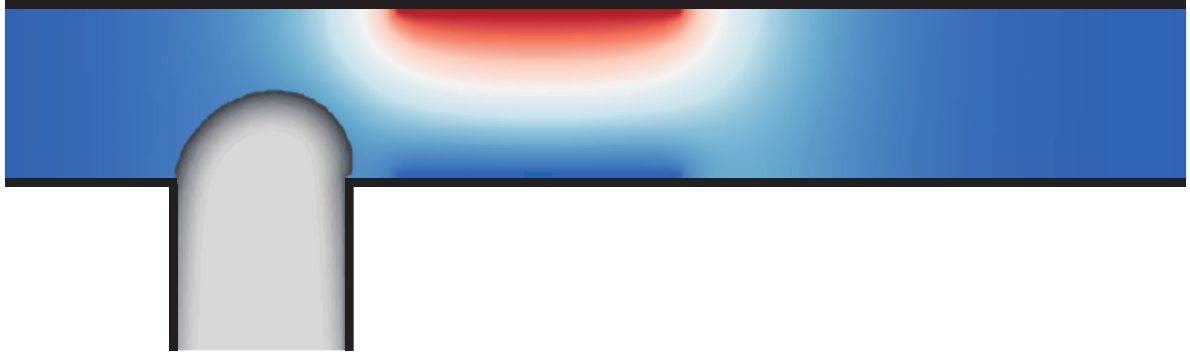}
	\put(-150, 35){\scriptsize \textit{(a)}}
	\put(-70, -5){\scriptsize \textit{(i)}}
	\quad
	\includegraphics[width=0.3\textwidth]{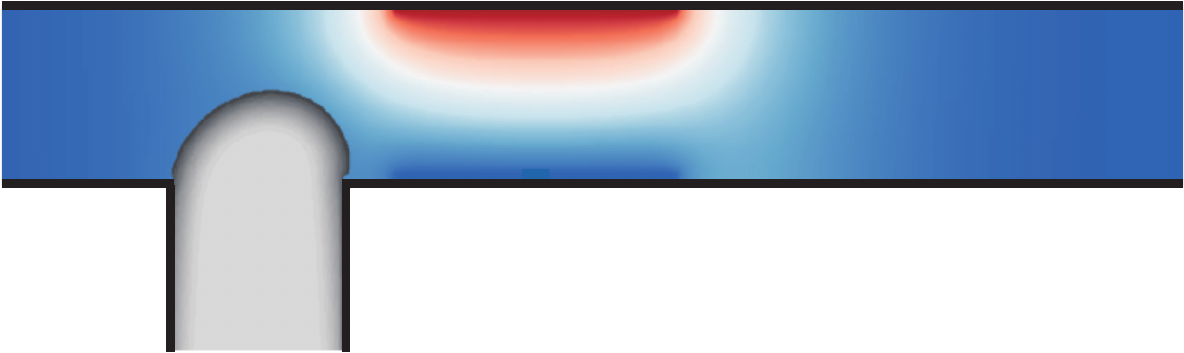}
	\put(-150, 35){\scriptsize \textit{(b)}}
	\put(-70, -5){\scriptsize \textit{(v)}}
	\quad
	\includegraphics[width=0.3\textwidth]{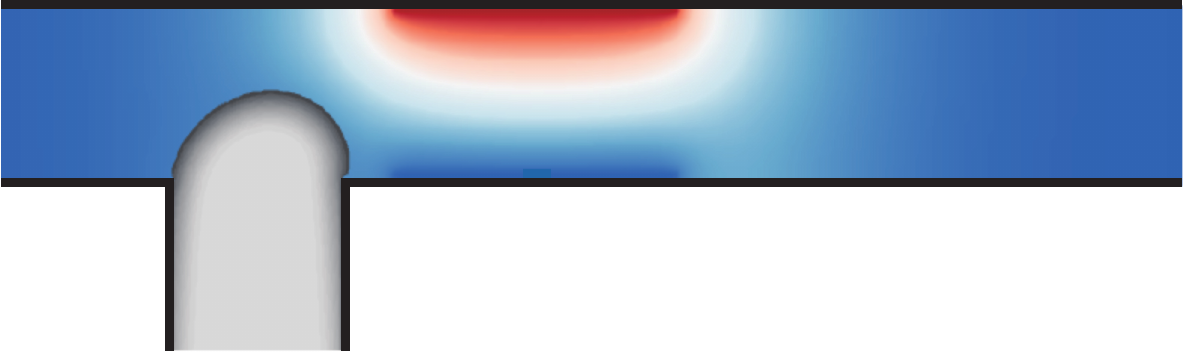}
	\put(-150, 35){\scriptsize \textit{(c)}}
	\put(-70, -5){\scriptsize \textit{(ix)}}
	
	\vspace{0.2cm}
	
	\includegraphics[width=0.3\textwidth]{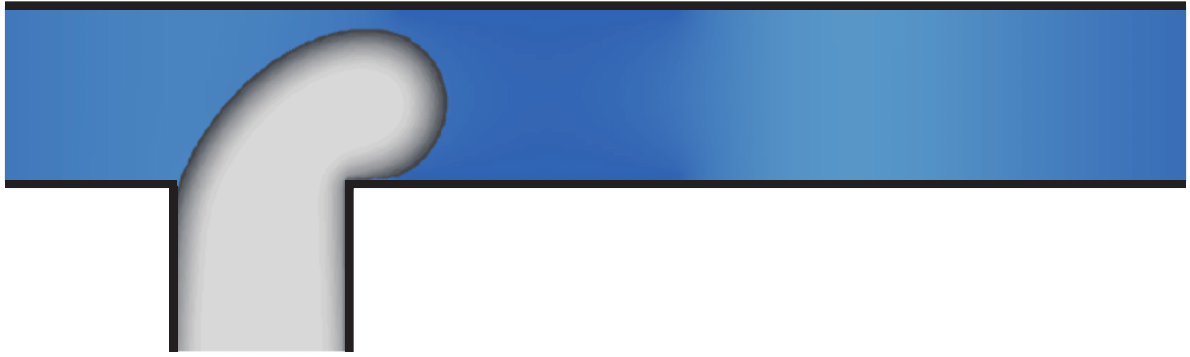}
	\put(-70, -5){\scriptsize \textit{(ii)}}
	\quad
	\includegraphics[width=0.3\textwidth]{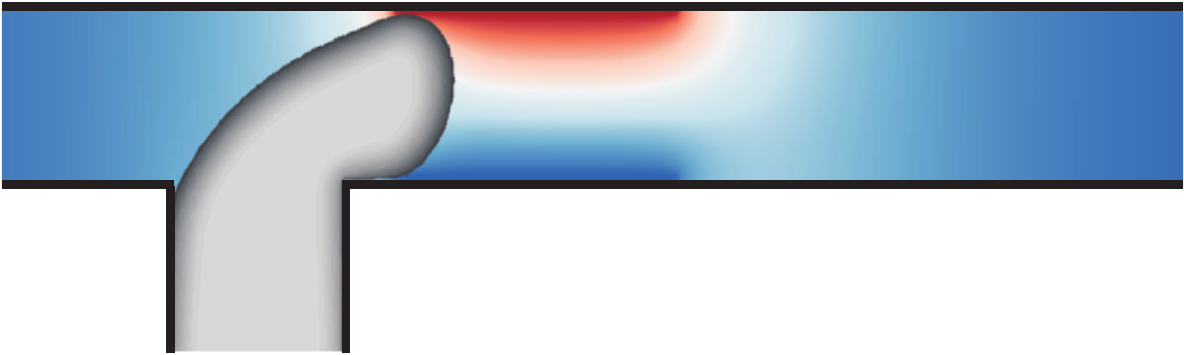}
	\put(-70, -5){\scriptsize \textit{(vi)}}
	\quad
	\includegraphics[width=0.3\textwidth]{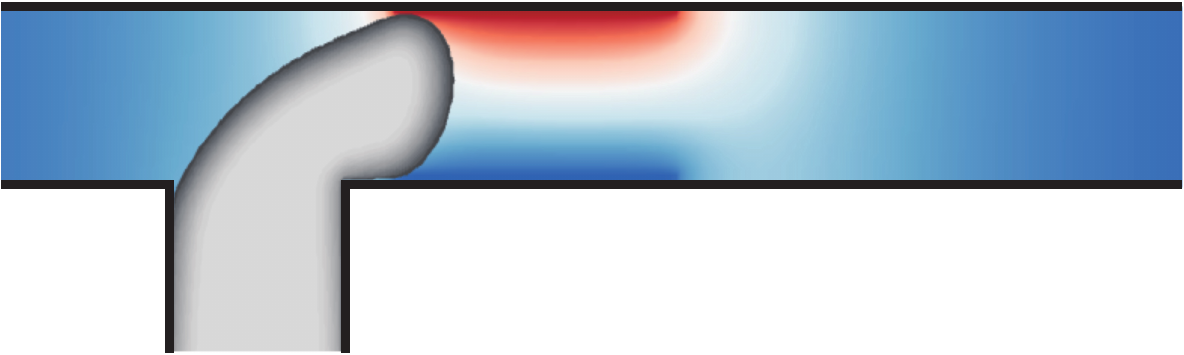}
	\put(-70, -5){\scriptsize \textit{(x)}}
	
	\vspace{0.2cm}
	
	\includegraphics[width=0.3\textwidth]{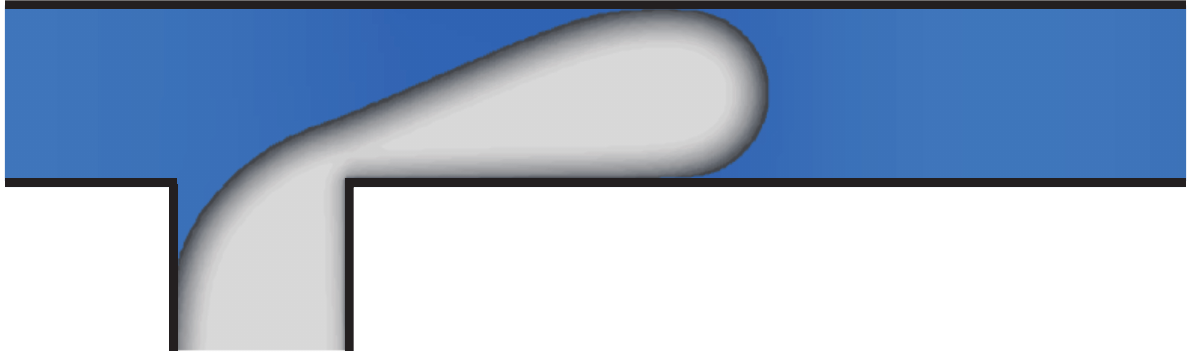}
	\put(-70, -5){\scriptsize \textit{(iii)}}
	\quad
	\includegraphics[width=0.3\textwidth]{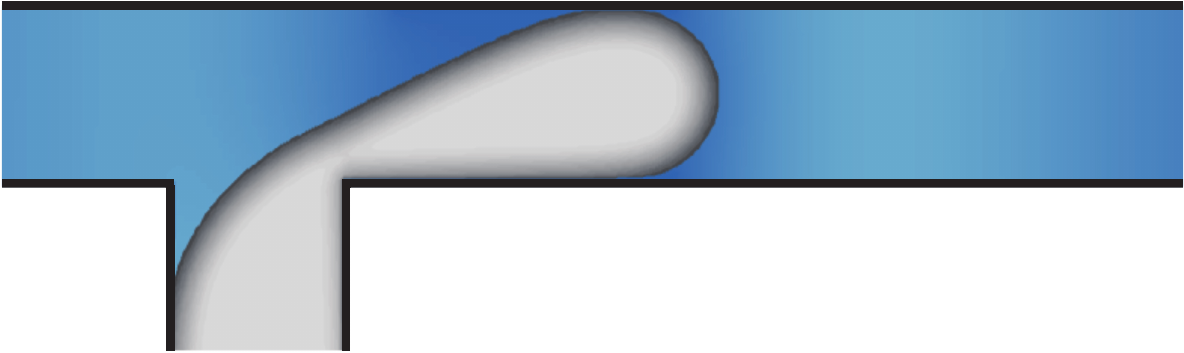}
	\put(-70, -5){\scriptsize \textit{(vii)}}
	\quad
	\includegraphics[width=0.3\textwidth]{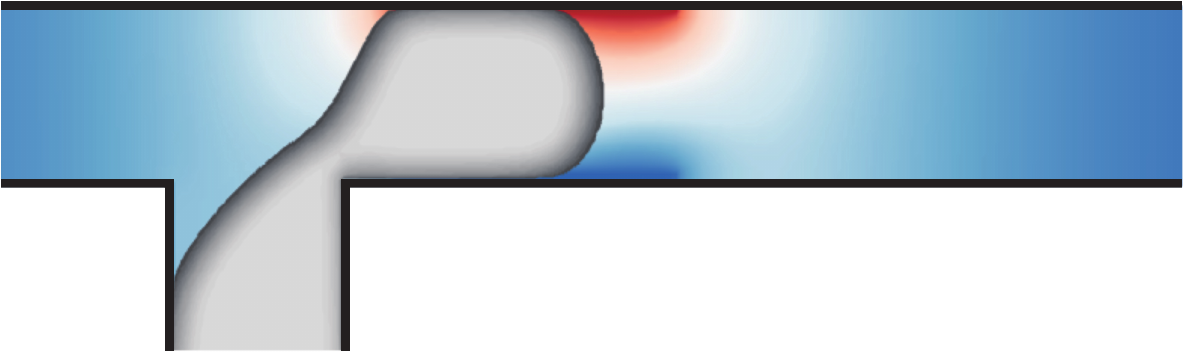}
	\put(-70, -5){\scriptsize \textit{(xi)}}
	
	\vspace{0.2cm}
	
	\includegraphics[width=0.3\textwidth]{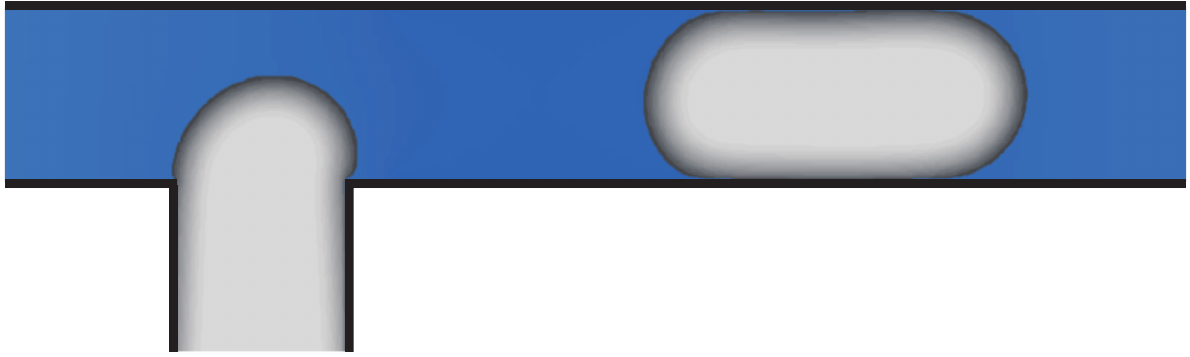}
	\put(-70, -5){\scriptsize \textit{(iv)}}
	\quad
	\includegraphics[width=0.3\textwidth]{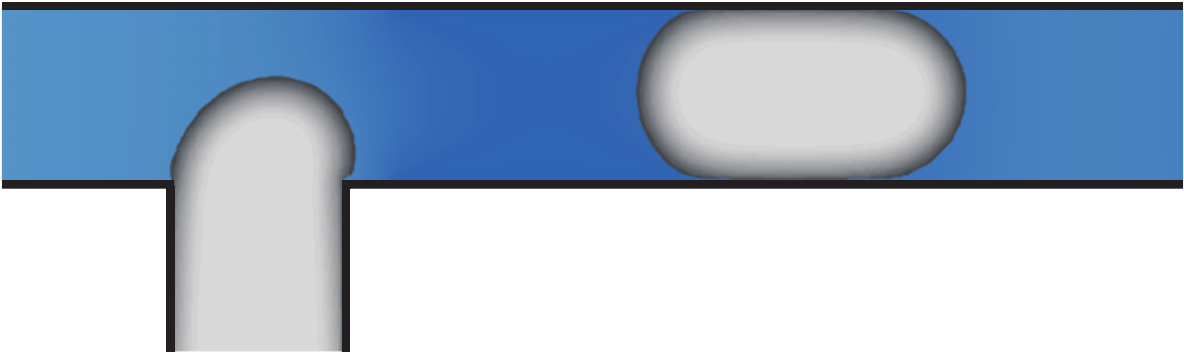}
	\put(-70, -5){\scriptsize \textit{(viii)}}
	\quad
	\includegraphics[width=0.3\textwidth]{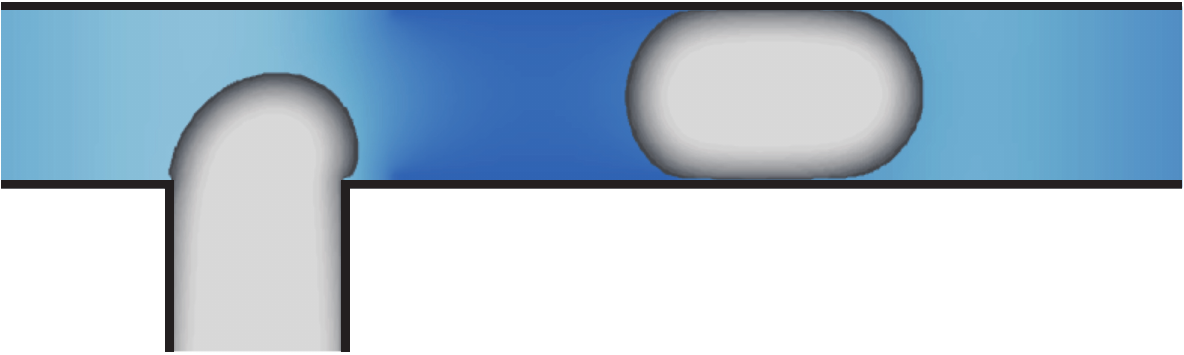}
	\put(-70, -5){\scriptsize \textit{(xii)}}
	
	\caption{Snapshots of droplet formation under top-biased PEF at different pulse durations: (a) ${t_d} = 0.1$, (b) ${t_d} = 0.4$, and (c) ${t_d} = 0.7$.}
	\label{fig6}
\end{figure}

\begin{figure}[H]
	\centering
	
	\includegraphics[width=0.5\textwidth]{fig4z-eps-converted-to.pdf}
	
	\vspace{0.2cm}
	
	\includegraphics[width=0.3\textwidth]{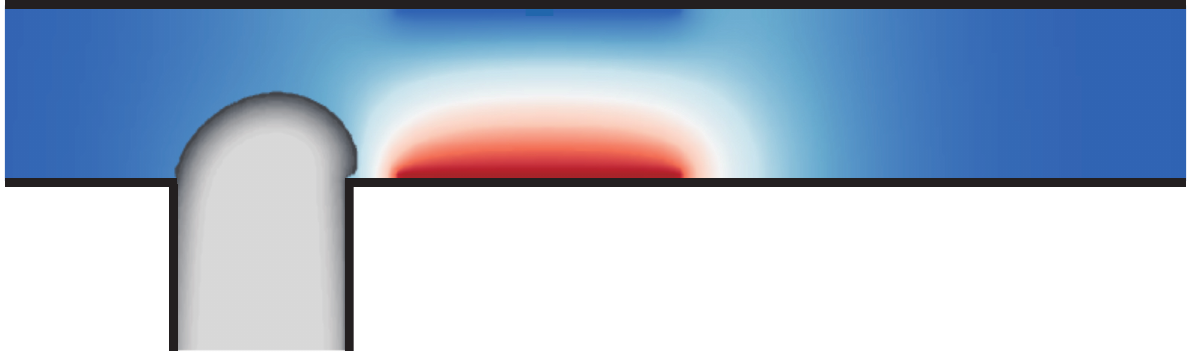}
	\put(-150, 35){\scriptsize \textit{(a)}}
	\put(-70, -5){\scriptsize \textit{(i)}}
	\quad
	\includegraphics[width=0.3\textwidth]{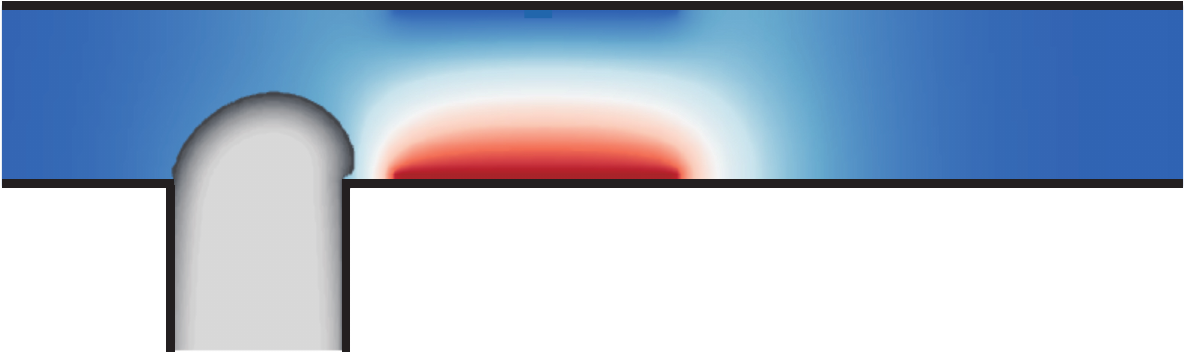}
	\put(-150, 35){\scriptsize \textit{(b)}}
	\put(-70, -5){\scriptsize \textit{(v)}}
	\quad
	\includegraphics[width=0.3\textwidth]{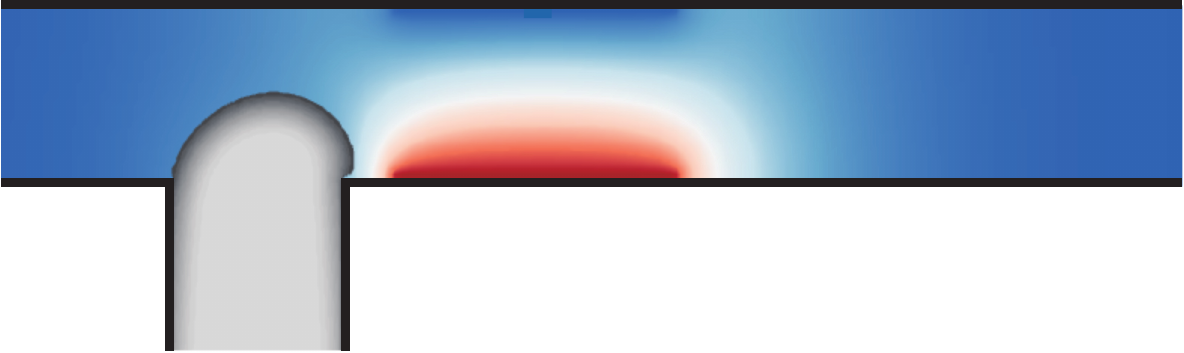}
	\put(-150, 35){\scriptsize \textit{(c)}}
	\put(-70, -5){\scriptsize \textit{(ix)}}
	
	\vspace{0.2cm}
	
	\includegraphics[width=0.3\textwidth]{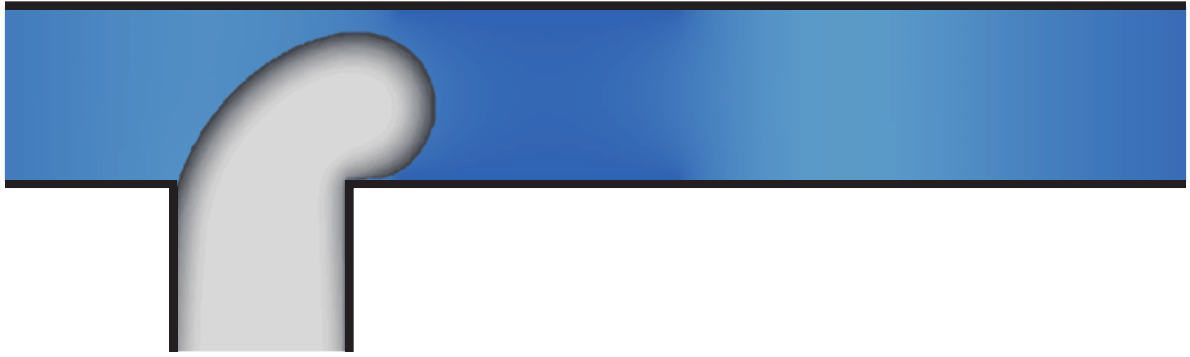}
	\put(-70, -5){\scriptsize \textit{(ii)}}
	\quad
	\includegraphics[width=0.3\textwidth]{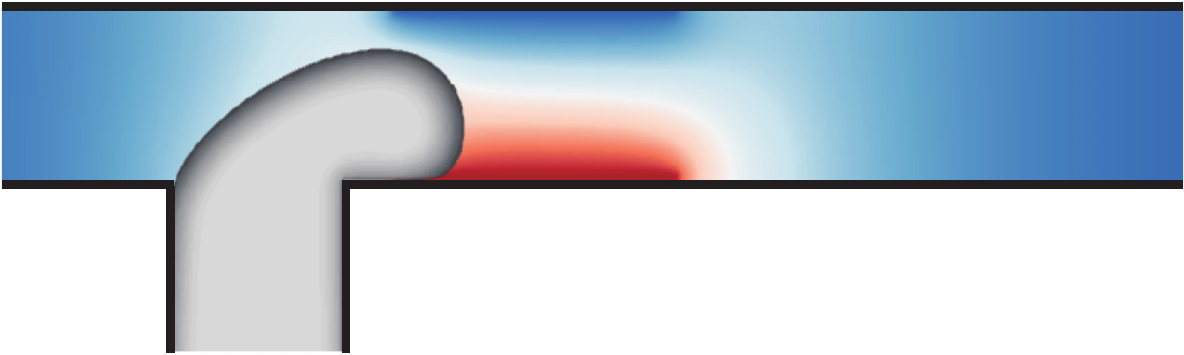}
	\put(-70, -5){\scriptsize \textit{(vi)}}
	\quad
	\includegraphics[width=0.3\textwidth]{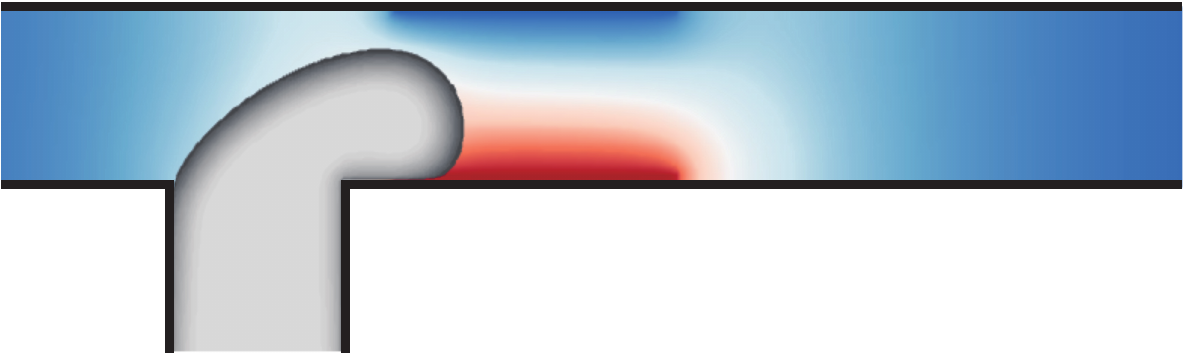}
	\put(-70, -5){\scriptsize \textit{(x)}}
	
	\vspace{0.2cm}
	
	\includegraphics[width=0.3\textwidth]{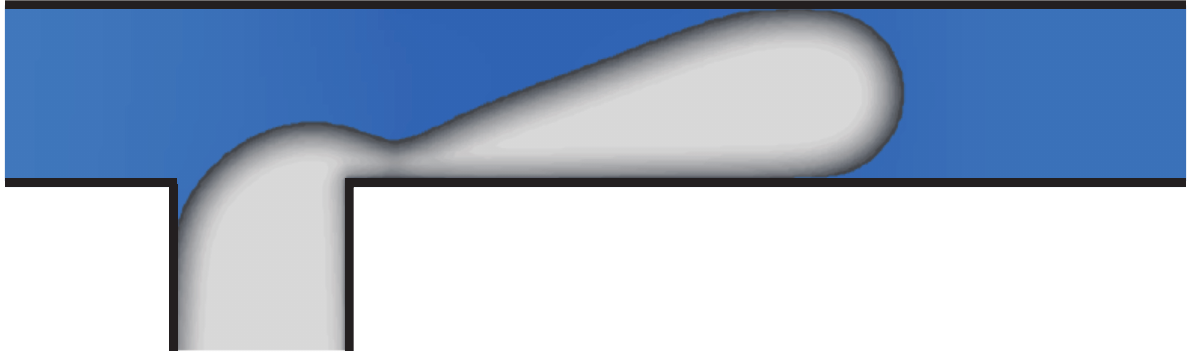}
	\put(-70, -5){\scriptsize \textit{(iii)}}
	\quad
	\includegraphics[width=0.3\textwidth]{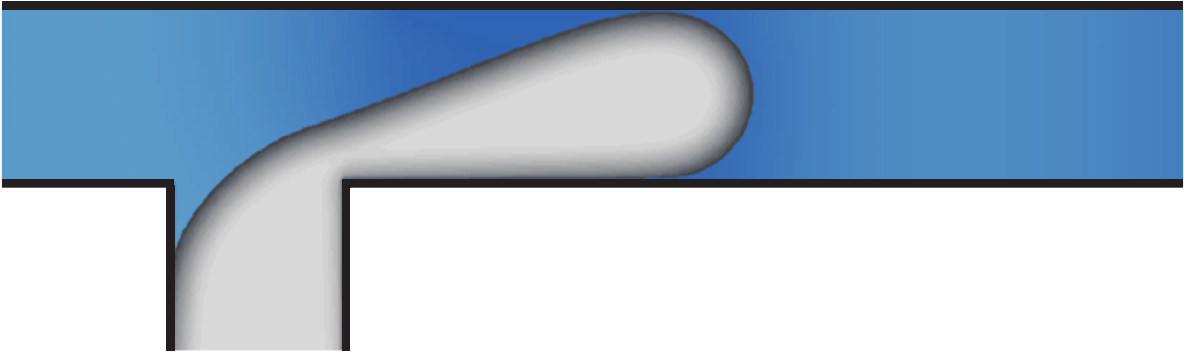}
	\put(-70, -5){\scriptsize \textit{(vii)}}
	\quad
	\includegraphics[width=0.3\textwidth]{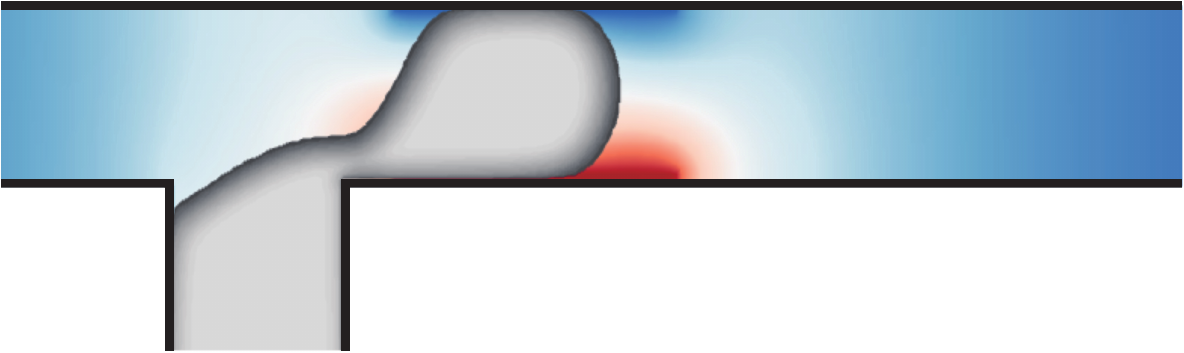}
	\put(-70, -5){\scriptsize \textit{(xi)}}
	
	\vspace{0.2cm}
	
	\includegraphics[width=0.3\textwidth]{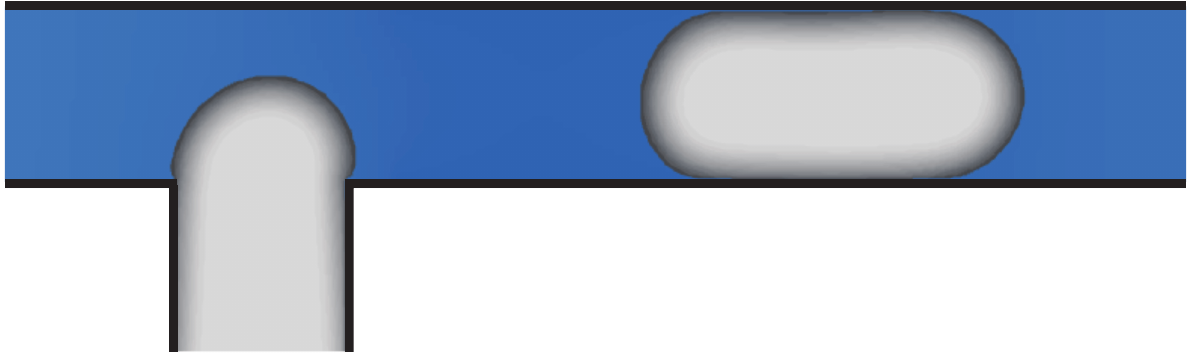}
	\put(-70, -5){\scriptsize \textit{(iv)}}
	\quad
	\includegraphics[width=0.3\textwidth]{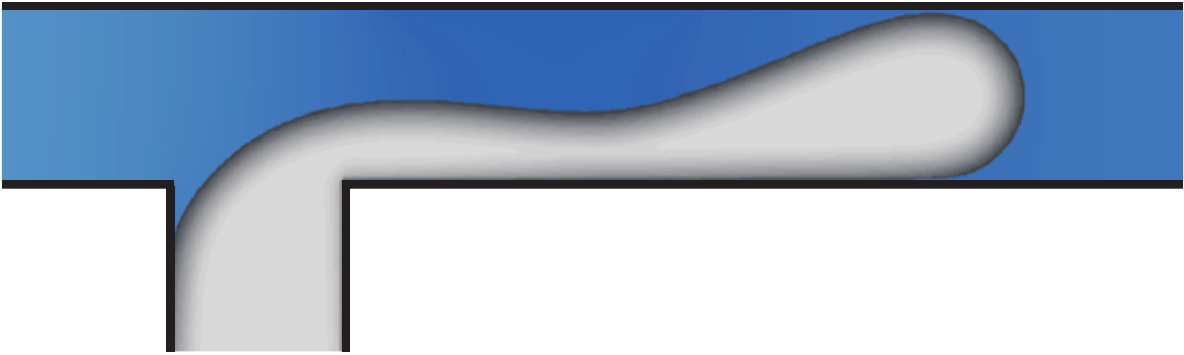}
	\put(-70, -5){\scriptsize \textit{(viii)}}
	\quad
	\includegraphics[width=0.3\textwidth]{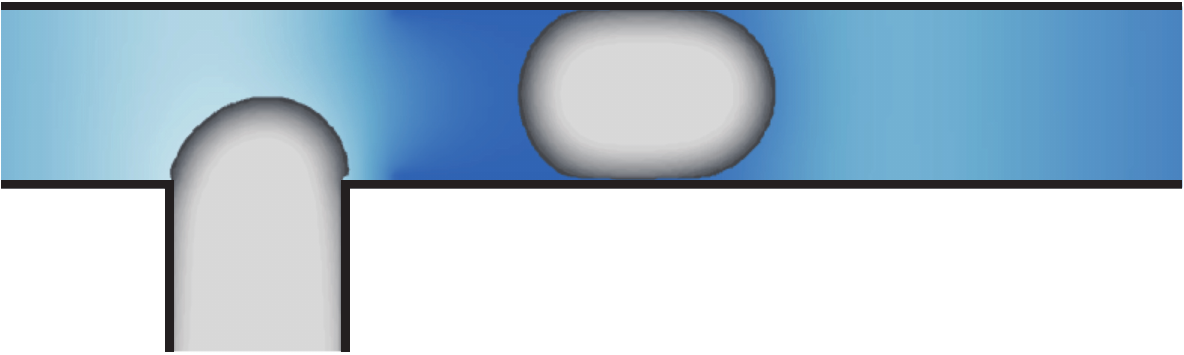}
	\put(-70, -5){\scriptsize \textit{(xii)}}
	
	\caption{Snapshots of droplet formation under bottom-biased PEF at different pulse durations: (a) ${t_d} = 0.1$, (b) ${t_d} = 0.4$, and (c) ${t_d} = 0.7$.}
	\label{fig7}
\end{figure}

In the previous section, we discussed the effect of CEF on droplet formation. In contrast, PEF provides an additional means of regulating the droplet formation process through temporal modulation of the applied electric field. This section focuses on the effect of the pulse duration, ${T_d}$, on droplet formation at ${Q_r}=0.4$, where breakup occurs at the channel junction in the squeezing regime, as shown in the previous section. In the present study, a single electric pulse is applied at the onset of droplet formation and is switched off at $t={T_d}$. For this fixed ${Q_r}$, the droplet formation time in the absence of an electric field is denoted by ${T_t}$, and the corresponding droplet area is denoted by ${a_0}$. The pulse duration is then normalized as ${t_d}={T_d}/{T_t}$. Figures \ref{fig6} and \ref{fig7} compare the droplet formation processes under top-biased and bottom-biased PEFs at ${t_d}=0.1$, $0.4$, and $0.7$. These three cases correspond to pulse actuation during the early filling stage, over nearly the entire filling stage, and from the filling stage into the squeezing stage, respectively.

For the top-biased PEF, the droplet area decreases monotonically with increasing ${t_d}$. When ${t_d}=0.1$, the behaviour of the dispersed phase remains close to that observed in the absence of an electric field, since the short pulse duration does not significantly alter the liquid-liquid interface (figure \ref{fig6}(a)). At ${t_d}=0.4$, the interface is drawn towards the top electrode under the action of the electric stress (figure \ref{fig6}(vi)). After the electric field is switched off, the dispersed phase continues to move downstream under the hydrodynamic stress imposed by the continuous phase. However, the longer pulse duration promotes earlier channel blockage and earlier neck formation, thereby leading to smaller droplets. When ${t_d}=0.7$, the electric field remains active during the squeezing stage and helps maintain the liquid-liquid interface near the top electrode (figure \ref{fig6}(x) and (xi)). Under the combined action of electric and hydrodynamic stresses, the interface evolves rapidly from a convex to a concave shape, which accelerates neck thinning and rupture. Consequently, the droplets formed in this case are the smallest.

In contrast, the bottom-biased PEF exhibits a different trend. Although the behaviour at ${t_d}=0.1$ is broadly similar to that under the top-biased PEF, the droplet does not break up within the computational domain when ${t_d}=0.4$. During the filling stage, the liquid-liquid interface deforms towards the bottom electrode. After the electric field is removed, the dispersed phase continues to fill the main channel and subsequently enters the squeezing stage. During this process, the neck shifts downstream of the T-junction (figure \ref{fig7}(vii) and (viii)), indicating that the applied pulse delays droplet breakup. However, when ${t_d}$ is increased to $0.7$, the dispersed phase continues to deform towards the high-potential region under the action of the electric field (figure \ref{fig7}(x)) and then stretches along the electric field lines until the channel is blocked. Once blockage occurs, the liquid-liquid interface has already developed a concave shape (figure \ref{fig7}(xi)), which promotes rapid neck thinning and rupture, ultimately producing smaller droplets.

\begin{figure}
	\centerline{\includegraphics[width=0.6\textwidth]{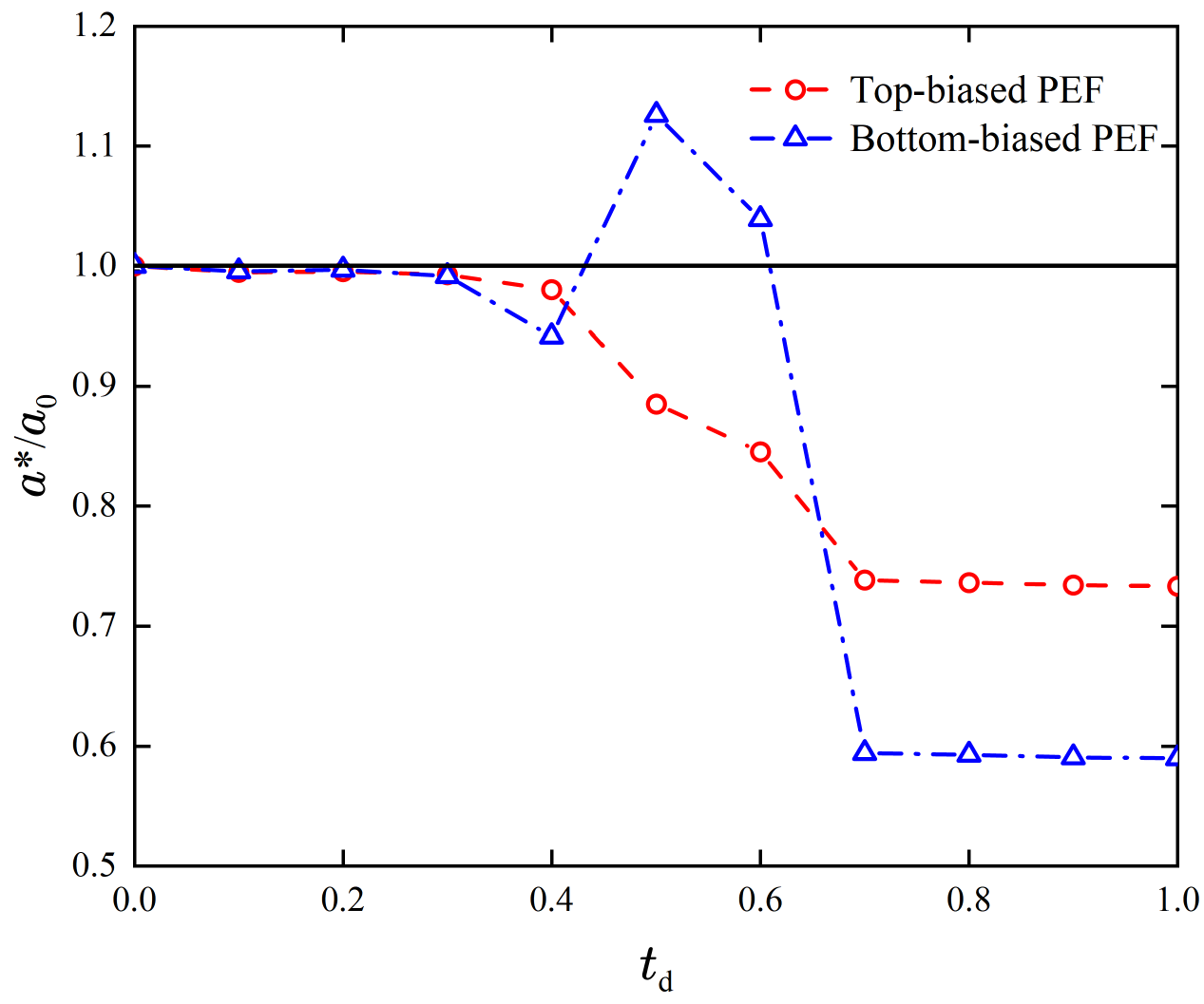}} 
	\caption{The relationship between the droplet's non-dimensional area (normalized by the droplet formation area ${a_0}$ under no electric field conditions) and the pulse duration.} 
	\label{fig8} 
\end{figure}

Figure \ref{fig8} quantitatively compares the droplet area as a function of ${t_d}$ under both electric-field configurations, normalized by the droplet area obtained in the absence of an electric field. For the top-biased PEF, the droplet area remains nearly unchanged when ${t_d}<0.3$, indicating that a short pulse acting only during the early filling stage has little influence on the subsequent breakup process. As ${t_d}$ increases further, the droplet area decreases monotonically and reaches a minimum at ${t_d}=0.7$, reflecting the increasing influence of the electric field on the later filling stage and the squeezing stage. By contrast, under the bottom-biased PEF, the droplet area exhibits a non-monotonic dependence on ${t_d}$. In the range $0.4<{t_d}\leq0.6$, the droplet area becomes larger than that obtained without an electric field, consistent with the downstream shift of the neck and the associated delay in breakup discussed above. This delayed-breakup behaviour is also consistent with that reported in microchannel experiments under alternating electric actuation \cite{Teo_MN2020}. When ${t_d}$ is increased further, however, the electric field persists into the squeezing stage and accelerates neck thinning and rupture, causing the droplet area to decrease sharply.

\subsubsection{Impact of PEF onset time}
\label{sec:4.2.2}

To further elucidate the role of PEF in controlling droplet formation, this section focuses on the onset time of the pulse, ${T_o}$. The flow-rate ratio is fixed at ${Q_r}=0.6$, for which droplet breakup occurs downstream of the T-junction in the absence of an electric field, as discussed in Sec.~\ref{sec:4.1}. Compared with the case of ${Q_r}=0.4$, where breakup already occurs at the junction, this condition allows the influence of pulse onset time on the breakup location and necking dynamics to be more clearly identified. As in Sec.~\ref{sec:4.2.1}, ${T_t}$ and ${a_0}$ denote the droplet formation time and droplet area in the absence of an electric field, respectively, and the normalized onset time is defined as ${t_o}={T_o}/{T_t}$. Unless otherwise stated, the pulse duration is fixed at ${t_d}=0.3$ throughout this subsection.

\begin{figure}[H]
	\centering
	
	\includegraphics[width=0.5\textwidth]{fig4z-eps-converted-to.pdf}
	
	\vspace{0.2cm}
	
	\includegraphics[width=0.3\textwidth]{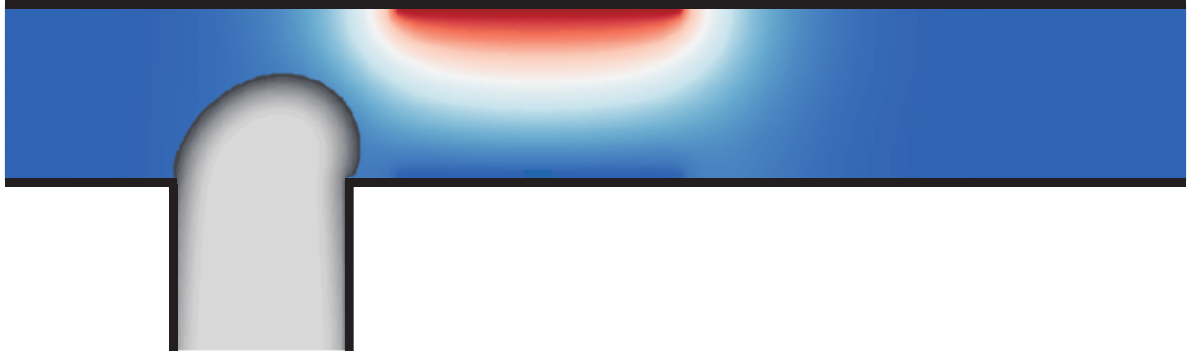}
	\put(-150, 35){\scriptsize \textit{(a)}}
	\put(-70, -5){\scriptsize \textit{(i)}}
	\quad
	\includegraphics[width=0.3\textwidth]{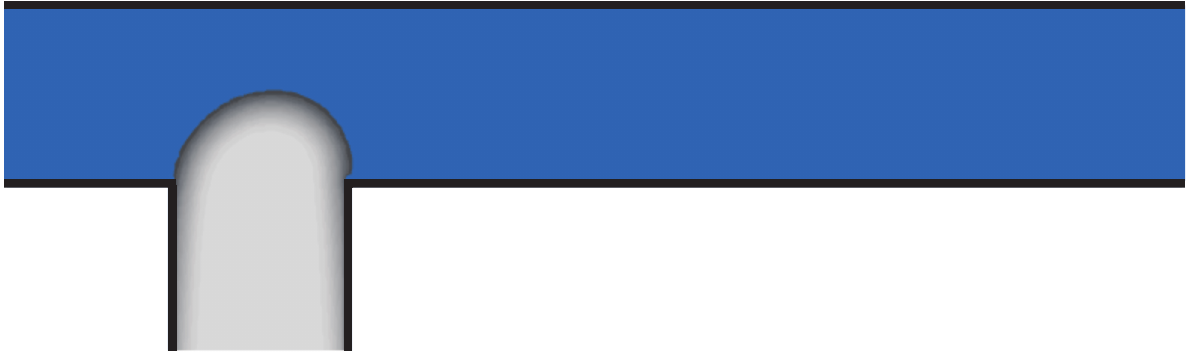}
	\put(-150, 35){\scriptsize \textit{(b)}}
	\put(-70, -5){\scriptsize \textit{(v)}}
	\quad
	\includegraphics[width=0.3\textwidth]{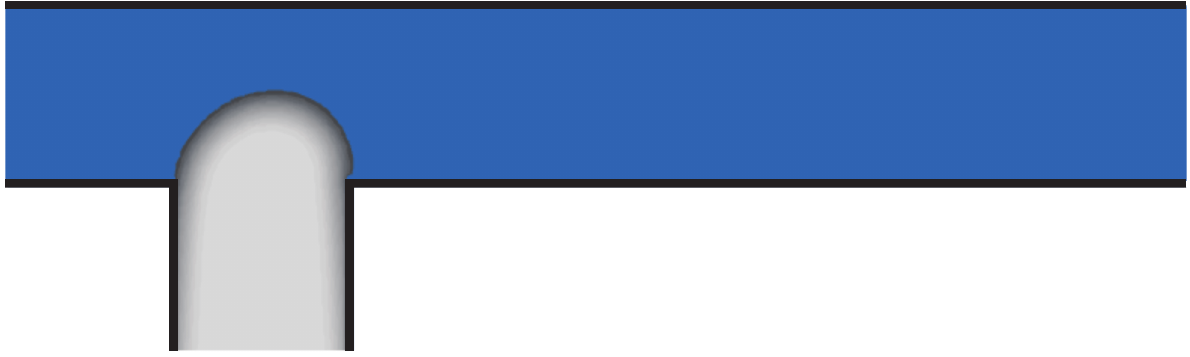}
	\put(-150, 35){\scriptsize \textit{(c)}}
	\put(-70, -5){\scriptsize \textit{(ix)}}
	
	\vspace{0.2cm}
	
	\includegraphics[width=0.3\textwidth]{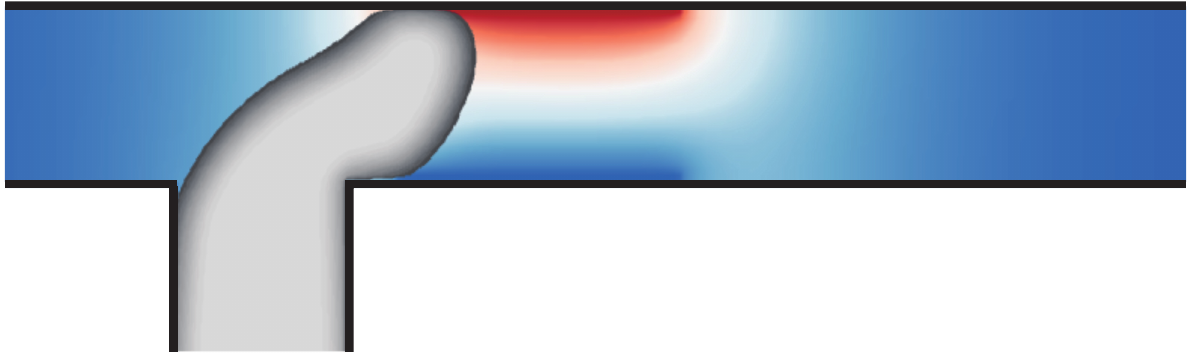}
	\put(-70, -5){\scriptsize \textit{(ii)}}
	\quad
	\includegraphics[width=0.3\textwidth]{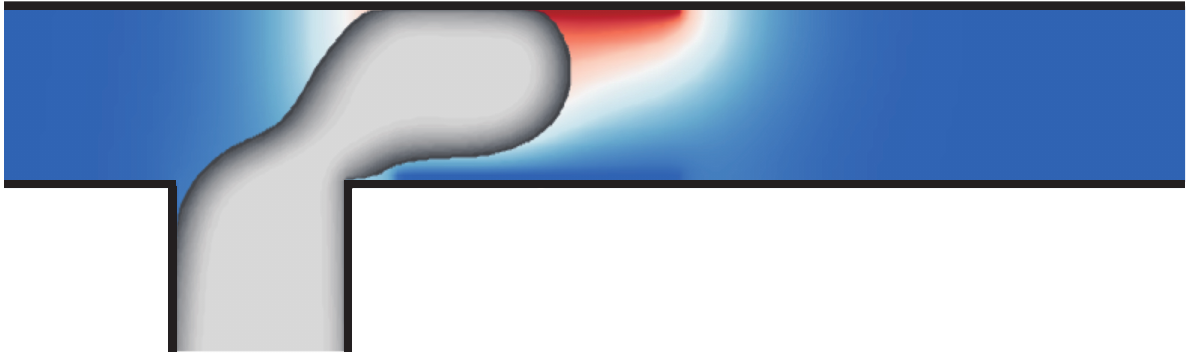}
	\put(-70, -5){\scriptsize \textit{(vi)}}
	\quad
	\includegraphics[width=0.3\textwidth]{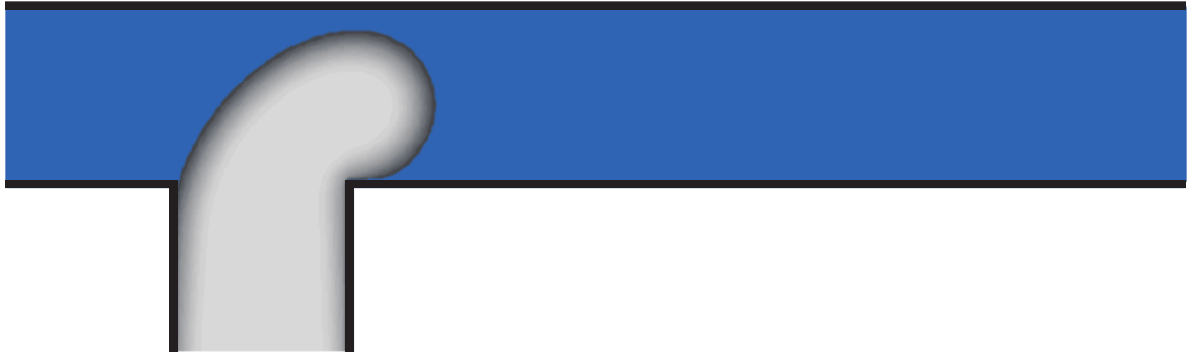}
	\put(-70, -5){\scriptsize \textit{(x)}}
	
	\vspace{0.2cm}
	
	\includegraphics[width=0.3\textwidth]{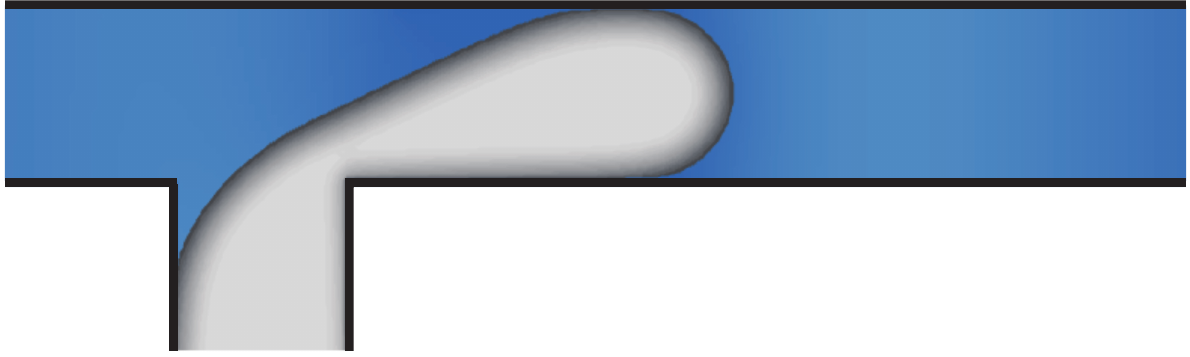}
	\put(-70, -5){\scriptsize \textit{(iii)}}
	\quad
	\includegraphics[width=0.3\textwidth]{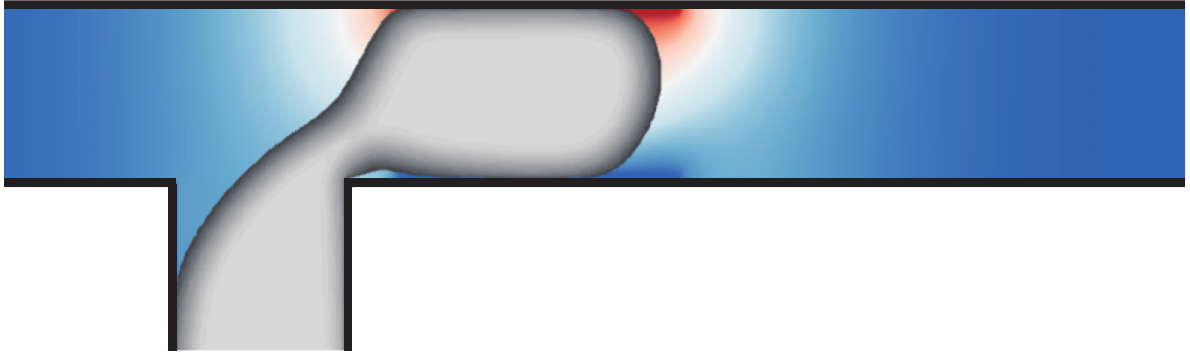}
	\put(-70, -5){\scriptsize \textit{(vii)}}
	\quad
	\includegraphics[width=0.3\textwidth]{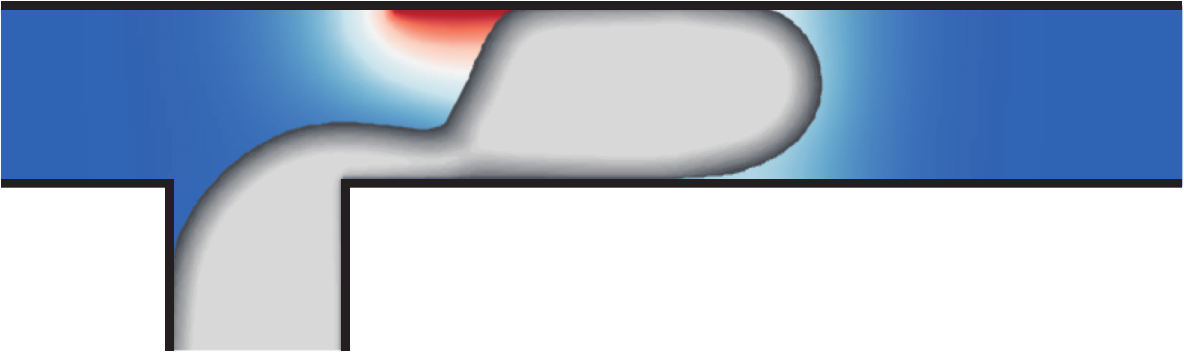}
	\put(-70, -5){\scriptsize \textit{(xi)}}
	
	\vspace{0.2cm}
	
	\includegraphics[width=0.3\textwidth]{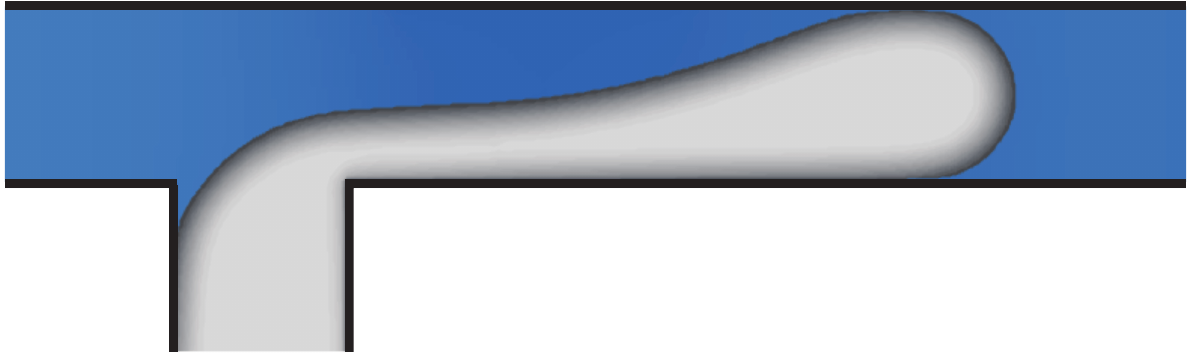}
	\put(-70, -5){\scriptsize \textit{(iv)}}
	\quad
	\includegraphics[width=0.3\textwidth]{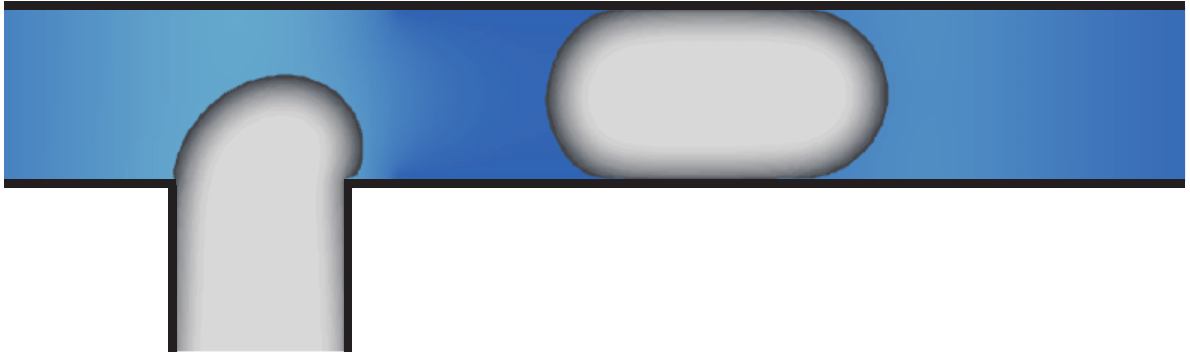}
	\put(-70, -5){\scriptsize \textit{(viii)}}
	\quad
	\includegraphics[width=0.3\textwidth]{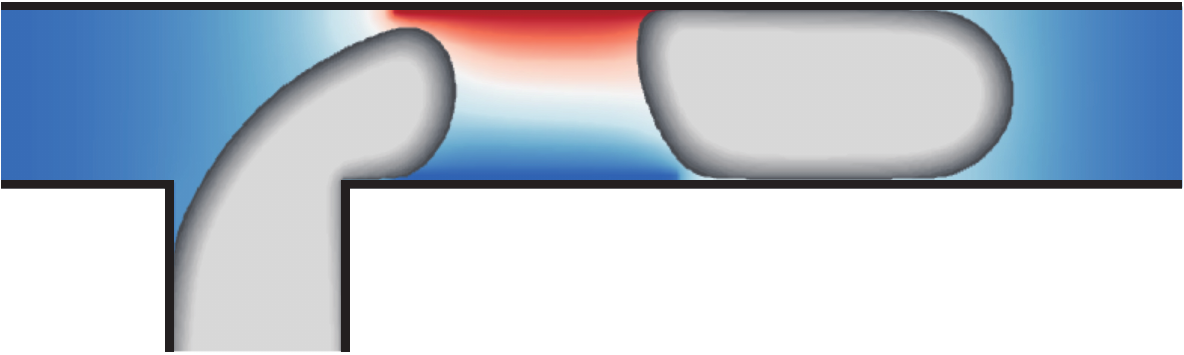}
	\put(-70, -5){\scriptsize \textit{(xii)}}
	
	\caption{Snapshots of droplet formation under top-biased PEF at different pulse onset time: (a) ${t_o} = 0.1$, (b) ${t_o} = 0.3$, and (c) ${t_o} = 0.5$.}
	\label{fig9}
\end{figure}

\begin{figure}
	\centerline{\includegraphics[width=0.6\textwidth]{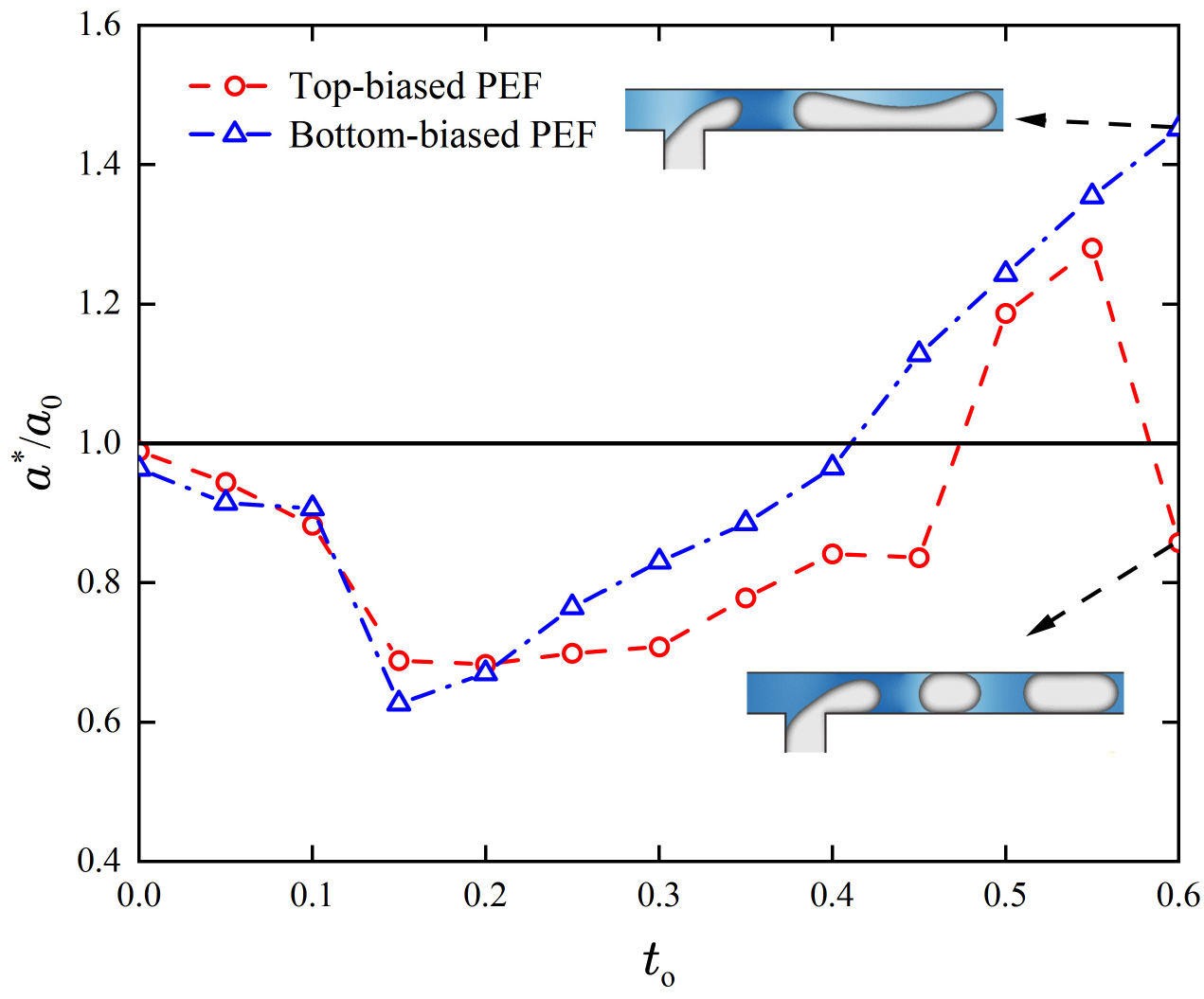}} 
	\caption{The relationship between the droplet's non-dimensional area (normalized by the droplet formation area ${a_0}$ under no electric field conditions) and the pulse onset time.} 
	\label{fig10} 
\end{figure}

Figure \ref{fig9} compares the droplet formation process at ${t_o}=0.1$, $0.3$, and $0.5$. When ${t_o}=0.1$, the liquid-liquid interface deforms under the electric stress during the filling stage and begins to block the channel (figure \ref{fig9}(ii)), consistent with the behaviour shown in figure \ref{fig6}(ii). However, once the electric field is switched off, the interface cannot remain pinned to the top wall, and necking is not triggered immediately. Owing to the relatively high flow rate, the incipient neck is then convected downstream together with the bulk flow (figure \ref{fig9}(iii) and (iv)). As the onset of PEF is delayed, the dispersed phase that has already entered the main channel is pulled towards the top electrode after the pulse is applied (figure \ref{fig9}(vi) and (xi)). The difference is that, at ${t_o}=0.3$, the neck does not form immediately after the pulse is applied, but develops only after a short period of squeezing, during which the dispersed phase continues to fill the main channel (figure \ref{fig9}(vii)). By contrast, when ${t_o}=0.5$, necking is initiated almost immediately upon pulse application (figure \ref{fig9}(xi)). At this stage, the dispersed phase already blocks the channel completely, and the liquid-liquid interface rapidly becomes concave under the combined action of electric and hydrodynamic stresses, which accelerates neck thinning and leads to droplet breakup.
Figure \ref{fig10} quantitatively compares the normalized droplet area, ${a}/{a_0}$, for different onset times. Overall, the droplet area exhibits a non-monotonic dependence on ${t_o}$, decreasing first and then increasing. When the pulse is applied too early (${t_o}<0.15$), it acts mainly during the filling stage and cannot sustain interfacial pinning long enough to trigger immediate breakup. As ${t_o}$ increases, the pulse acts closer to the onset of neck formation and therefore promotes neck thinning more effectively, leading to smaller droplets. However, when the pulse is applied at a later stage of filling, a larger volume of dispersed phase has already entered the main channel before the electric field is turned on. Although the electric stress can still induce rapid breakup near the electrode, the larger pre-breakup volume results in larger droplets. It is worth noting that, at ${t_o}=0.6$, the top-biased PEF generates two smaller droplets instead of one. This is because the interface spanning the electrode region undergoes strong deformation under the electric stress and develops two distinct thinning locations near the electrode edges, leading to almost simultaneous rupture at both sites.

\subsection{Roles of shear stress and electric force in droplet deformation}
\label{sec:4.3}
\begin{figure}[H]
	\centering
	\subfigure[]{\includegraphics[width=0.42\textwidth]{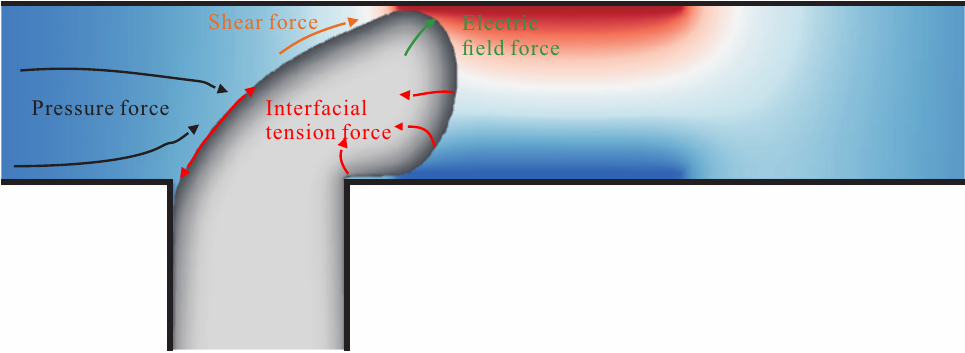}}
	\hspace{0.4cm}
	\subfigure[]{\includegraphics[width=0.42\textwidth]{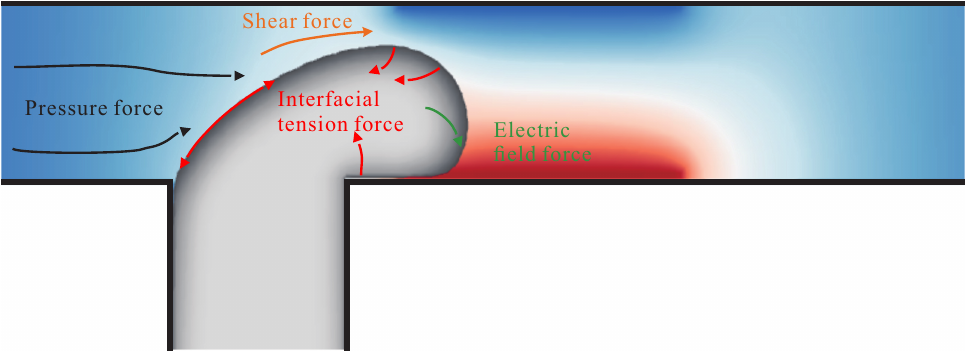}}
	\caption{Force analysis diagram of droplet in microchannels: (a) top-biased PEF, (b) bottom-biased PEF.}
	\label{fig:fig11}
\end{figure}

As mentioned above, the break-up of droplets is primarily governed by the combined effect of hydrodynamic stress and electric field stress. Under the influence of electric stress, the dispersed phase exhibits different flow characteristics, leading to changes in droplet size. However, the underlying hydrodynamic mechanisms behind this phenomenon are still unclear. Therefore, this section will further investigate the reasons behind the observed flow dynamics behaviors of the dispersed phase. The process of the dispersed phase filling the main channel until droplet break-up is considered as a control volume. The flow behaviour of this control volume is influenced by fluid shear stress, electric field stress, pressure, and its own surface tension, as shown in figure \ref{fig:fig11}.
For a given continuous and dispersed phase, the dimensionless droplet area is determined by the interplay of various forces. Initially, assuming a fully developed flow of the continuous phase within the channel, the interlayer velocity and pressure difference between inlet and outlet can be described using the Hagen-Poiseuille equation as \cite{Zhu_CEJ2024}
\begin{equation}
	u\left( y \right) = 6{u_c}\frac{y}{w}\left( {1 - \frac{y}{w}} \right),0 \le y \le w,
\end{equation}
\begin{equation}
	{{\bf{F}}_P} \approx \frac{{12{\mu _c}{{\bf{u}}_c}}}{{{w^3}}}wL,
\end{equation}
where $y$ and $L$ represents the height and length of the control body in the main channel, respectively. Additionally, since the two-phase fluids filling the main channel are in relative motion, the shear force exerted by the continuous phase on the control volume can be determined based on Newton's law of viscosity and previous studies \cite{Zhu_CEJ2024}
\begin{equation}
	{{\bf{F}}_\tau } \approx {\mu _c}\frac{{du}}{{dy}} \approx {\mu _c}\frac{{{{\bf{u}}_{\bf{c}}}}}{w},
\end{equation}
Additionally, interfacial tension is crucial for maintaining the stability of the two-phase interface. Due to the different curvatures at the neck and head of the control volume, the Laplace pressure acting on the control volume is calculated as \cite{Zhu_CEJ2024}
\begin{equation}
	{{\bf{F}}_{\bf{\gamma }}} = ({\bf{\Delta }}{{\bf{p}}_{{\bf{\gamma ,neck}}}} + {\bf{\Delta }}{{\bf{p}}_{{\bf{\gamma ,tip}}}}) \cdot {w^2} =  - \gamma w
\end{equation}
Finally, the electric force acting on the control volume follows the form given in (\ref{eq2.11}). Since $R=S$ in the present study, the tangential electric stress is relatively weak and is therefore neglected in the following analysis. Accordingly, the control volume is assumed to be primarily subjected to the normal electric stress. Based on previous studies \cite{Garstecki_Labonchip2006,Singh_JFM2020}, we divide the droplet area into a linear combination of the area ${a_{fill}}$ during the filling stage, where the dispersed phase enters the main channel, and the area ${a_{squeeze}}$ during the squeezing stage, as follows
\begin{equation}
	a = {a_{fill}} + {a_{squeeze}}.
\end{equation}

During the filling stage, the progress of the dispersed phase is chiefly governed by its own flow rate, as well as the shear forces exerted by the surrounding flow. This is due to the velocity difference between the dispersed phase and the continuous phase in the horizontal direction, with the shear force accelerating the filling of the dispersed phase. At the same time, the presence of surface tension hinders the flow of the control volume. Therefore, the filling area of the dispersed phase can be calculated as
\begin{equation}
	{m_{fill}} = \int_{{t_{fill}}} {\left( {{{\bf{u}}_{\bf{d}}} + {{\bf{u}}^{\bf{*}}}} \right)wdt} ,
\end{equation}
where ${{\bf{u}}^{\bf{*}}}$ is the velocity of the dispersed phase entering the main channel due to the combined effects of shear and surface tension. It can be defined as
\begin{equation}
	{{\bf{u}}^*} = {{\bf{F}}_\tau }w\frac{{\left( {1 - {k_\gamma }} \right)}}{{{\mu _c} + {\mu _d}}} = {{\bf{u}}_{\bf{c}}} \cdot \frac{{{\mu _c}}}{{{\mu _c} + {\mu _d}}} \cdot \left( {1 - {k_\gamma }} \right),
\end{equation}
since the electric stress, as an interfacial force, does not directly influence the filling of the dispersed phase, here $k_\gamma$ serves as a correction factor that accounts for the effect of surface tension in resisting the motion of the dispersed phase. Therefore, during the filling stage, the velocity of the droplet is governed solely by these physical parameters. To introduce a dimensionless time, ${t^*} = {{ut} \mathord{\left/
		{\vphantom {{ut} w}} \right.
		\kern-\nulldelimiterspace} w}$, where $u$ represents the combined velocity of both the continuous and dispersed phases, the relationship for the rate of change in mass is expressed as follows

\begin{equation}
     \frac{{d{m_{fill}}}}{{d{t^*}}} = \frac{{\left( {{{\bf{u}}_{\bf{d}}} + {{\bf{u}}^*}} \right){w^2}}}{{{{\bf{u}}_{\bf{c}}} + {{\bf{u}}_{\bf{d}}}}} = \frac{{\left[ {{{\bf{u}}_{\bf{d}}} + {\textstyle{{{\mu _c}} \over {{\mu _c} + {\mu _d}}}}{{\bf{u}}_{\bf{c}}}\left( {1 - {k_\gamma }} \right)} \right]{w^2}}}{{{{\bf{u}}_{\bf{c}}} + {{\bf{u}}_{\bf{d}}}}}
\end{equation}
Thus, the dimensionless area of the droplet filling the main channel ${a_{fill}}^* = {m_{fill}}/\left( {\rho {w^2}} \right)$ is related to the flow rate ratio ${Q_r}$ as follows
\begin{equation}
	\frac{{d{a_{fill}}^*}}{{d{t^*}}} = \frac{{\left[ {{Q_r} + {\textstyle{{{\mu _c}} \over {{\mu _c} + {\mu _d}}}}\left( {1 - {k_\gamma }} \right)} \right]}}{{\rho \left( {1 + {Q_r}} \right)}}.
	\label{eq4.9}
\end{equation}

\begin{figure}[H]
	\centering
	\subfigure[]{\includegraphics[width=0.42\textwidth]{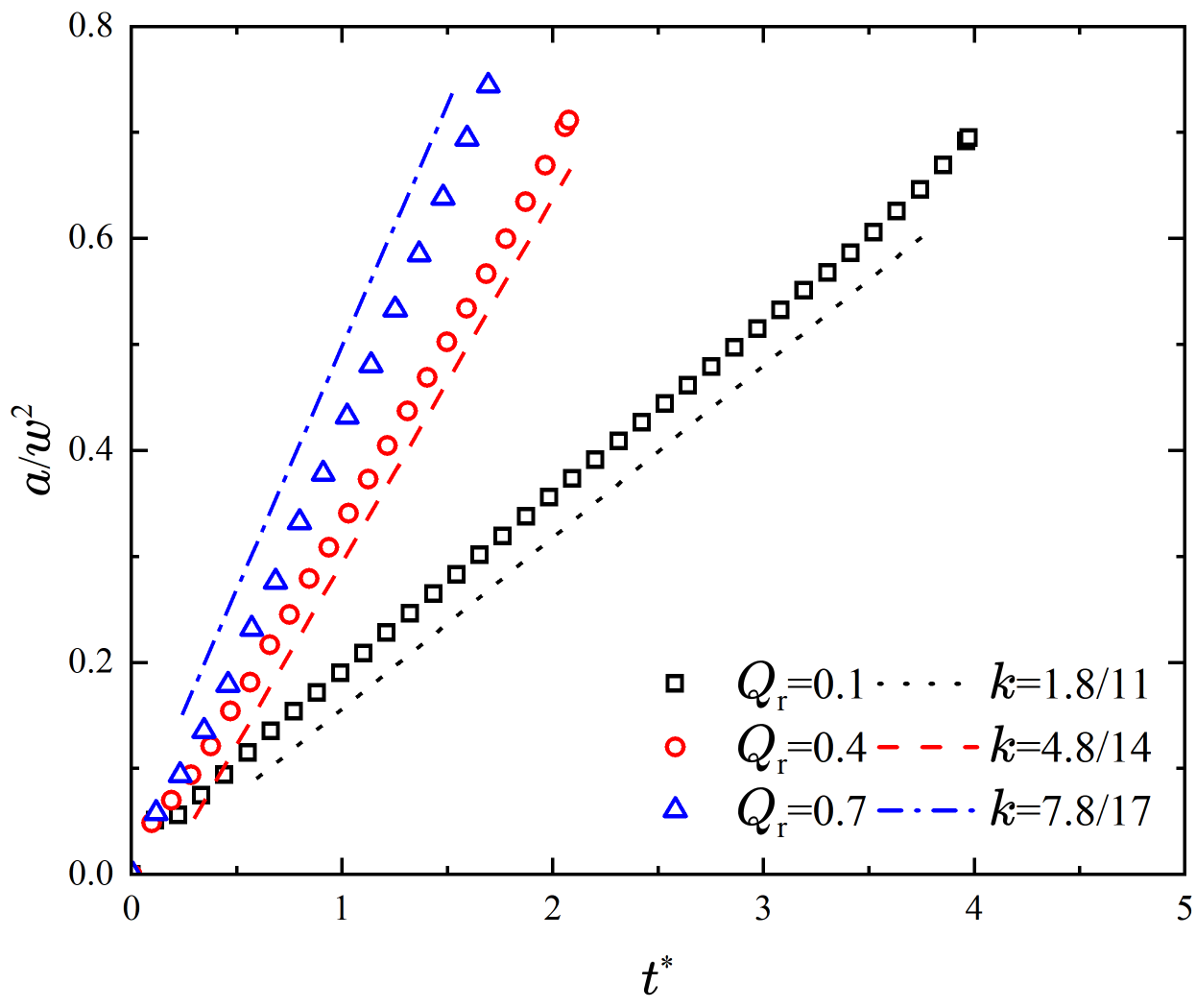}}
	\hspace{0.4cm}
	\subfigure[]{\includegraphics[width=0.42\textwidth]{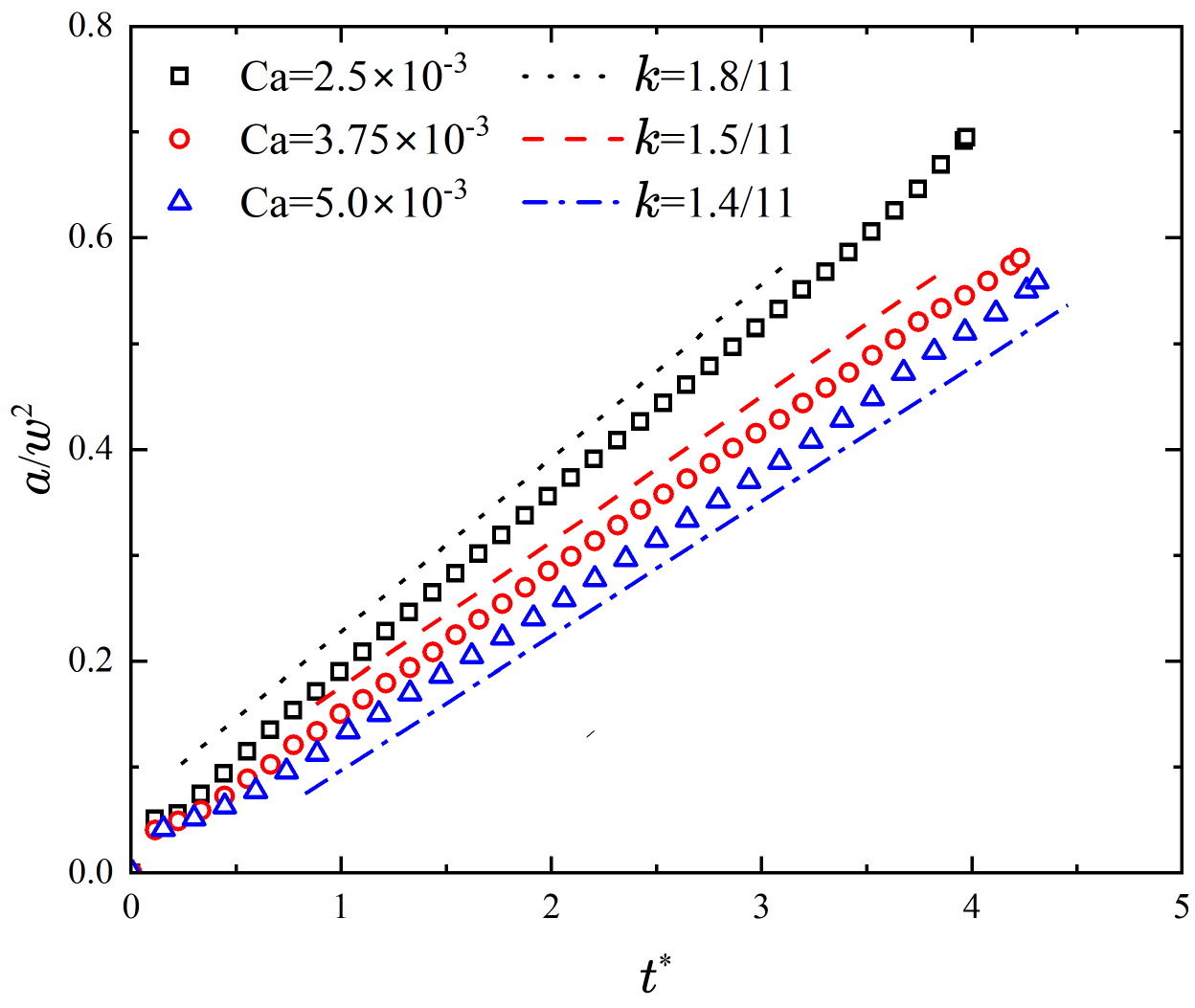}}
	
	\vspace{0.4cm}
	
	\subfigure[]{\includegraphics[width=0.42\textwidth]{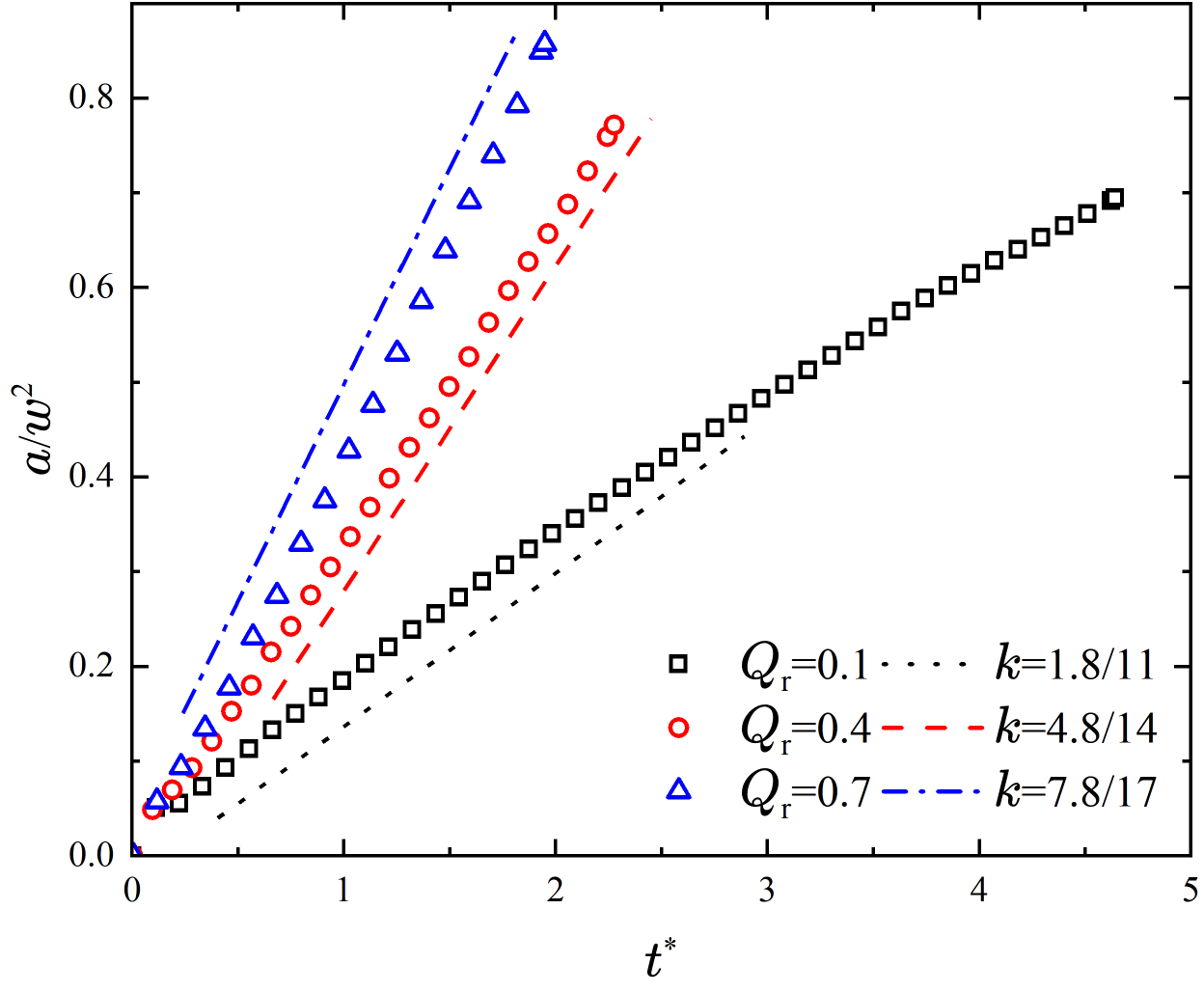}}
	\hspace{0.4cm}
	\subfigure[]{\includegraphics[width=0.42\textwidth]{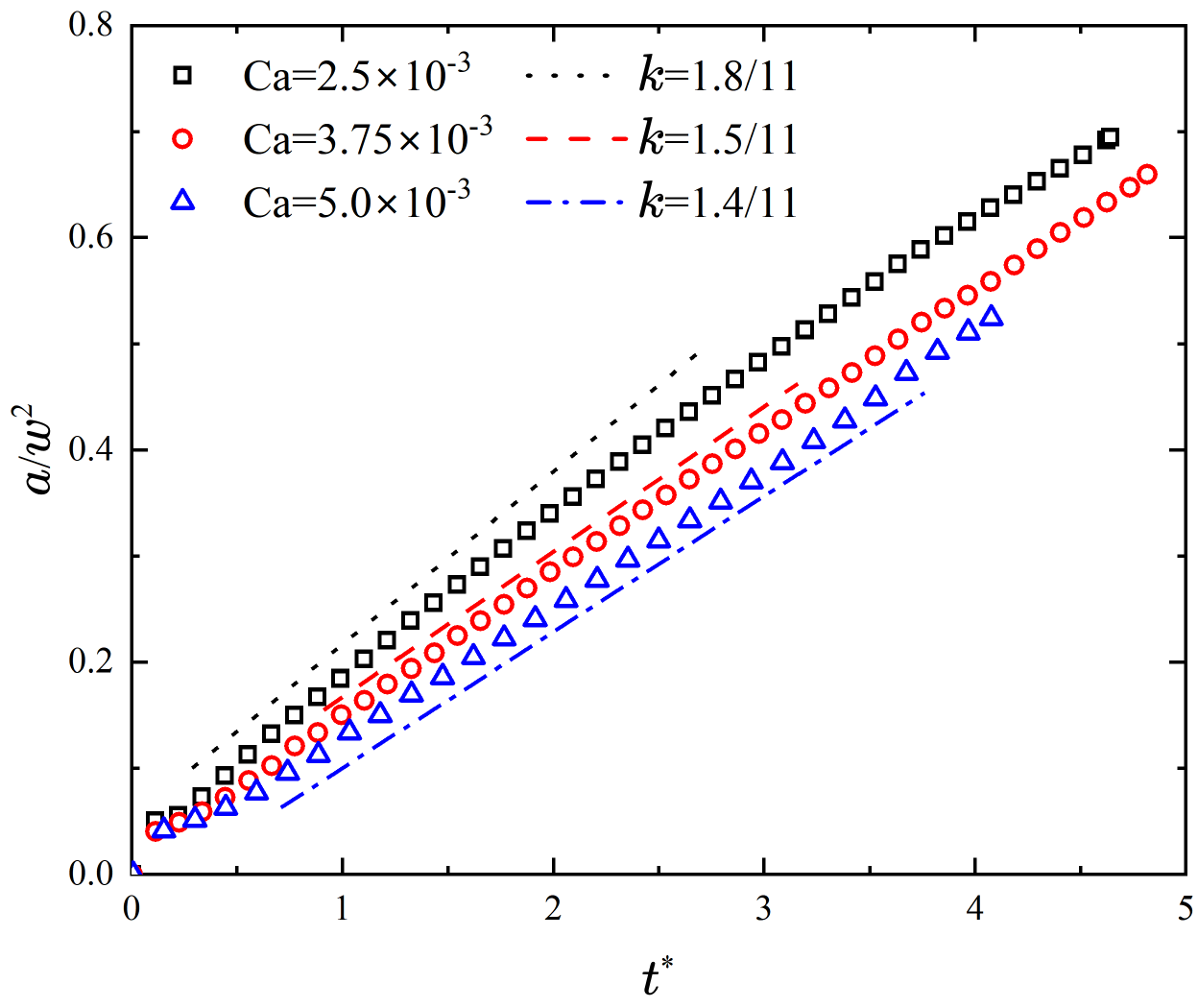}}
	\caption{Evolution of non-dimensional area as a function of non-dimensional time during the filling stage under different conditions: (a) top-biased PEF with fixed $Ca = 2.5 \times {10^{ - 3}}$, (b) top-biased PEF with fixed ${Q_r} = 0.1$, (c) bottom-biased PEF with fixed $Ca = 2.5 \times {10^{ - 3}}$, (d) bottom-biased PEF with fixed ${Q_r} = 0.1$.}
	\label{fig12}
\end{figure}

\begin{figure}[H]
	\centering
	\subfigure[]{\includegraphics[width=0.32\textwidth]{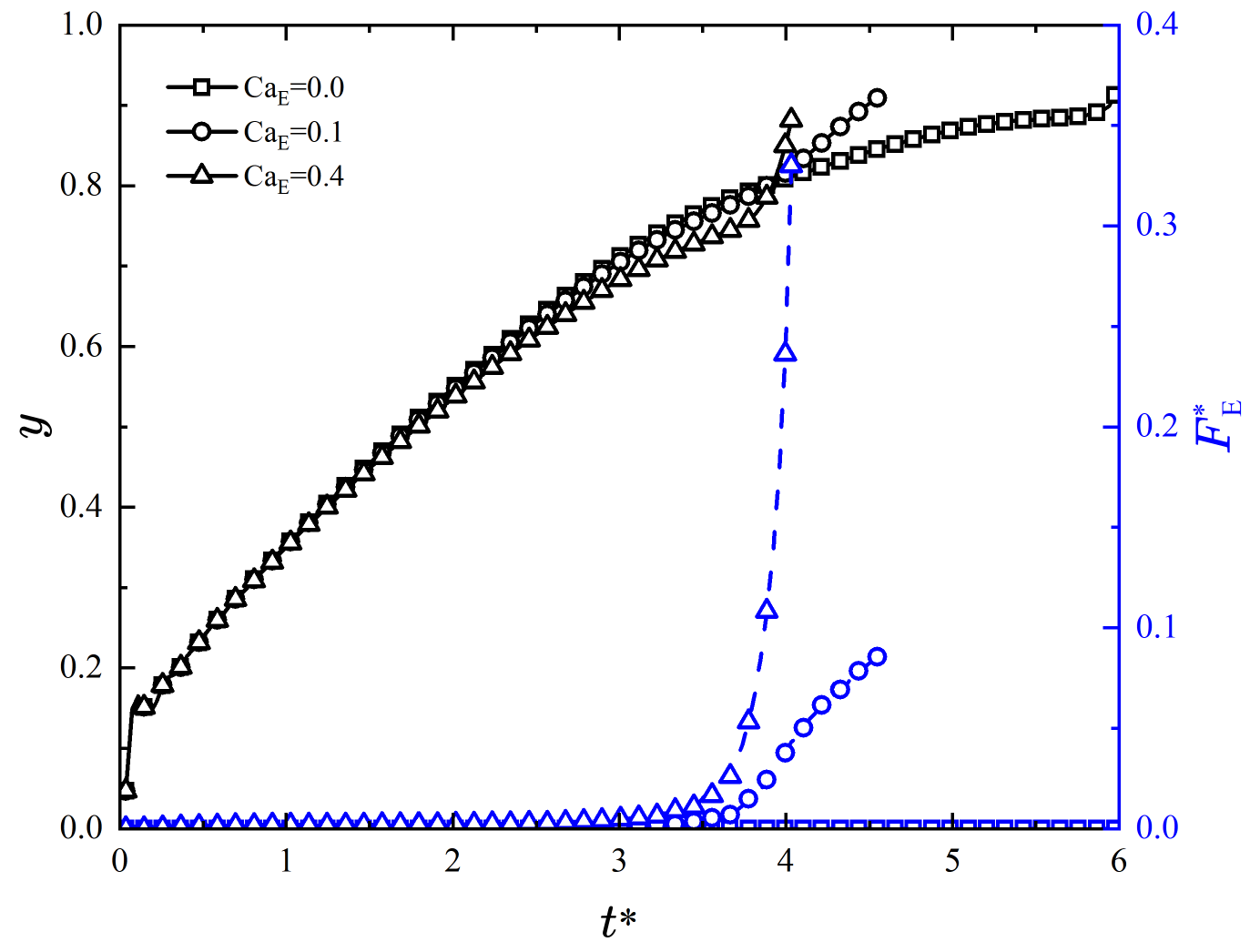}}
	\hspace{0.1cm}
	\subfigure[]{\includegraphics[width=0.32\textwidth]{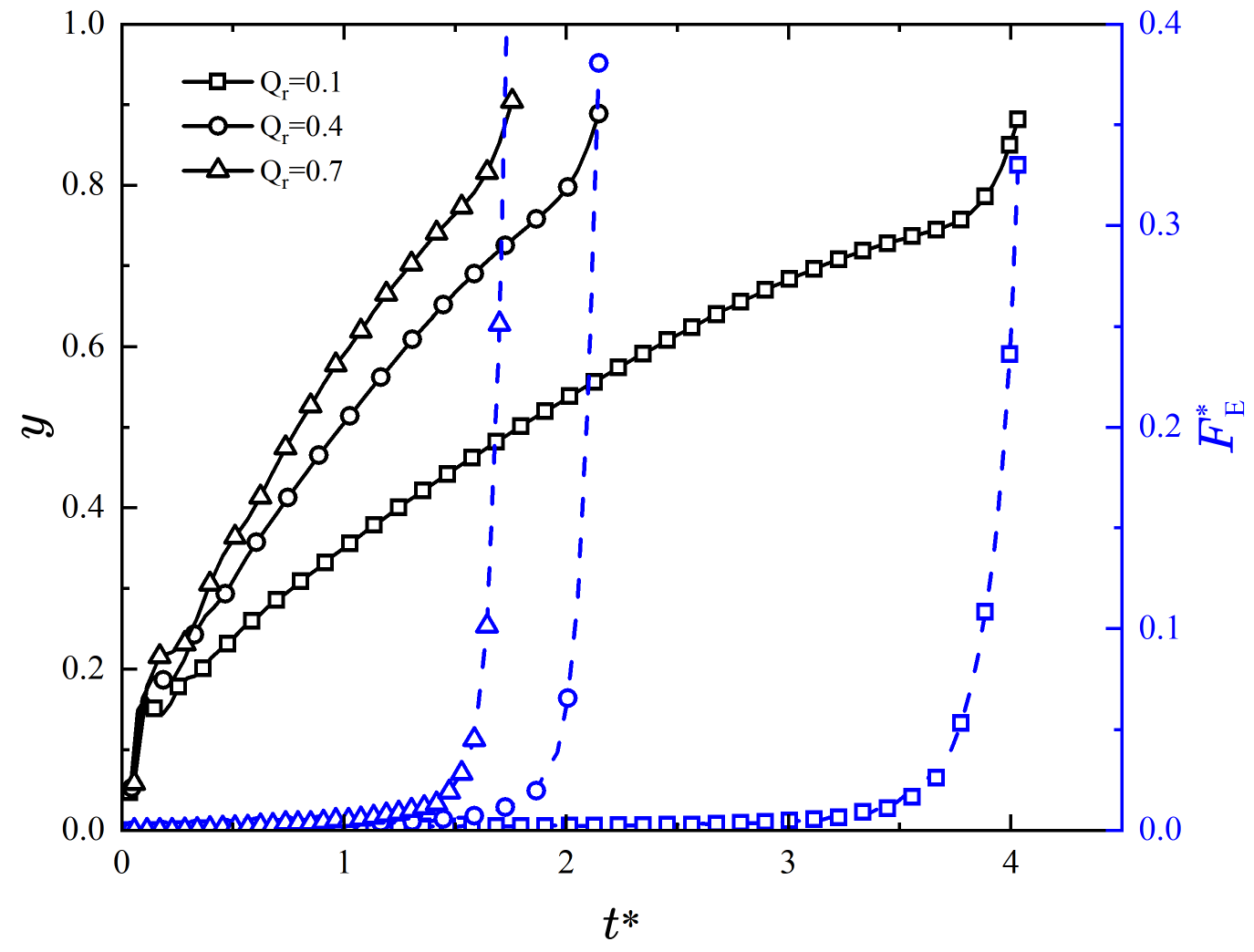}}
	\hspace{0.1cm}
	\subfigure[]{\includegraphics[width=0.32\textwidth]{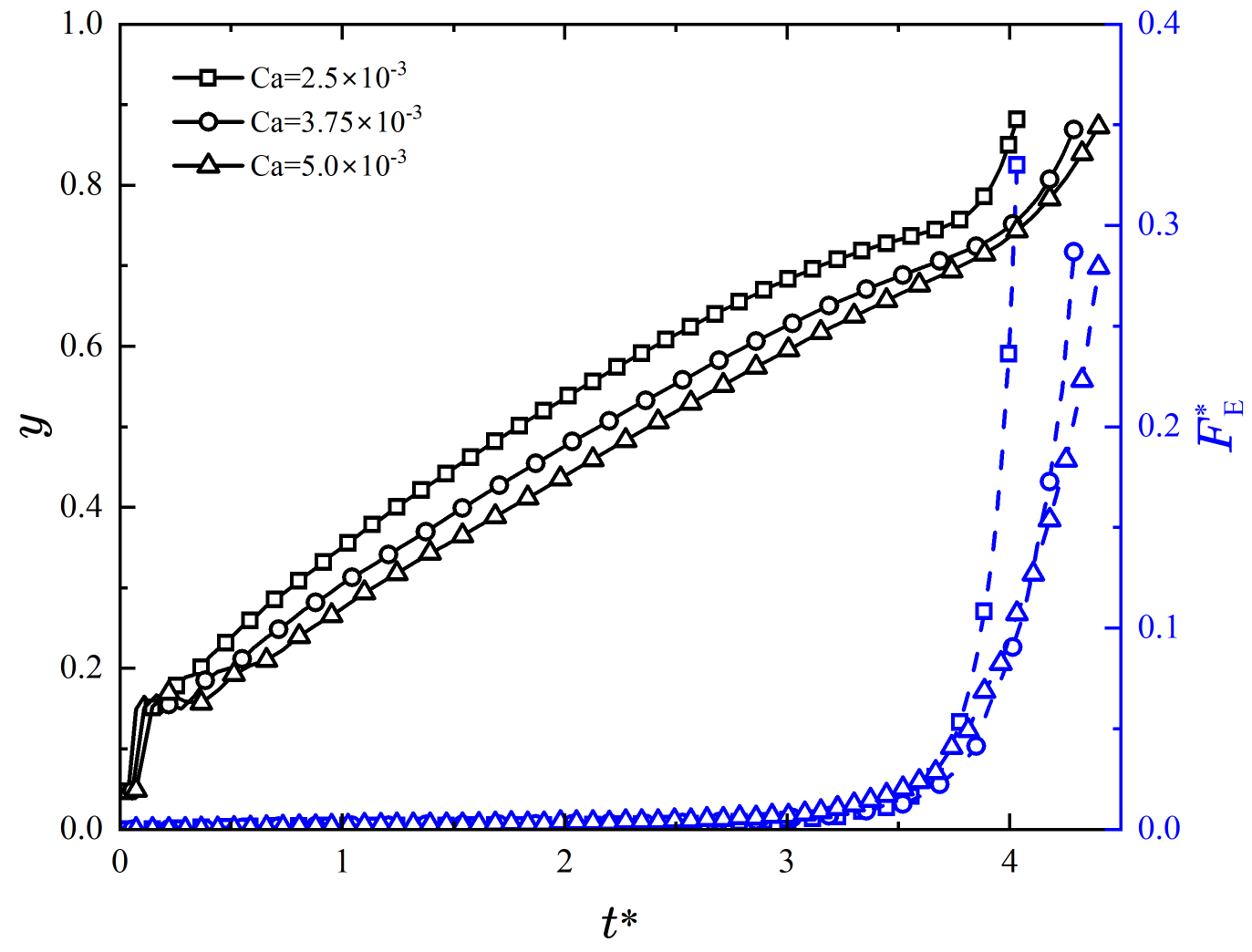}}
	
	\vspace{0.2cm}
	
	\subfigure[]{\includegraphics[width=0.32\textwidth]{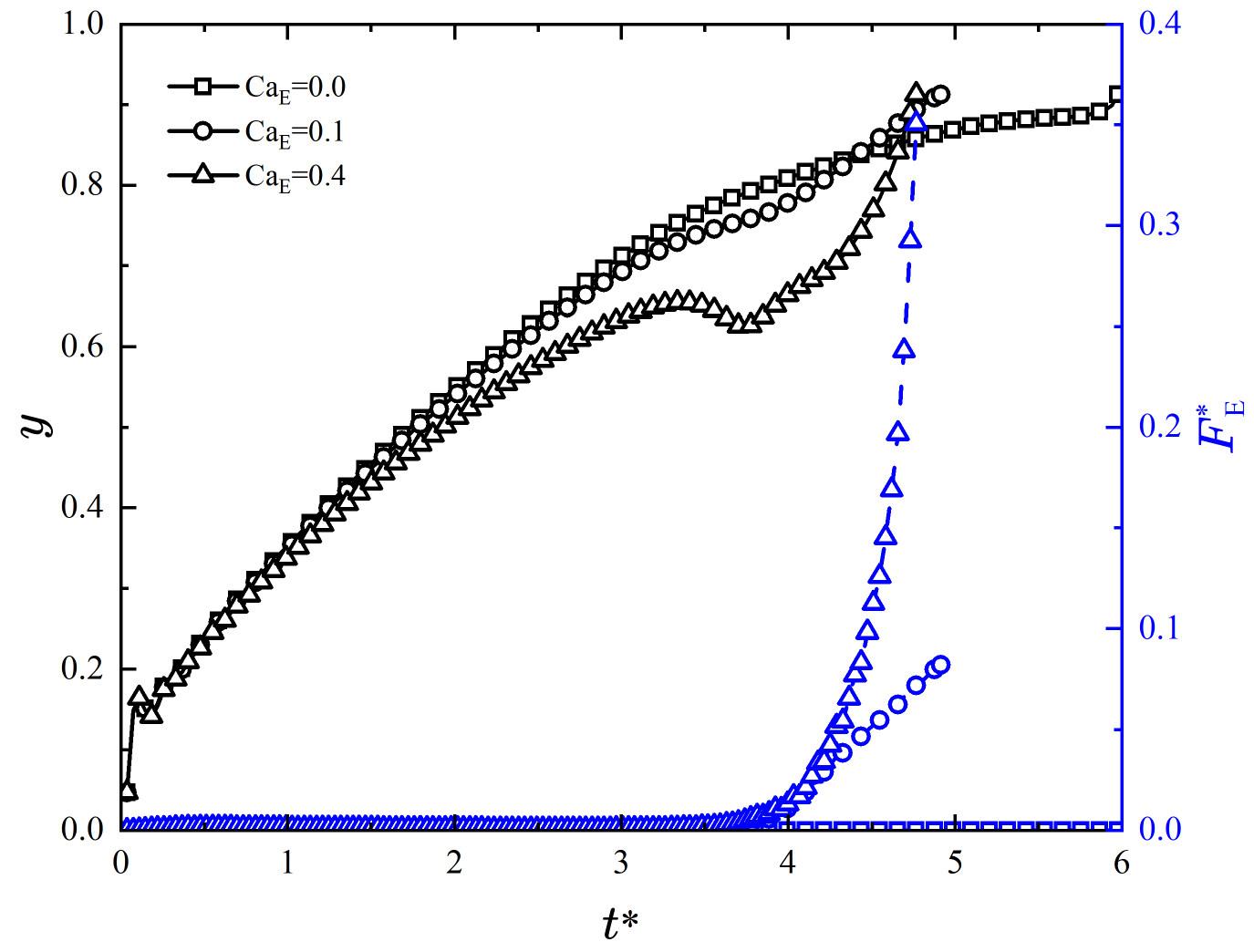}}
	\hspace{0.1cm}
	\subfigure[]{\includegraphics[width=0.32\textwidth]{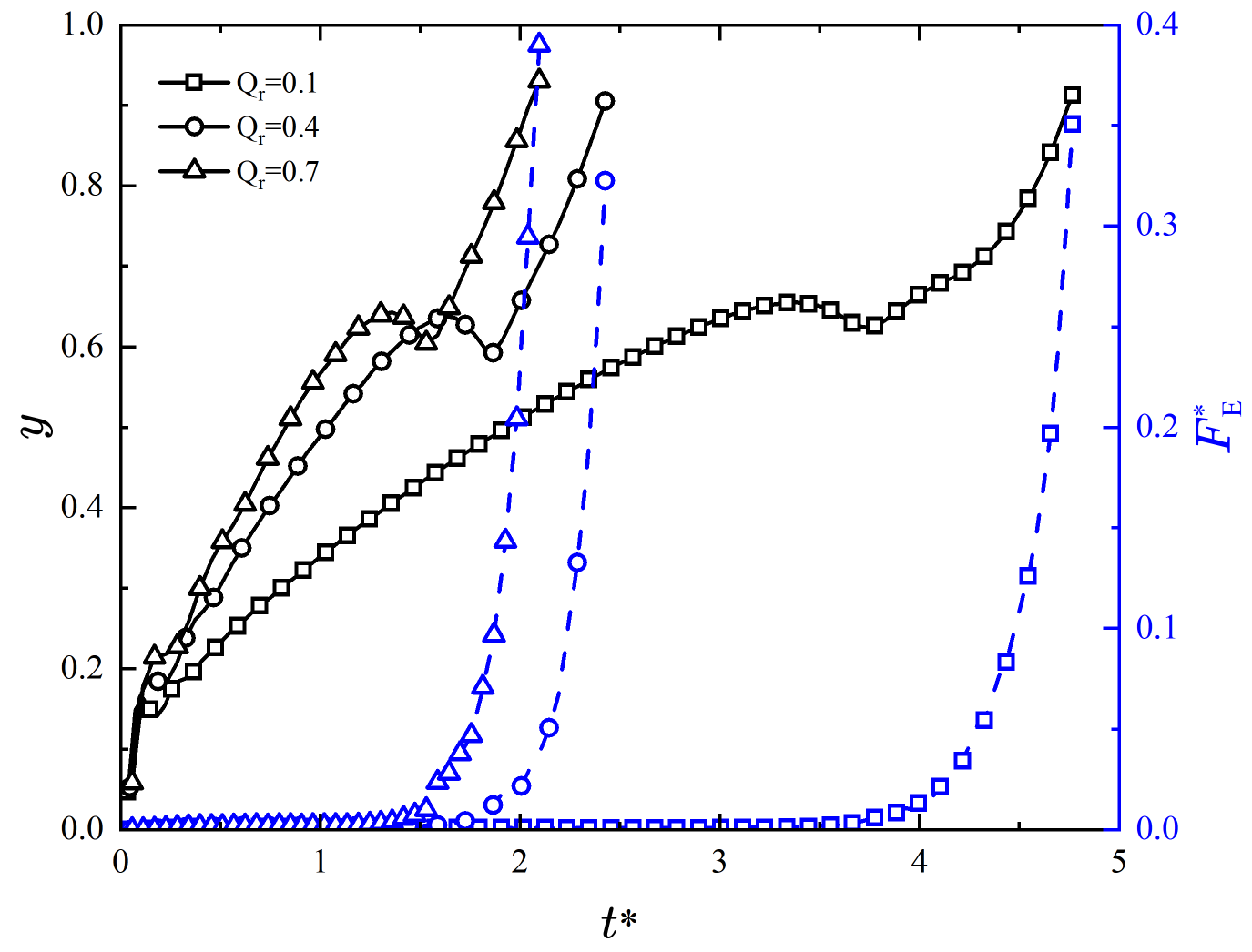}}
	\hspace{0.1cm}
	\subfigure[]{\includegraphics[width=0.32\textwidth]{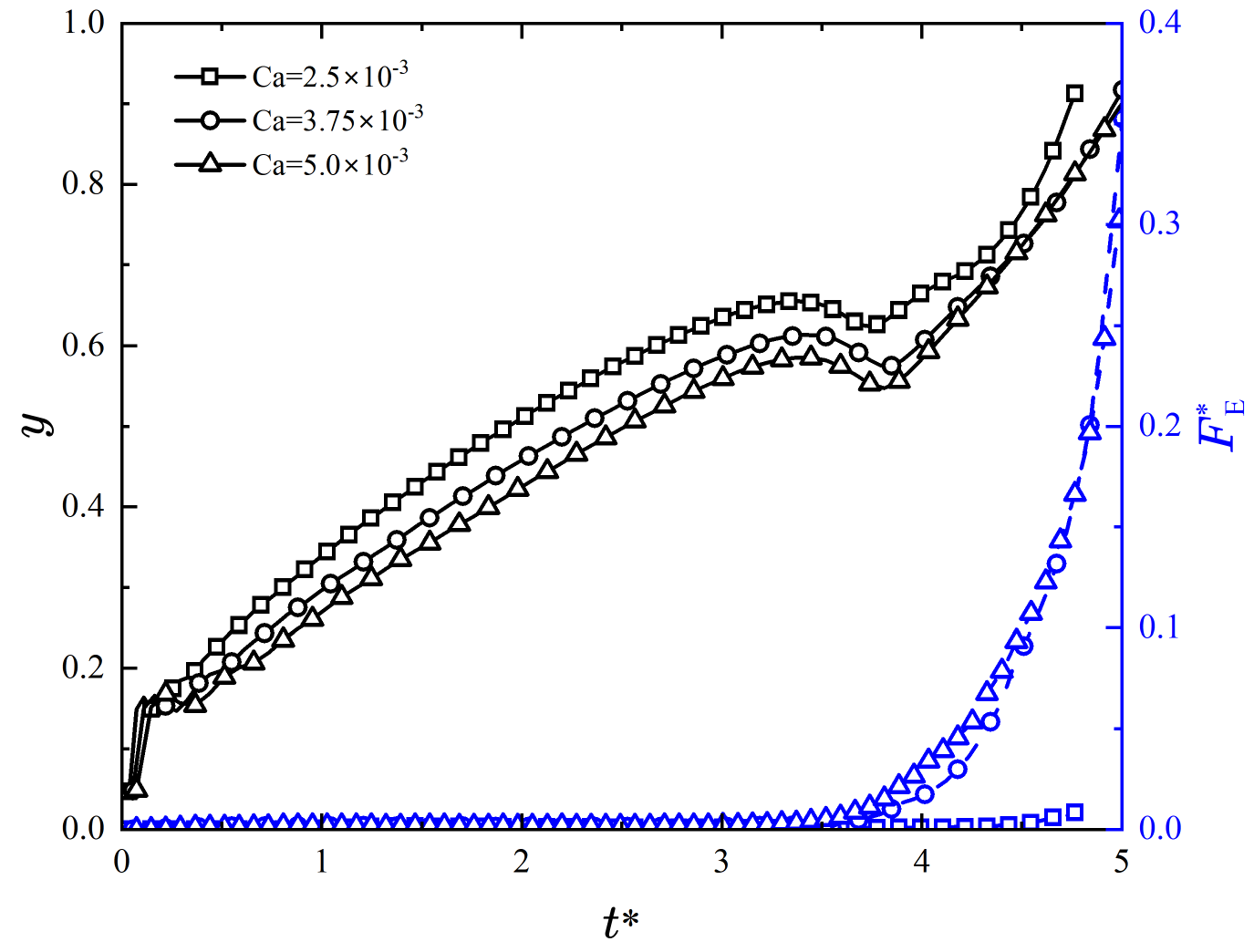}}
	\caption{The non-dimensional height of the droplet (normalized by the channel width) on the left black axis and the electric field force acting on the top of the droplet (right blue axis) as a function of non-dimensional time under different electric field conditions. (a), (b), (c) top-biased PEF; (d), (e), (f) bottom-biased PEF. Except for the parameters specified in the legends, the other parameters are fixed as $C{a_E} = 0.4$, ${Q_r} = 0.1$, and $Ca = 2.5 \times {10^{ - 3}}$.}
	\label{fig13}
\end{figure}

Figure~\ref{fig12} shows the evolution of the filling area of the dispersed phase with time for different $Ca$ and $Q_r$ at $Ca_E=0.4$. As indicated by (\ref{eq4.9}), the non-dimensional filling area $a_{fill}^*$ varies approximately linearly with the non-dimensional time $t^*$ during the filling stage, and the corresponding slope is denoted by $k$. Here, $k_\gamma$ is introduced to account for the modifying effect of surface tension on the shear-induced forward filling. A value of $k_\gamma$ closer to unity indicates a more pronounced suppression of the forward motion by surface tension, with a smaller shear-induced correction. For a fixed $Ca$, the characteristic flow state of the continuous phase is essentially unchanged, while the streamwise velocity of the dispersed phase remains small during filling, so that $k_\gamma$ may be treated approximately as a constant within this stage. When $Ca=2.5\times10^{-3}$, the continuous-phase velocity is low and the suppressing effect of surface tension is correspondingly more evident; under this condition, taking $k_\gamma=0.9$ yields $k=(Q_r+0.08)/(1+Q_r)$. As shown in figures~\ref{fig12}(a) and~\ref{fig12}(c), this relation captures the overall variation of the filling-stage slope with $Q_r$ at fixed $Ca$. By contrast, at fixed $Q_r$, increasing $Ca$ raises the continuous-phase velocity and alters the interfacial force balance, thereby weakening the relative contribution of surface-tension-induced suppression and reducing the corresponding $k_\gamma$, as reflected in figures~\ref{fig12}(b) and~\ref{fig12}(d). Moreover, for the same $Ca$ and $Q_r$, the filling speeds under top-biased and bottom-biased PEF are nearly identical, indicating that the electric field primarily modifies the interfacial morphology and the filling duration, while having little influence on the filling flux itself.
Although the previous discussion has indicated that the filling velocity remains unaffected by the application of PEF, Section \ref{sec:4.2} reveals that applying PEF at different stages does influence the droplet formation process. As such, figure \ref{fig13} illustrates how the droplet height and the electric force acting on the droplet's apex evolve over time, expressed as a function of dimensionless time $t^*$, for various combinations of $Ca$, $Q_r$, and $C{a_E}$.

\begin{figure}[H]
	\centering
	\subfigure[]{\includegraphics[width=0.42\textwidth]{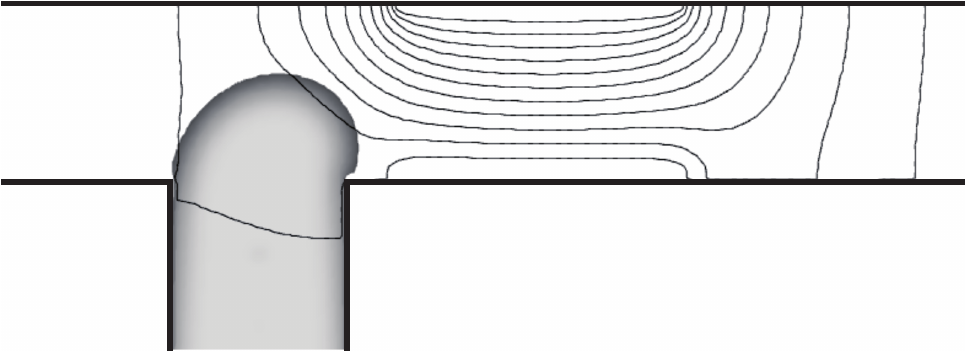}}
	\hspace{0.2cm}
	\subfigure[]{\includegraphics[width=0.42\textwidth]{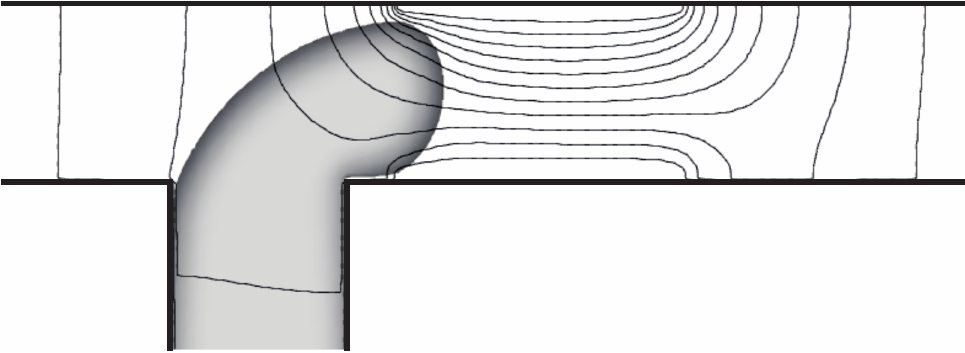}}
	
	\vspace{0.2cm}
	
	\subfigure[]{\includegraphics[width=0.42\textwidth]{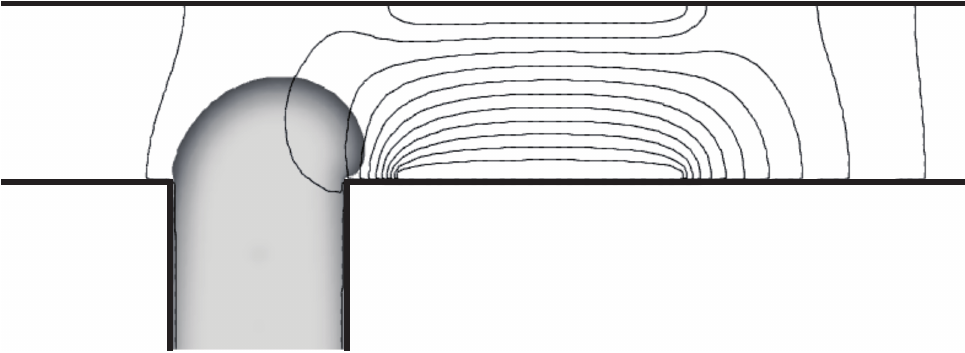}}
	\hspace{0.2cm}
	\subfigure[]{\includegraphics[width=0.42\textwidth]{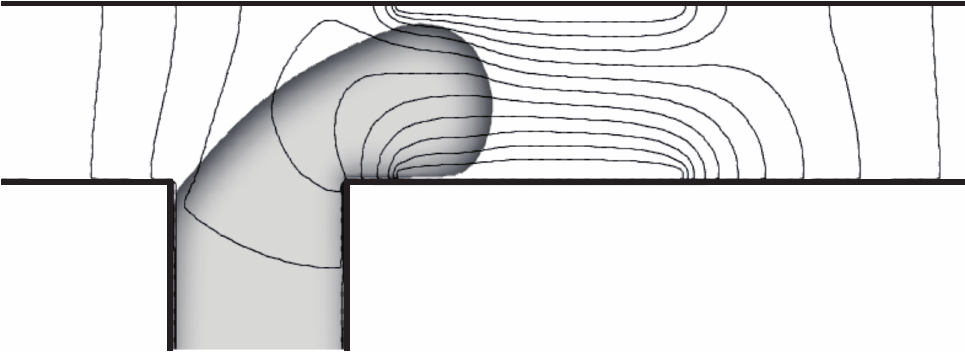}}
	\caption{The shape of the dispersed phase entering the main channel and the variation of the equipotential line distribution under different electric field conditions: (a), (b) top-biased PEF; (c), (d) bottom-biased PEF.}
	\label{fig14}
\end{figure}

Regarding top-biased PEF, when the dispersed phase first enters the main channel, the equipotential lines are sparse, resulting in a relatively weak electrical force on the dispersed phase(figures \ref{fig13}(a) and \ref{fig14}(a)). In the initial filling stage, the filling behaviour of the dispersed phase is almost identical across different $C{a_E}$ values. As the dispersed approaches the electrode, however, the equipotential lines become denser near the top(figure \ref{fig13}(b)), leading to an increase in the electric force acting on the interface. This caused the control volume to block the channel under the combined influence of hydrodynamic stress and electrical force. This behaviour is consistent for all values of ${Q_r}$ and $Ca$; in each case, the electric force only becomes significant once the dispersed phase is near the electrode, after which the control volume rapidly blocks the main channel (see figures \ref{fig13}(b) and (c)).

In the case of bottom-biased PEF, the dispersed phase is initially drawn towards the bottom electrode by the lateral electric stress as it enters the main channel (figure \ref{fig14}(c)). As filling continues and the dispersed phase moves between the electrodes, the electric force on the control volume reorients towards the top electrode. Subsequently, the control volume deforms towards the top electrode under the electric field's influence, eventually blocking the channel (figure \ref{fig14}(d)). As depicted in figure \ref{fig13}(d), the height of the droplet's apex initially rises, then decreases, and finally increases again until blockage occurs, reflecting the electric field's impact. This intermediate decrease is influenced by the strength of the electric field. For $C{a_E} = 0.1$, the control volume remains in a growth regime throughout the filling stage, although the rise in height is slightly less pronounced compared to the no electric field case ($C{a_E} = 0$). In contrast, for $C{a_E} = 0.4$, the dispersed phase exhibits a period of decreasing apex height for all $Q_r$ and $Ca$ values. This behaviour explains why, as shown in figure \ref{fig7}(b), the droplet does not break up within the computational domain when a pulse with a duration of $0.4 {t_d}$ is applied: during the PEF's duration, the electric field inhibits the full blockage of the control volume, thereby delaying neck formation.

\begin{figure}[H]
	\centering
	\subfigure[]{\includegraphics[width=0.3\textwidth]{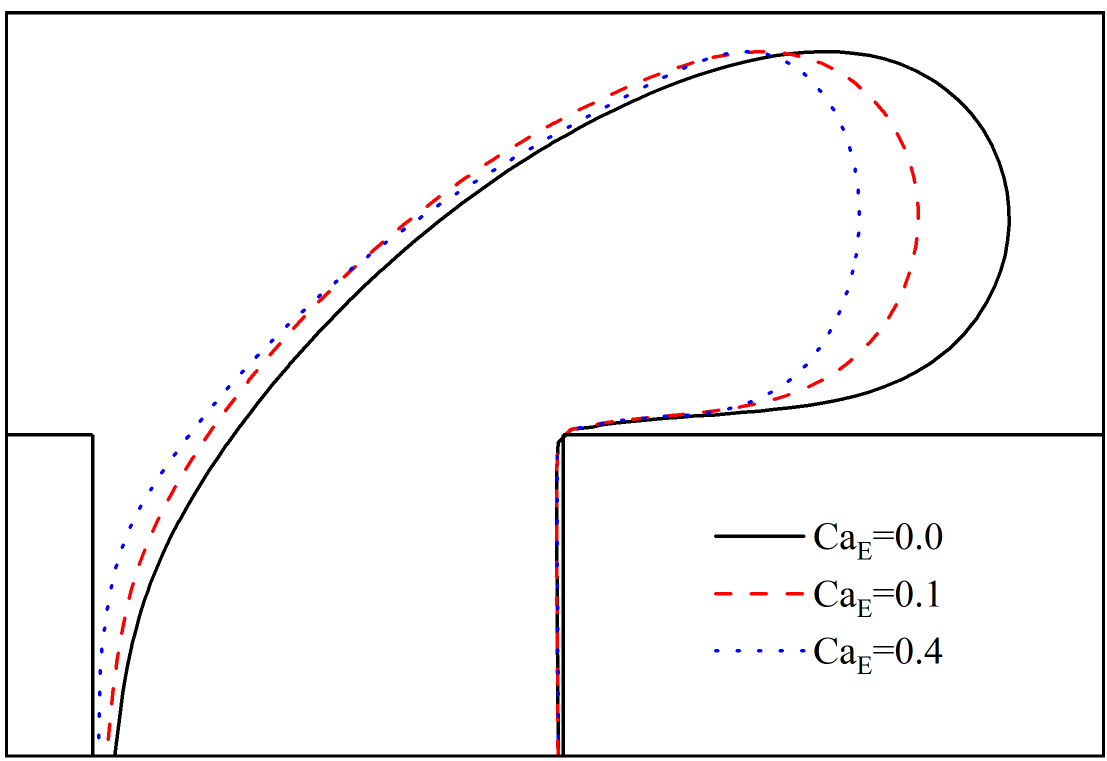}}
	\hspace{0.2cm}
	\subfigure[]{\includegraphics[width=0.3\textwidth]{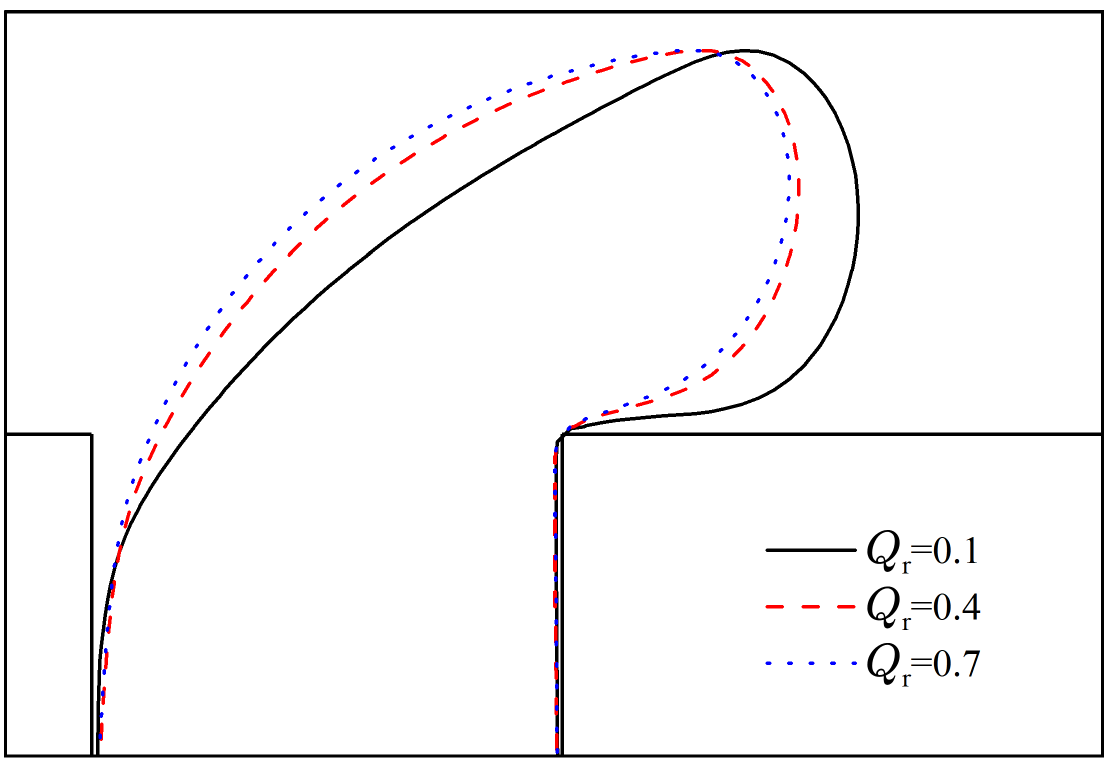}}
	\hspace{0.2cm}
	\subfigure[]{\includegraphics[width=0.3\textwidth]{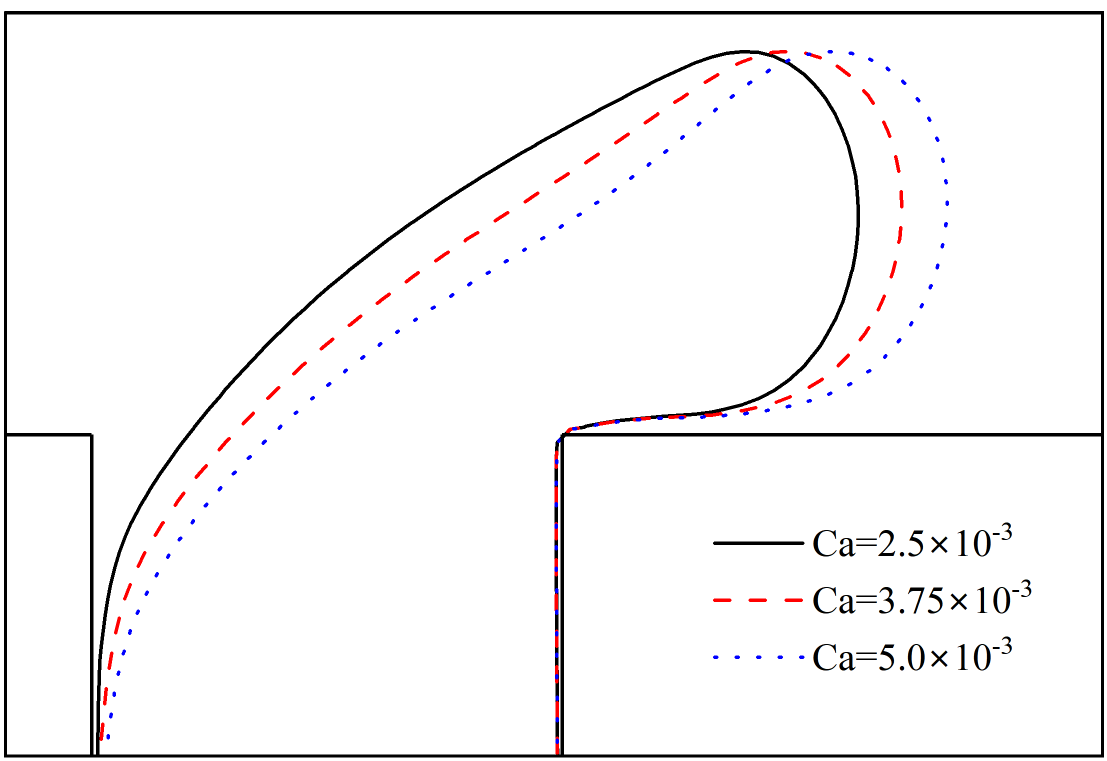}}
	
	\vspace{0.2cm}
	
	\subfigure[]{\includegraphics[width=0.3\textwidth]{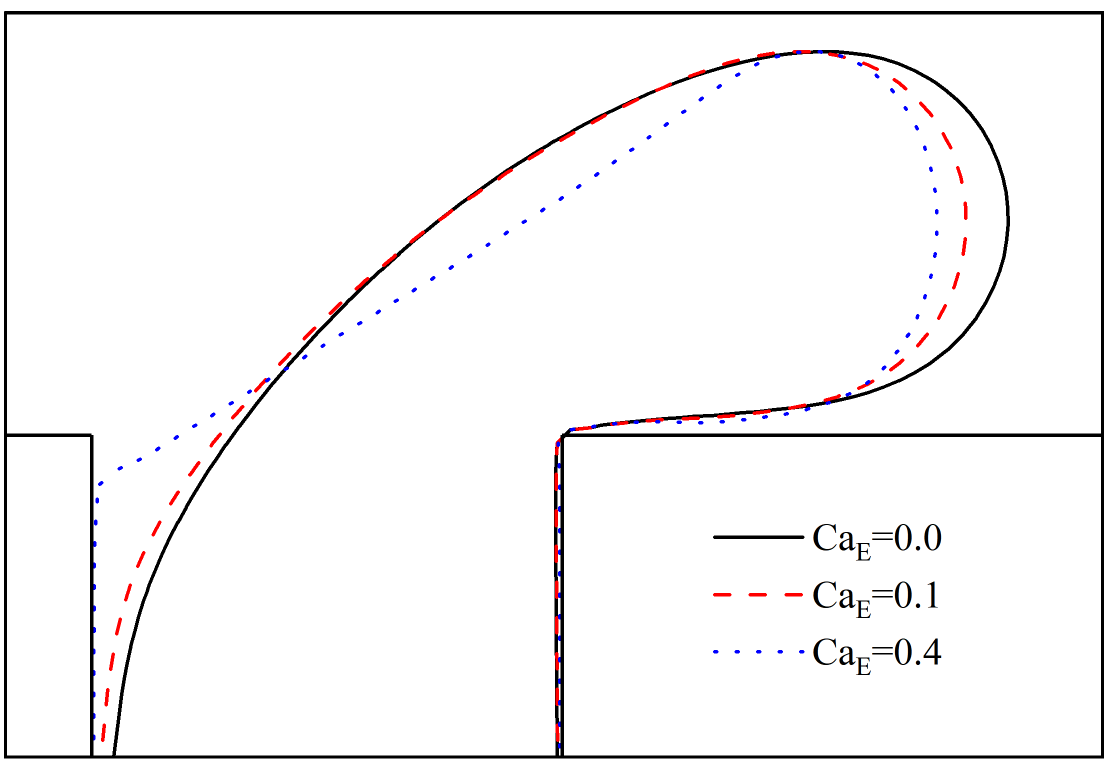}}
	\hspace{0.2cm}
	\subfigure[]{\includegraphics[width=0.3\textwidth]{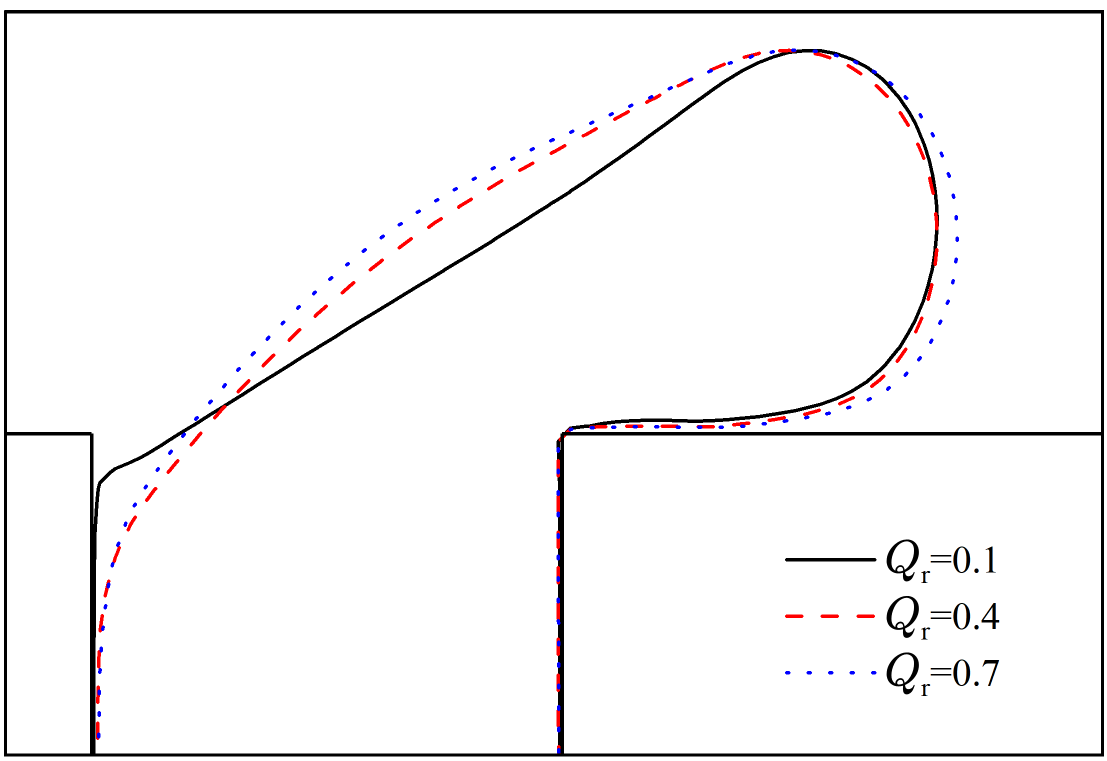}}
	\hspace{0.2cm}
	\subfigure[]{\includegraphics[width=0.3\textwidth]{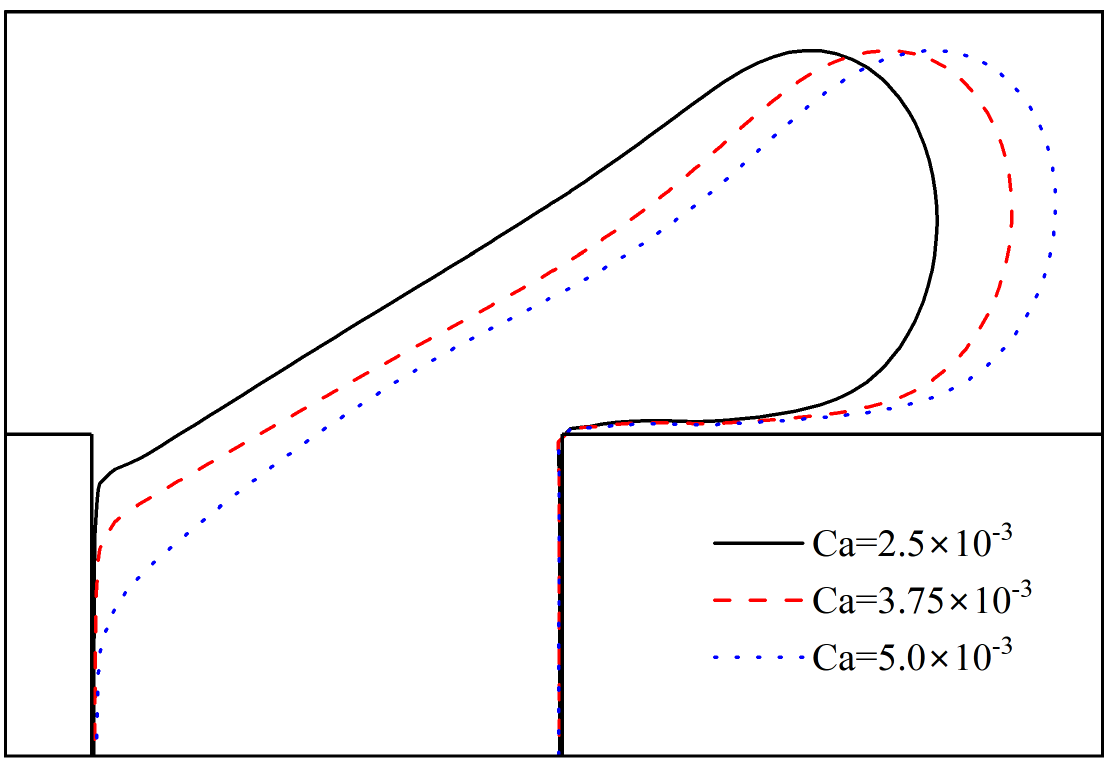}}
	\caption{The droplet profiles at the end of the filling stage under different conditions. (a), (b), (c) top-biased PEF; (d), (e), (f) bottom-biased PEF. Except for the parameters specified in the legends, the other parameters are fixed as $C{a_E} = 0.4$, ${Q_r} = 0.1$, and $Ca = 2.5 \times {10^{ - 3}}$.}
	\label{fig15}
\end{figure}

\begin{figure}[H]
	\centering
	\subfigure[]{\includegraphics[width=0.42\textwidth]{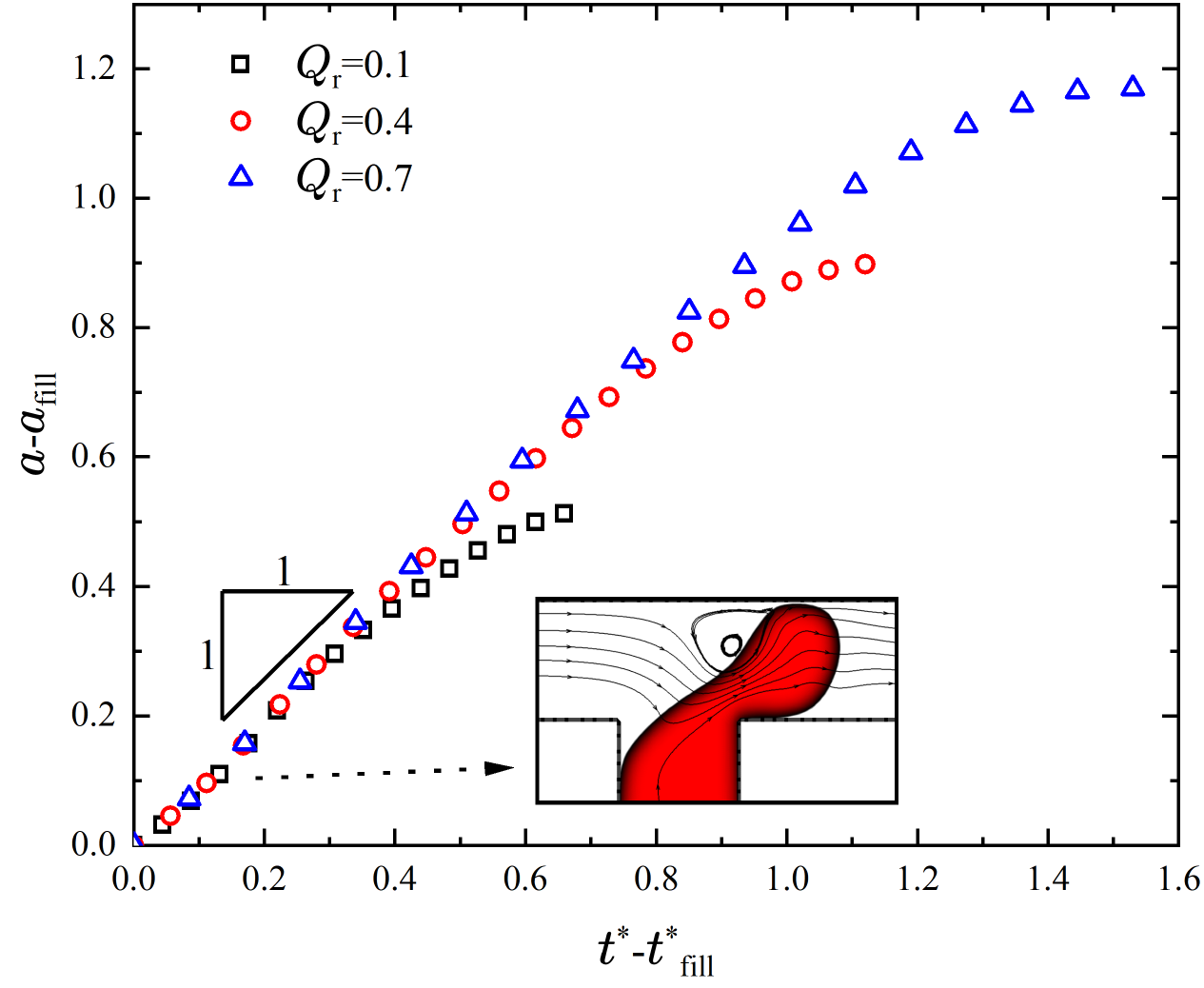}}
	\hspace{0.4cm}
	\subfigure[]{\includegraphics[width=0.42\textwidth]{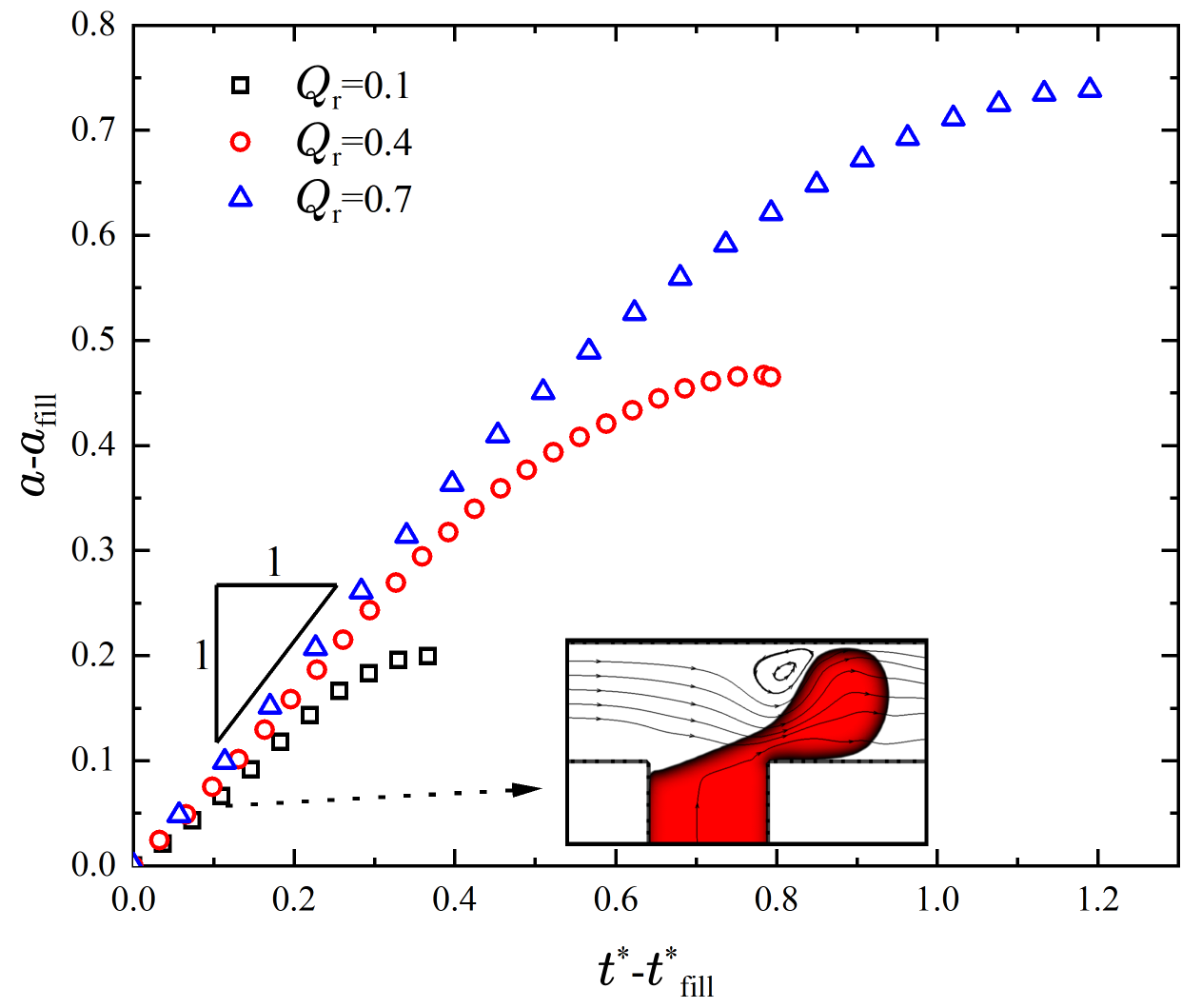}}
	
	\vspace{0.4cm}
	
	\subfigure[]{\includegraphics[width=0.42\textwidth]{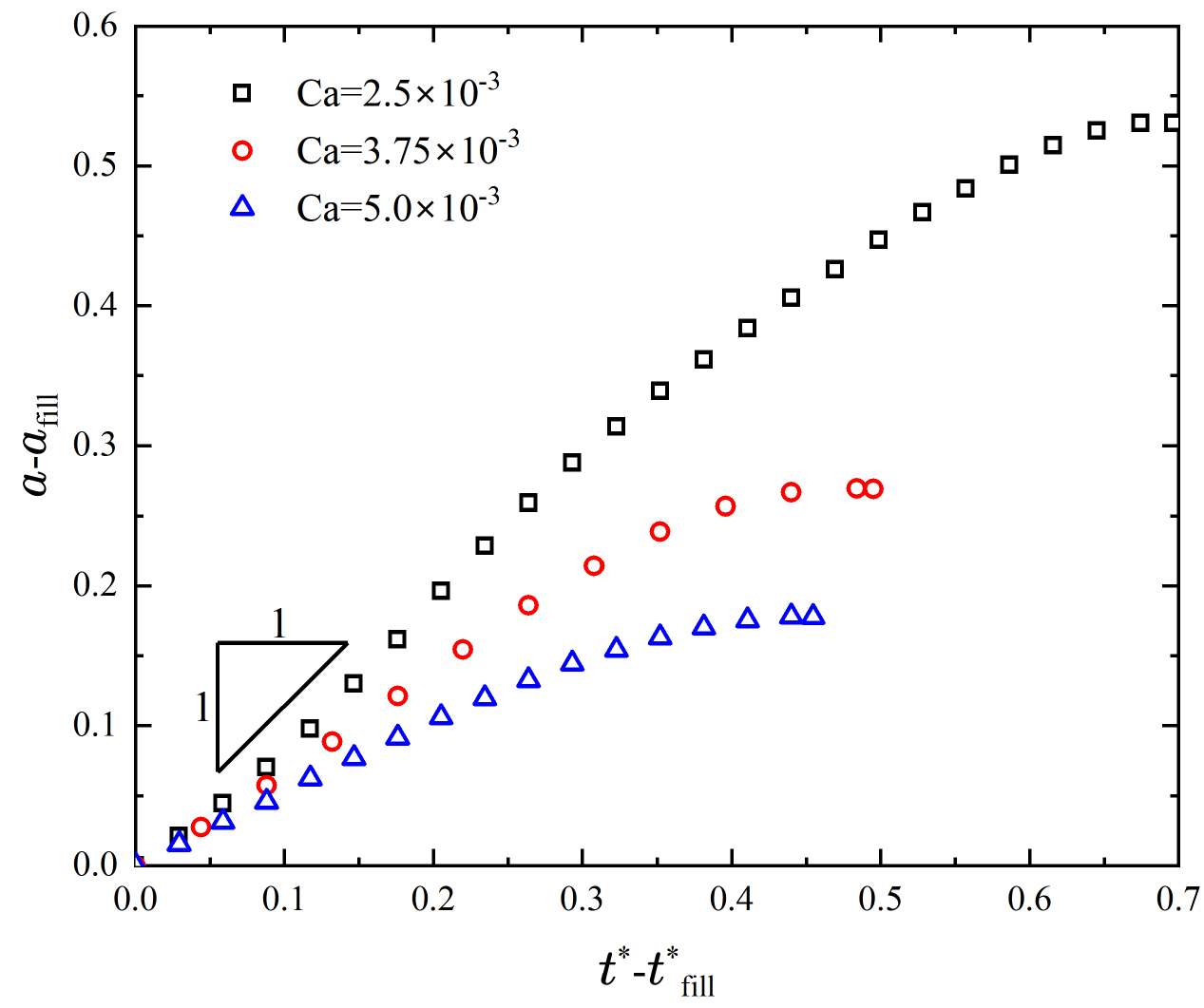}}
	\hspace{0.4cm}
	\subfigure[]{\includegraphics[width=0.42\textwidth]{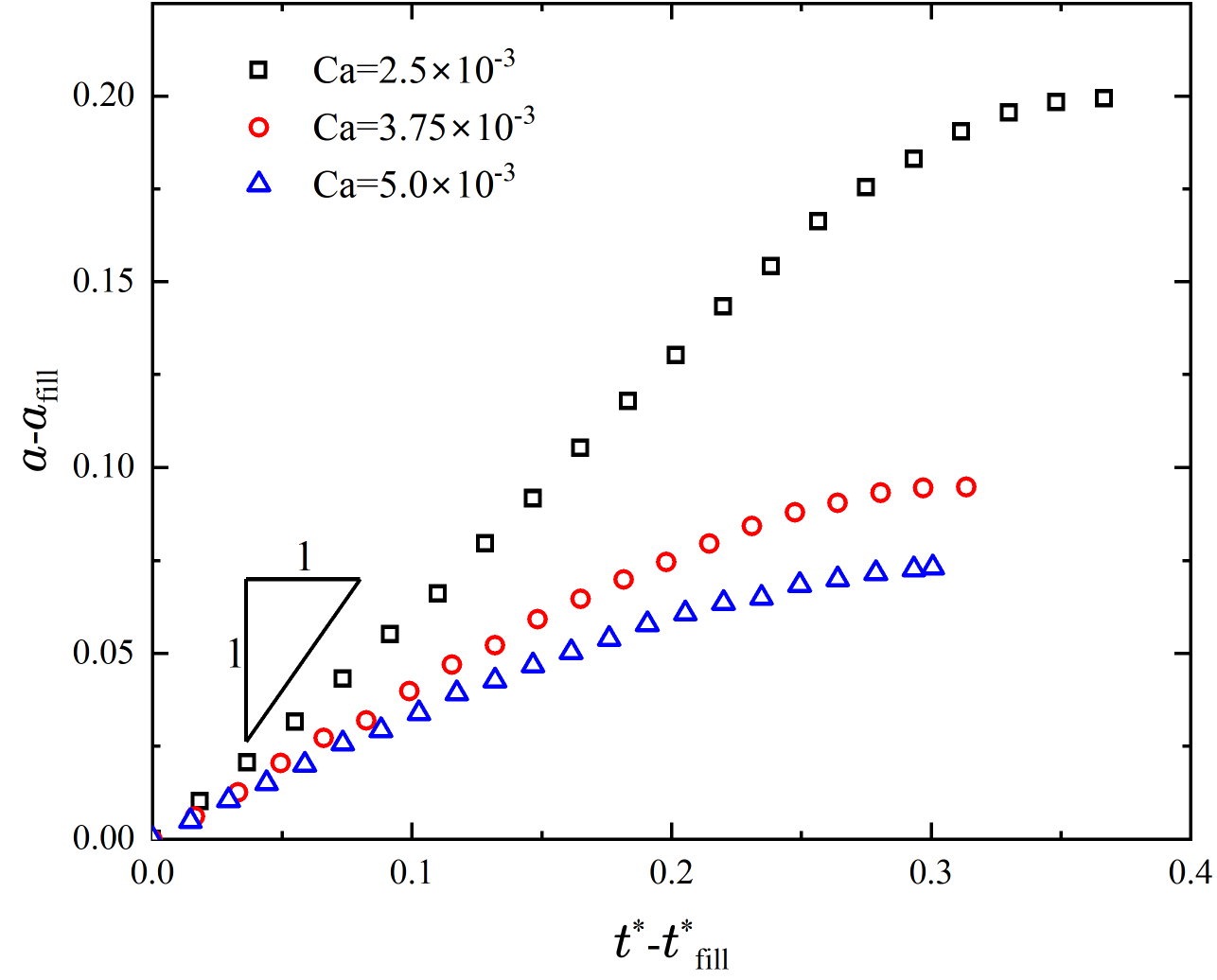}}
	\caption{Evolution of non-dimensional area as a function of non-dimensional time during the squeezing stage under different conditions: (a) top-biased PEF with fixed $Ca = 2.5 \times {10^{ - 3}}$, (b) top-biased PEF with fixed ${Q_r} = 0.1$, (c) bottom-biased PEF with fixed $Ca = 2.5 \times {10^{ - 3}}$, (d) bottom-biased PEF with fixed ${Q_r} = 0.1$.}
	\label{fig16}
\end{figure}

In figure \ref{fig15}, we present the morphological diagrams at the end of the droplet filling stage under different conditions of $C{a_E}$, ${Q_r}$, and $Ca$ for top-biased PEF and bottom-biased PEF. From the figure, it is evident that as ${Ca_E}$ increases, the position at which the droplet blocks the channel shifts earlier (figures \ref{fig15}(a) and \ref{fig15}(d)). However, when a top-biased PEF is applied to the microchannel, the curvature of the dispersed phase does not show significant changes as the electric field force increases, and it remains convex. In contrast, when a high voltage is applied to the bottom electrode, the liquid-liquid interface first moves closely along the bottom of the channel during the filling process, and then, under the influence of the electric field force, deforms toward the top electrode. As a result, at the end of the filling stage, the liquid-liquid interface becomes concave. Therefore, when a high potential is applied to the bottom electrode, the filling area of the droplet during the squeezing stage decreases sharply with increasing voltage. Furthermore, as the flow rate ratio increases (figures \ref{fig15} (b) and \ref{fig15}(e)), the curvature of the liquid-liquid interface increases synchronously. Notably, for the same ${Q_r}$ and $Ca$, the curvature of the liquid-liquid interface when applying bottom-biased PEF is smaller than that with top-biased PEF. This also explains why smaller droplets can be generated under bottom-biased PEF \cite{Singh_JFM2020}. As $Ca$ increases, while keeping ${Qr}$ and ${Ca_E}$ constant, it can be observed that the liquid-liquid interface begins to shift from a convex to a concave shape. At this point, the effect of the electric field on the control volume weakens. A comparison between figures \ref{fig15}(b, e) and figures \ref{fig15}(c, f) reveals that for a low capillary number, the blocking point of the dispersed phase remains at the top electrode and does not change with ${Q_r}$. However, when $Ca$ increases while keeping ${Q_r}$ constant, the blocking point of the dispersed phase shifts downstream.
At the squeezing stage, the electric field force causes the liquid-liquid interface to pin at the top electrode \cite{Singh_JFM2020}. The dispersed phase at the electrode hinders the flow of the continuous phase in the main channel, which leads to part of the continuous phase fluid pushing the dispersed phase to fill what should have flowed into the downstream of the main channel. Therefore, during this process, the filling speed of the dispersed phase is influenced by the combined effects of the electric field force, pressure, and shear flow force. This can be expressed as
\begin{equation}
	{m_{squeeze}} = \int_{{t_{squeeze}}} {\rho \left( {{{\bf{u}}_d} + g\left( {{{\bf{F}}_e},{{\bf{F}}_p},{{\bf{F}}_\tau }} \right){{\bf{u}}_c}} \right)wdt} ,
\end{equation}
since the pinning of the liquid-liquid interface at the top electrode depends on the competitive relationship between these three forces, we can simplify ${g\left( {{{\bf{F}}_e},{{\bf{F}}_p},{{\bf{F}}_\tau }} \right)}$ as
\begin{equation}
	g\left( {{{\bf{F}}_e},{{\bf{F}}_p},{{\bf{F}}_\tau }} \right) = \frac{{{{\bf{F}}_e}}}{{{{\bf{F}}_p} + {{\bf{F}}_\tau }}}.
\end{equation}
Thus, the droplet filling rate can be derived as
\begin{equation}
	k = \frac{{d{a_{squeeze}}}}{{dt}} = \frac{{\frac{{{{\bf{F}}_e}}}{{{{\bf{F}}_p} + {{\bf{F}}_\tau }}} + {Q_r}}}{{1 + {Q_r}}}.
\end{equation}
For $Ca = 2.5 \times {10^{ - 3}}$, the lower flow rate prevents the hydrodynamic stress from causing the liquid-liquid interface pinned at the top electrode from moving downstream. At this point, ${{\bf{F}}_e} = {{\bf{F}}_p} + {{\bf{F}}_\tau }$, so the filling rate of the dispersed phase, $k=1$(see figure \ref{fig16}($a$) and ($c$)). The insert figure also shows that, under these conditions, a vortex forms at the top electrode, which hinders part of the continuous phase flow and accelerates the flow rate of the remaining continuous phase. figures \ref{fig16}(b) and \ref{fig16}(d) shows the relationship between the filling area and time for different $Ca$ values while keeping ${Q_r} = 0.1$ constant. As discussed in figure \ref{fig14}, increasing the flow rate leads to the downstream shift of the blocking point of the dispersed phase. At this point, the dispersed phase is less affected by the horizontal electric field force, making pinning more difficult. In this process, ${{{\bf{F}}_e}}$ decreases while ${{{\bf{F}}_p} + {{\bf{F}}_\tau }}$ increases. Therefore, for $Ca = 3.75 \times {10^{ - 3}}$ and $5 \times {10^{ - 3}}$, the filling rate of the dispersed phase is less than $1$.

\subsection{Manipulating Droplet Sequences}
\label{sec:4.4}
\begin{figure}[H]
	\centering
	\subfigure[]{\includegraphics[width=0.9\textwidth]{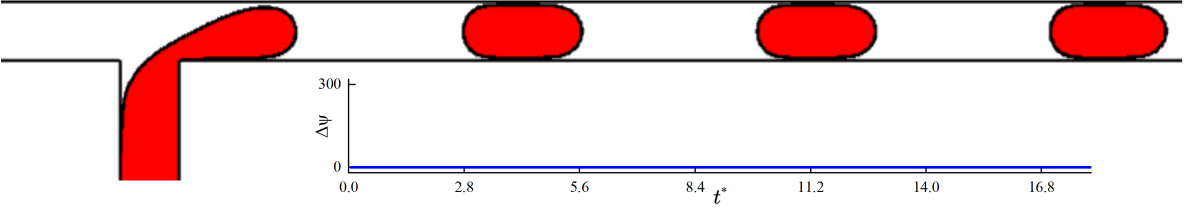}}
	
	\vspace{0.05cm}
	
	\subfigure[]{\includegraphics[width=0.9\textwidth]{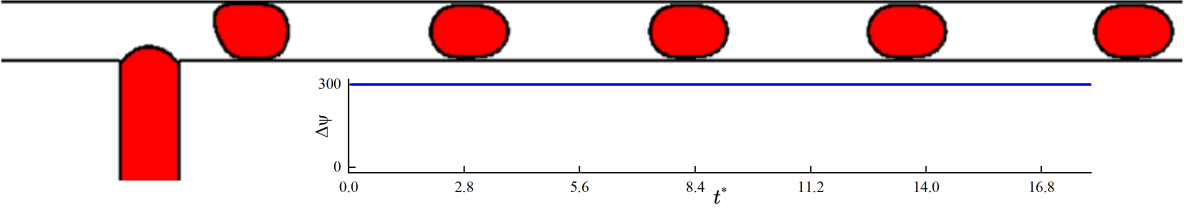}}
	
	\vspace{0.05cm}
	
	\subfigure[]{\includegraphics[width=0.9\textwidth]{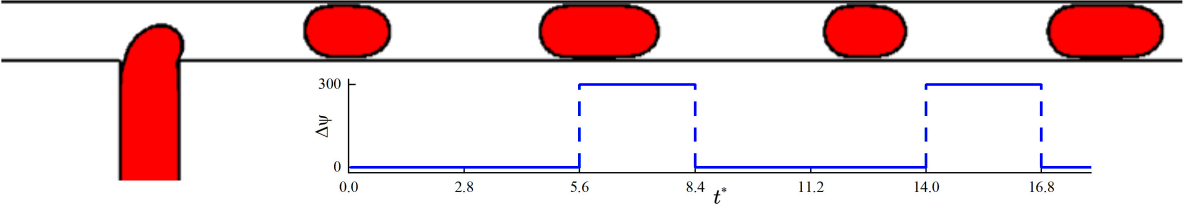}}
	
	\vspace{0.05cm}
	
	\subfigure[]{\includegraphics[width=0.9\textwidth]{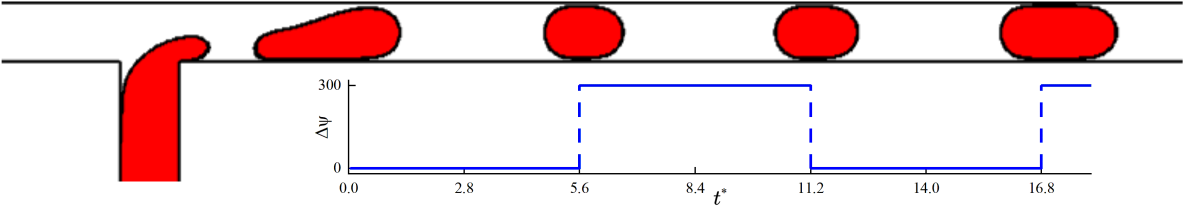}}
	
	\vspace{0.05cm}
	
	\subfigure[]{\includegraphics[width=0.9\textwidth]{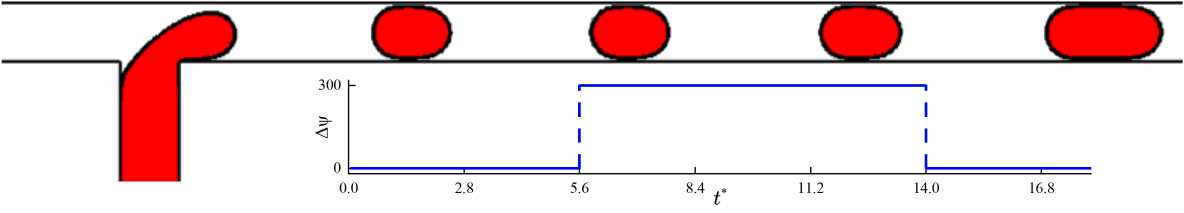}}
	
	\caption{Droplet sequences generated under varying electric field pulse durations at ${Q_r} = 0.6$. The top panel of each subfigure displays the distribution of the dispersed phase (red droplets), while the bottom panel shows the corresponding electric pulse signal ($\Delta\psi$) over non-dimensional time ($t^*$). The specific sequences are: (a) lll, (b) sss, (c) lsls, (d) lss, and (e) lsss.}
	\label{fig17}
\end{figure}

Following the experimental strategy of Shojaeian and Hardt \cite{Shojaeian_APL2018}, a controlled direct-current electric field was applied to regulate droplet size and sequence. Figure \ref{fig17} demonstrates the control of successive droplet sizes by applying electric field pulses at a flow rate ratio of $0.4$. At this flow rate, the dispersed phase can break at the junction under the influence of only hydrodynamic stress, meaning that the application of PEF primarily adjusts the size of individual droplets. Figures \ref{fig17}(c), \ref{fig17}(d) and \ref{fig17}(e) show the sequences $lsls$, $lss$, and $lsss$, corresponding to the application of zero voltage for $t^* = 5.6$, followed by rectangular voltage pulses of $300~\text{V}$ with durations of $t^* = 2.8, 5.6$, and $8.4$, respectively. Therefore, by controlling the PEF, we can achieve on-demand adjustment of the sizes of consecutive droplets.

\begin{figure}[H]
	\centering
	\subfigure[]{\includegraphics[width=0.9\textwidth]{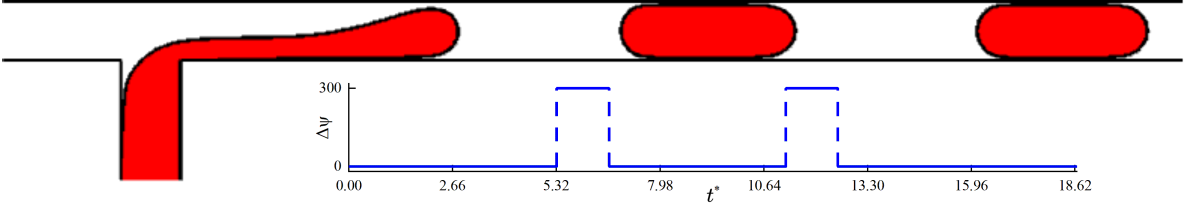}}
	
	\vspace{0.05cm}
	
	\subfigure[]{\includegraphics[width=0.9\textwidth]{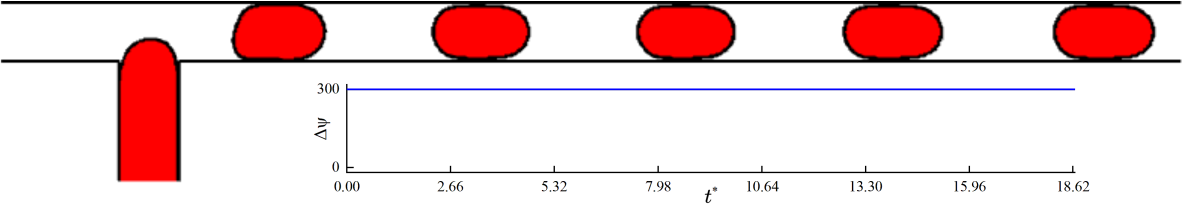}}
	
	\vspace{0.05cm}
	
	\subfigure[]{\includegraphics[width=0.9\textwidth]{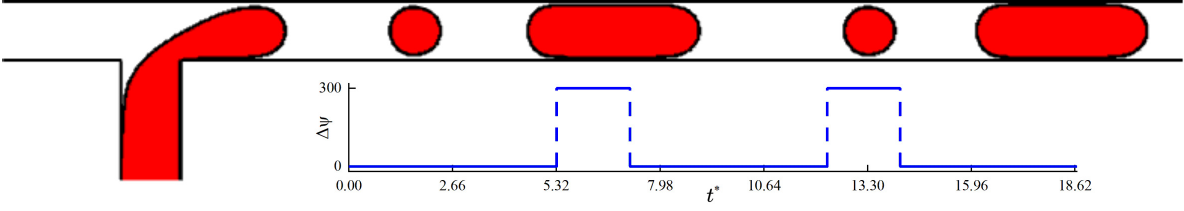}}
	
	\vspace{0.05cm}
	
	\subfigure[]{\includegraphics[width=0.9\textwidth]{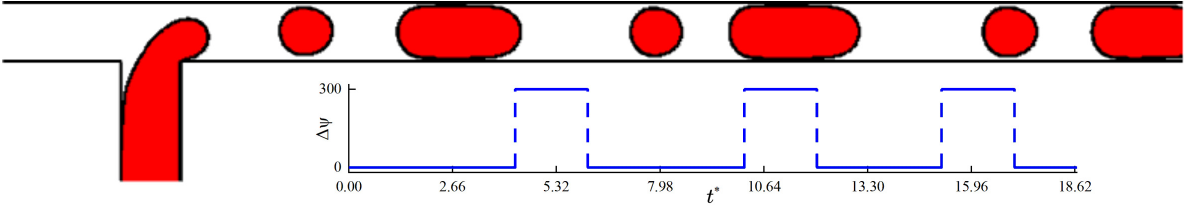}}
	
	\vspace{0.05cm}
	
	\subfigure[]{\includegraphics[width=0.9\textwidth]{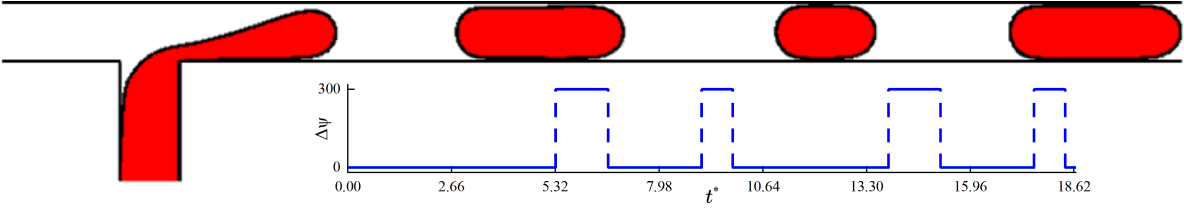}}
	
	\caption{Droplet sequences generated under varying electric field pulse durations at ${Q_r} = 0.6$. The top panel of each subfigure displays the distribution of the dispersed phase (red droplets), while the bottom panel shows the corresponding electric pulse signal ($\Delta\psi$) over non-dimensional time ($t^*$). The specific sequences are: (a) lll, (b) mmm, (c) lsls, (d) msms, and (e) lmlm.}
	\label{fig18}
\end{figure}

Moreover, by varying the flow rate ratio (${Q_r}$) and the duration of the PEF, droplets may no longer break at the junction of the channel. When the dispersed phase passes through the electrode and the electric field is applied afterward, the break-off point shifts to the middle of the electrode. At this point, due to the residual dispersed phase in the main channel, the subsequent droplets no longer undergo the complete filling process. As a result, the formation of the latter droplets is influenced not only by the coupling of hydrodynamic stress and electric field stress but also by the residual phase in the channel. Figure \ref{fig18} shows the droplet sequences for different electric field pulse durations at ${Q_r} = 0.6$. For simplicity, we introduce a medium-sized droplet, denoted as "$m$". Therefore, the droplet sequences shown in figure \ref{fig18} are $lll$, $mmm$, $lsls$, $msms$, and $lmlm$. It can be observed that the size of the small droplets here is smaller than that of droplets formed under the CEF conditions (see figures \ref{fig18}(b), \ref{fig18}(c), and \ref{fig18}(d)). This is because, when the previous droplet breaks at the electrode, there is more residual dispersed phase in the channel. The application of the electric field causes this residual phase to directly block the channel, leading to droplet break-off (figure \ref{fig9}). Alternatively, a period of zero voltage can be applied to allow more dispersed phase to fill the channel, with the filling speed following the derivation in Section \ref{sec:4.3}. Subsequently, an electric field is applied to deform and block the dispersed phase, which increases the droplet size (figure \ref{fig18}(e)).

\begin{figure}[H]
	\centering
	
	\includegraphics[height=5.5cm]{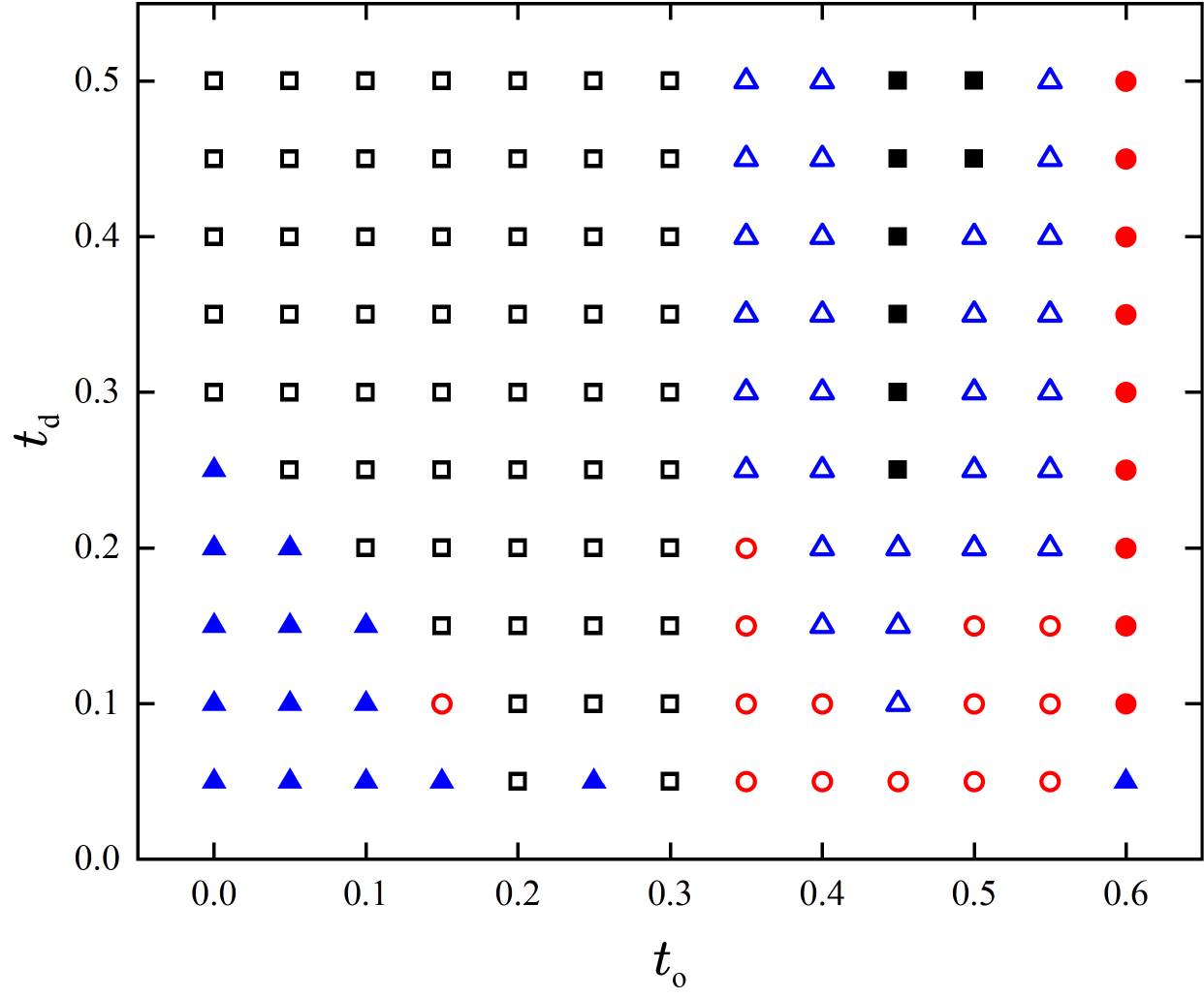}
	\put(-200, 145){\scriptsize\textit{(a)}}
	\hspace{0.3cm}
	\includegraphics[height=5.5cm]{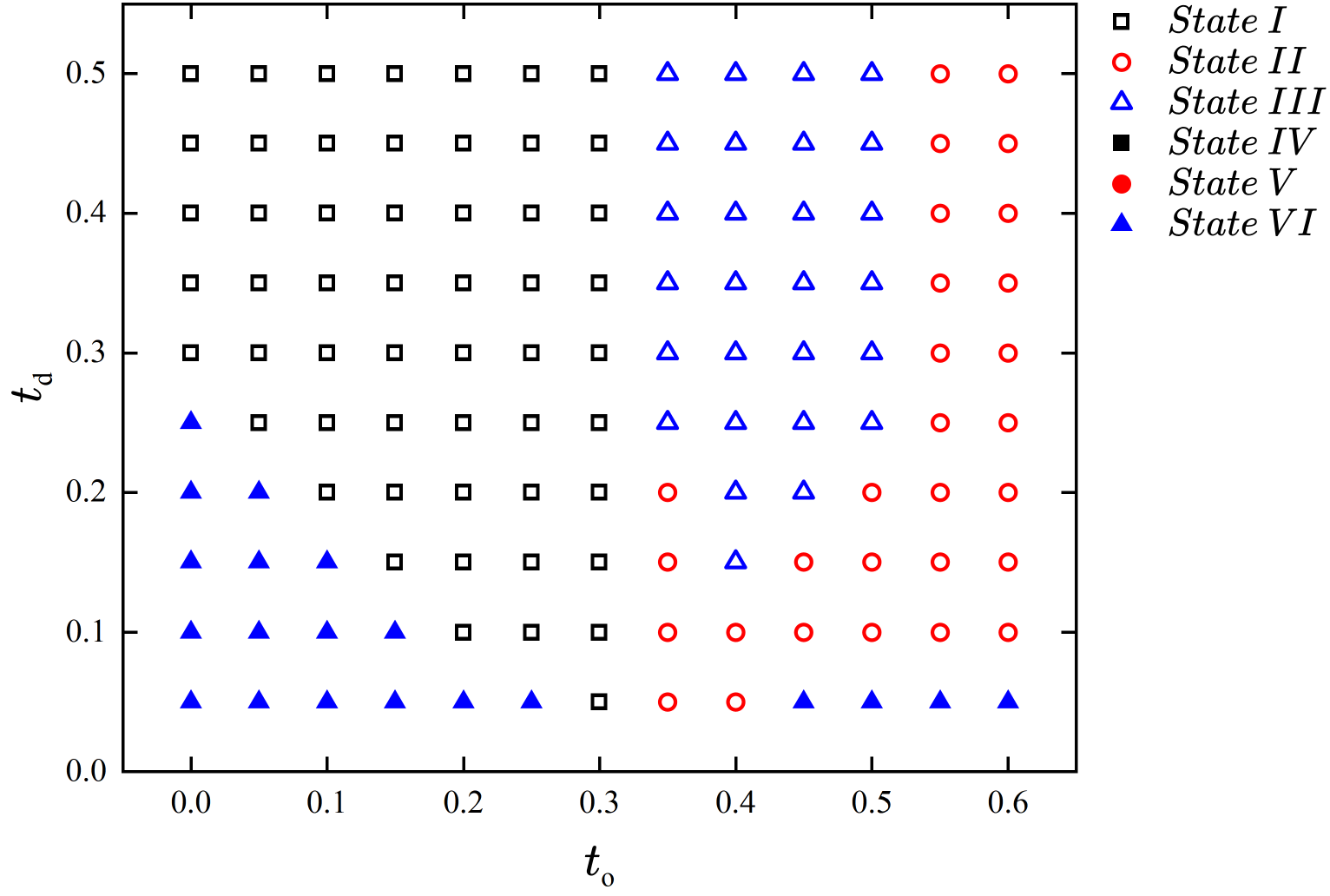}
	\put(-240, 145){\scriptsize\textit{(b)}}
	
	\vspace{0.1cm}
	
	\includegraphics[width=0.3\textwidth]{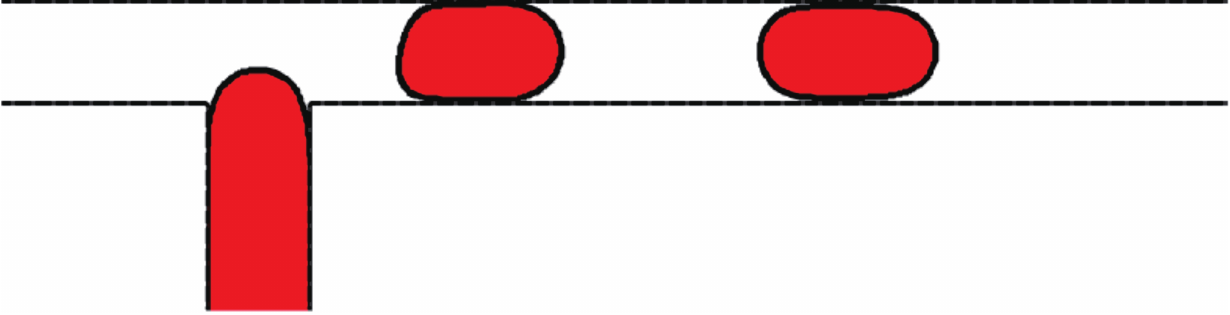}
	\put(-80, 38){\scriptsize\textit{State I}}
	\hspace{0.1cm}
	\includegraphics[width=0.3\textwidth]{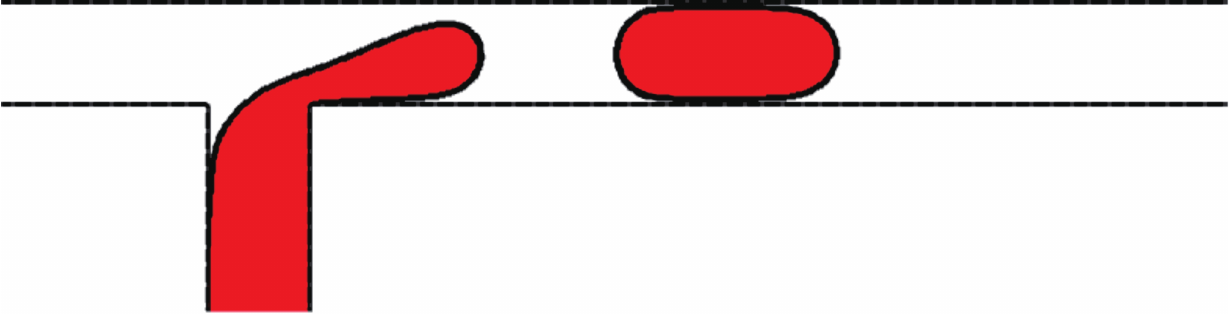}
	\put(-80, 38){\scriptsize\textit{State II}}
	\hspace{0.1cm}
	\includegraphics[width=0.3\textwidth]{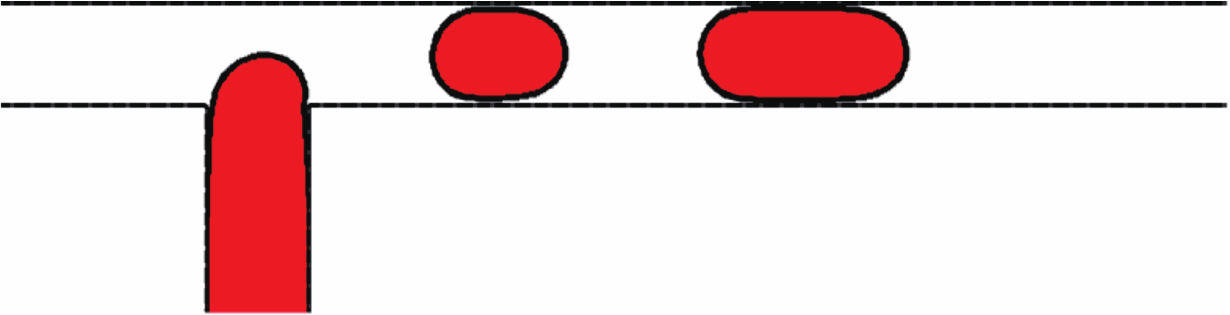}
	\put(-80, 38){\scriptsize\textit{State III}}
	
	\vspace{0.1cm}
	
	\includegraphics[width=0.3\textwidth]{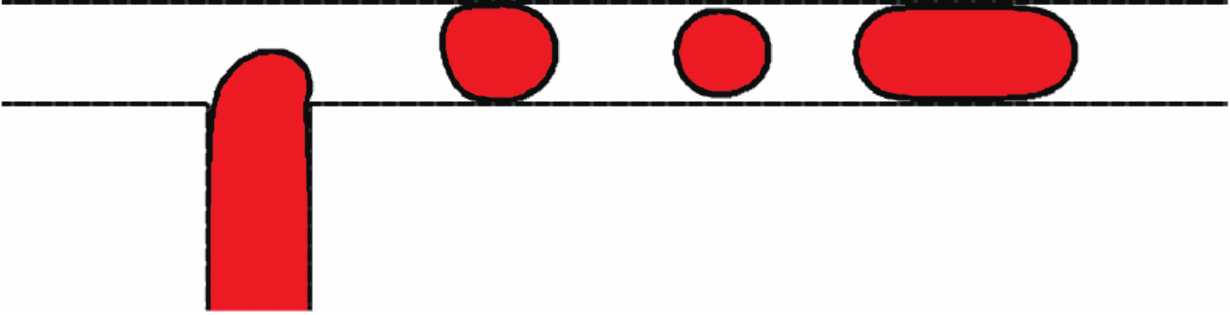}
	\put(-80, 38){\scriptsize\textit{State IV}}
	\hspace{0.1cm}
	\includegraphics[width=0.3\textwidth]{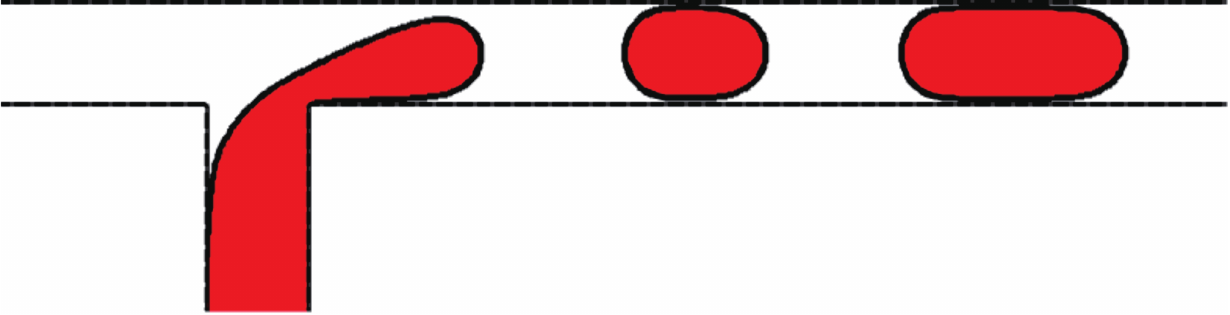}
	\put(-80, 38){\scriptsize\textit{State V}}
	\hspace{0.1cm}
	\includegraphics[width=0.3\textwidth]{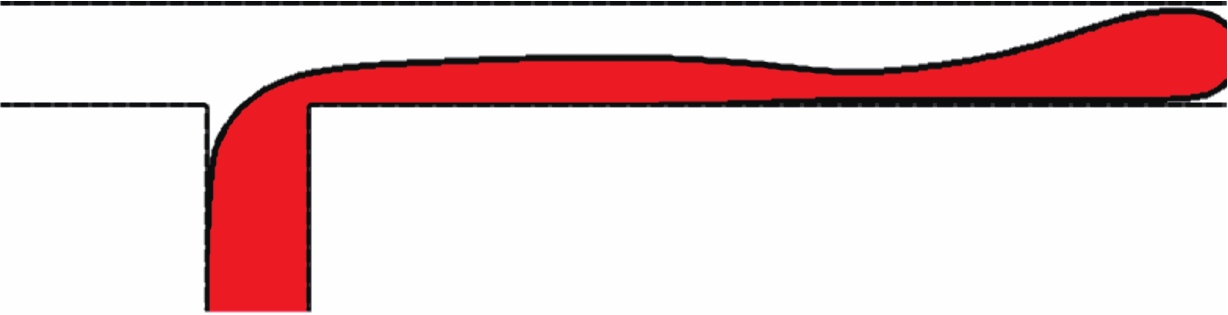}
	\put(-80, 38){\scriptsize\textit{State VI}}
	
	\caption{The phase diagram of different droplet sequences observed in a T-junction under different pulse conditions: (a) top-biased PEF, (b) bottom-biased PEF. State I: A single droplet is generated during the pulse period, and the droplet breaks at the junction of the channel. State II: A single droplet is generated during the pulse period, and the droplet breaks downstream of the junction. State III: Two droplets are generated during the pulse period, and the droplets break at the junction of the channel. State IV: Three droplets are generated during the pulse period, and the droplets break at the junction of the channel. State V: Two droplets are generated during the pulse period, and the droplets break downstream of the junction. State VI: No droplet is generated during the pulse period.}
	\label{fig19}
\end{figure}

Figure \ref{fig19} shows the phase diagram of droplet sequences generated at different electrode positions when ${Q_r}=0.6$. The diagram is divided based on the droplet break-off points at the pipe junction (State I), between the electrodes (State II-V), and downstream of the electrodes (State VI). Among them, the case where droplets break at the electrodes is further subdivided into single droplet formation (State II), two droplets (State III), three droplets (State IV), and the simultaneous formation of two droplets at the electrodes (State V). As ${t_o}$ increases, a shorter ${t_d}$ of the PEF allows the droplets to break earlier. Meanwhile, the break-off point of the droplets shifts downstream with increasing ${t_o}$, transitioning from State I to State II. As the break-off point moves further downstream, increasing ${t_d}$ leads to changes in the formation of one to two droplets, causing the droplet sequence to shift from State II to State III, and then to State IV. Due to the short electrodes and the sustained electric field, the break-off point of the droplet moves forward, causing no residual dispersed phase in the main channel after the formation of the third droplet. It is noteworthy that delaying ${t_o}$ does not cause the break-off point of the first droplet to continue shifting downstream, but instead, it reaches its maximum when ${t_o} = 0.45$. At this point, the curved liquid-liquid interface is just in the middle of the electrode, quickly blocking the channel, and the neck is formed at the electrode edge under the influence of the electric field. For bottom-biased PEF and when ${t_o} > 0.5$, the electric field force is insufficient to form the neck directly at the electrode edge. This results in PEF only affecting the formation of the second droplet under these two conditions. As ${t_o}$ increases to $0.6$, the dispersed phase is about to form the neck under the shear force. The dispersed phase in the middle of the electrode begins to deform and block under the influence of the electric field, leading to the simultaneous formation of two necks, and consequently, two droplets are formed at the same time.

\section{Conclusion}
\label{sec:conclusion}

In this work, a phase-field lattice Boltzmann model was employed to investigate droplet formation in a T-junction microchannel under pulsed electric fields. Three electrical configurations were considered, namely no electric field, a top-biased electrode, and a bottom-biased electrode. The results show that the electric field modifies the force balance governing droplet formation mainly through two mechanisms, namely transverse migration and contact-line pinning. Under top-biased actuation, the dispersed phase rapidly migrates towards the upper wall, which accelerates channel blockage. Under bottom-biased actuation, the dispersed phase first attaches to the lower wall and subsequently deforms upward under the electric force, thereby delaying the transition to the squeezing stage. As a result, the electric field changes the filling duration while exerting little influence on the filling flux. In addition, the presence of the electric field shifts the breakup point upstream to the T-junction for cases in which droplets would otherwise break downstream, leading to more robust control of droplet formation \cite{Zhu_CEJ2024}.

The results further show that the squeezing dynamics are governed by pinning of the droplet interface at the electrode edges. This pinning counteracts the downstream advection driven by hydrodynamic shear and thereby localises the breakup process near the T-junction. In contrast to constant electric forcing, pulsed electric forcing introduces an additional temporal control parameter. By adjusting the pulse width and onset time relative to the hydrodynamic cycle, the interfacial evolution and breakup process can be modulated in a stage-dependent manner. Under selected pulse conditions, pinch-off can be delayed, consistent with the trend reported in the AC-electric-field experiments of Teo \textit{et al.} \cite{Teo_MN2020}. Finally, regime maps and droplet sequence diagrams were constructed to illustrate the generation of programmable droplet trains under different pulse parameters. These results provide physical insight and numerical guidance for the design of electrically actuated microfluidic systems for droplet generation and manipulation.

\section*{Acknowledgements}
\label{sec:acknowledge}
This work is financially supported by the National Natural Science Foundation of China (Grant No. 12472297)

\section*{Declaration of competing interest}
\label{Declaration of interests}
The authors report no conflict of interest.



\begin{thebibliography}{99}
	
\bibitem{Schneider}
T. Schneider, J. Kreutz, and D. T. Chiu, The potential impact of droplet microfluidics in biology, Anal. Chem. 85 (2013) 3476.

\bibitem{Whitesides_nat2006}
G. M. Whitesides, The origins and the future of microfluidics, Nature 442 (2006) 368.

\bibitem{Song_ACIEE2003}
H. Song, J. D. Tice, and R. F. Ismagilov, A microfluidic system for controlling reaction networks in time, Angew. Chem. Int. Ed. 42 (2003) 768.

\bibitem{Korczyk_NC2019}
P. M. Korczyk, V. van Steijn, S. Błoński, D. Zaremba, D. A. Beattie, and P. Garstecki, Accounting for corner flow unifies the understanding of droplet formation in microfluidic channels, Nat. Commun. 10 (2019) 2528.

\bibitem{Bontoux2006}
N. Bontoux, A. Pépin, Y. Chen, A. Ajdari, and H. A. Stone, Experimental characterization of hydrodynamic dispersion in shallow microchannels, Lab Chip 6 (2006) 930.

\bibitem{Tice2003}
J. D. Tice, H. Song, A. D. Lyon, and R. F. Ismagilov, Formation of droplets and mixing in multiphase microfluidics at low values of the Reynolds and the capillary numbers, Langmuir 19 (2003) 9127.

\bibitem{Zhang_NC2024}
H. Zhang, R. Gupte, Y. Li, C. Huang, A. R. Guzman, J. J. Han, H. Jung, R. Sabnis, P. de Figueiredo, and A. Han, NOVAsort for error-free droplet microfluidics, Nat. Commun. 15 (2024) 9444.

\bibitem{Elvira_NC2013}
K. S. Elvira, X. C. i Solvas, R. C. Wootton, and A. J. Demello, The past, present and potential for microfluidic reactor technology in chemical synthesis, Nat. Chem. 5 (2013) 905.

\bibitem{Shang_Labonchip2016}
L. Shang, Y. Cheng, J. Wang, Y. Yu, Y. Zhao, Y. Chen, and Z. Gu, Osmotic pressure-triggered cavitation in microcapsules, Lab Chip 16 (2016) 251.

\bibitem{Shi2021}
N. Shi, M. Mohibullah, and C. J. Easley, Active flow control and dynamic analysis in droplet microfluidics, Annu. Rev. Anal. Chem. 14 (2021) 133.

\bibitem{Umbanhowar_langmuir2000}
P. B. Umbanhowar, V. Prasad, and D. A. Weitz, Monodisperse emulsion generation via drop break off in a coflowing stream, Langmuir 16 (2000) 347.

\bibitem{Thorsen_PRL2001}
T. Thorsen, R. W. Roberts, F. H. Arnold, and S. R. Quake, Dynamic pattern formation in a vesicle-generating microfluidic device, Phys. Rev. Lett. 86 (2001) 4163.

\bibitem{Haringa_PRF2019}
C. Haringa, C. de Jong, D. A. Hoang, L. M. Portela, C. R. Kleijn, M. T. Kreutzer, and V. van Steijn, Breakup of elongated droplets in microfluidic T-junctions, Phys. Rev. Fluids 4, 024203 (2019).


\bibitem{Anna_APL2003}
S. L. Anna, N. Bontoux, and H. A. Stone, Formation of dispersions using ‘flow focusing’ in microchannels, Appl. Phys. Lett. 82 (2003) 364.

\bibitem{Glawdel2012a}
T. Glawdel, C. Elbuken, and C. L. Ren, Droplet formation in microfluidic T-junction generators operating in the transitional regime. I. Experimental observations, Phys. Rev. E 85 (2012) 016323.

\bibitem{Glawdel2012b}
T. Glawdel, C. Elbuken, and C. L. Ren, Droplet formation in microfluidic T-junction generators operating in the transitional regime. II. Modeling, Phys. Rev. E 85 (2012) 016323.

\bibitem{Demench_JFM2008}
M. De Menech, P. Garstecki, F. Jousse, and H. A. Stone, Transition from squeezing to dripping in a microfluidic T-shaped junction, J. Fluid Mech. 595 (2008) 141.

\bibitem{Garstecki_Labonchip2006}
P. Garstecki, M. J. Fuerstman, H. A. Stone, and G. M. Whitesides, Formation of droplets and bubbles in a microfluidic T-junction—scaling and mechanism of break-up, Lab Chip 6 (2006) 437.

\bibitem{Lee_JAP2014}
C. P. Lee, T. S. Lan, and M. F. Lai, Fabrication of two-dimensional ferrofluid microdroplet lattices in a microfluidic channel, J. Appl. Phys. 115 (2014) 17B527.

\bibitem{SAHORE_AM2018}
V. Sahore, S. R. Doonan, and R. C. Bailey, Droplet microfluidics in thermoplastics: device fabrication, droplet generation, and content manipulation using integrated electric and magnetic fields, Anal. Methods 10 (2018) 4264.

\bibitem{SHYAM_JFM2022}
S. Shyam, B. Dhapola, and P. K. Mondal, Magnetofluidic-based controlled droplet breakup: Effect of non-uniform force field, J. Fluid Mech. 944 (2022) A51.

\bibitem{Miralles_LabonChip2015}
V. Miralles, A. Huerre, H. Williams, B. Fournié, and M. C. Jullien, A versatile technology for droplet-based microfluidics: thermomechanical actuation, Lab Chip 15 (2015) 2133.

\bibitem{JUNG_Labonchip2016}
J. H. Jung, G. Destgeer, B. Ha, J. Park, and H. J. Sung, On-demand droplet splitting using surface acoustic waves, Lab Chip 16 (2016) 3235.

\bibitem{Huang_Labchip2017}
Y. Huang, Y. L. Wang, and T. N. Wong, AC electric field controlled non-Newtonian filament thinning and droplet formation on the microscale, Lab Chip 17 (2017) 2969.

\bibitem{Singh_JFM2020}
R. Singh, S. S. Bahga, and A. Gupta, Electrohydrodynamic droplet formation in a T-junction microfluidic device, J. Fluid Mech. 905 (2020) A29.

\bibitem{Park2016}
S. Y. Park and Y. Nam, Single-sided digital microfluidic (SDMF) devices for effective coolant delivery and enhanced two-phase cooling, Micromachines 8 (2016) 3.

\bibitem{Taylor1966}
G. I. Taylor, Studies in electrohydrodynamics. I. The circulation produced in a drop by an electric field, Proc. R. Soc. London A 291 (1966) 159.


\bibitem{Yariv_JFM2016}
E. Yariv and Y. Almog, The effect of surface-charge convection on the settling velocity of spherical drops in a uniform electric field, J. Fluid Mech. 797 (2016) 536.

\bibitem{Papageorgiou_ARFM2019}
D. T. Papageorgiou, Film flows in the presence of electric fields, Annu. Rev. Fluid Mech. 51 (2019) 155.

\bibitem{Vlahovska_PRF2016}
P. M. Vlahovska, Electrohydrodynamic instabilities of viscous drops, Phys. Rev. Fluids 1, 060504 (2016).

\bibitem{Vlahovska2019}
P. M. Vlahovska, Electrohydrodynamics of drops and vesicles, Annu. Rev. Fluid Mech. 51 (2019) 305.

\bibitem{Link_ACIEE2006}
D. R. Link, E. Grasland-Mongrain, A. Duri, F. Sarrazin, Z. Cheng, G. Cristobal, M. Marquez, and D. A. Weitz, Electric control of droplets in microfluidic devices, Angew. Chem. Int. Ed. 45 (2006) 2556.

\bibitem{Yeh_MN2012}
C. H. Yeh, M. H. Lee, and Y. C. Lin, Using an electro-spraying microfluidic chip to produce uniform emulsions under a direct-current electric field, Microfluid. Nanofluid. 12 (2012) 475.

\bibitem{Jiang_JMF2021}
L. Jiang, L. Yu, P. Premaratne, Z. Zhang, and H. Qin, CFD-based numerical modeling to predict the dimensions of printed droplets in electrohydrodynamic inkjet printing, J. Manuf. Process. 66 (2021) 125.

\bibitem{Shojaeian_APL2018}
M. Shojaeian and S. Hardt, Fast electric control of the droplet size in a microfluidic T-junction droplet generator, Appl. Phys. Lett. 112 (2018) 194102.


\bibitem{Agarwal_PRF2020}
V. G. Agarwal, Dynamics of droplet formation and flow regime transition in a T-junction microchannel with a shear-thinning continuous phase, Phys. Rev. Fluids 5, 044203 (2020).

\bibitem{Zhu_CEJ2024}
H. Zhu, S. Lu, X. Xu, M. Lei, W. Xu, and Q. Li, Numerical investigation on flow regime transition mechanism and length prediction method of droplet in T-junction by electric field, Chem. Eng. J. 485 (2024) 149874.

\bibitem{Wang_IJHMT2016}
B. B. Wang, X. D. Wang, T. H. Wang, G. Lu, and W. M. Yan, Electro-coalescence of two charged droplets under constant and pulsed DC electric fields, Int. J. Heat Mass Transfer 98 (2016) 10.

\bibitem{Huang_AICHE2020}
X. Huang, L. He, X. Luo, and H. Yin, Droplet dynamic response in low-viscosity fluid subjected to a pulsed electric field and an alternating electric field, AIChE J. 66 (2020) e16869.

\bibitem{Sengupta_PRF2020}
R. Sengupta and A. S. Khair, Deformation of a conducting drop in a randomly fluctuating electric field, Phys. Rev. Fluids 5, 063701 (2020).

\bibitem{Teo_MN2020}
A. J. Teo, M. Yan, J. Dong, H. D. Xi, Y. Fu, S. H. Tan, and N. T. Nguyen, Controllable droplet generation at a microfluidic T-junction using AC electric field, Microfluid. Nanofluid. 24 (2020) 21.

\bibitem{Liu_POF2022}
Z. Liu, F. Cai, Y. Pang, Y. Ren, N. Zheng, R. Chen, and S. Zhao, Enhanced droplet formation in a T-junction microchannel using electric field: A lattice Boltzmann study, Phys. Fluids 34 (2022).

\bibitem{Liu_POF2009}
H. Liu and Y. Zhang, Droplet formation in a T-shaped microfluidic junction, J. Appl. Phys. 106 (2009) 034701.

\bibitem{Graaf_langmuir2006}
S. van der Graaf, T. Nisisako, C. G. P. H. Schr\"oen, R. G. M. van der Sman, and R. M. Boom, Lattice Boltzmann simulations of droplet formation in a T-shaped microchannel, Langmuir 22 (2006) 4144.

\bibitem{Ha2000}
J. W. Ha and S. M. Yang, Deformation and breakup of Newtonian and non-Newtonian conducting drops in an electric field, J. Fluid Mech. 405 (2000) 131.

\bibitem{Badalassi_jcp2003}
V. E. Badalassi, H. D. Ceniceros, and S. Banerjee, Computation of multiphase systems with phase field models, J. Comput. Phys. 190 (2003) 371.

\bibitem{Chiu_jcp2011}
P. H. Chiu and Y. T. Lin, A conservative phase field method for solving incompressible two-phase flows, J. Comput. Phys. 230 (2011) 185.

\bibitem{SunJCP2007}
Y. Sun and C. Beckermann, Sharp interface tracking using the phase-field equation, J. Comput. Phys. 220 (2007) 626.

\bibitem{FolchPRE1999}
R. Folch, J. Casademunt, and A. Hernández-Machado, Phase-field model for Hele-Shaw flows with arbitrary viscosity contrast. I. Theoretical approach, Phys. Rev. E 60 (1999) 1724.

\bibitem{Unverdi_JCP1992}
S. O. Unverdi and G. Tryggvason, A front-tracking method for viscous, incompressible, multi-fluid flows, J. Comput. Phys. 100 (1992) 25.

\bibitem{Liu_pof2019}
X. Liu, Z. Chai, and B. Shi, A phase-field-based lattice Boltzmann modeling of two-phase electro-hydrodynamic flows, Phys. Fluids 31 (2019).

\bibitem{Singh_PRFluids2019}
R. Singh, S. S. Bahga, and A. Gupta, Electric field induced droplet deformation and breakup in confined shear flows, Phys. Rev. Fluids 4 (2019) 033701.

\bibitem{Hua_pof2008}
J. Hua, L. K. Lim, and C. H. Wang, Numerical simulation of deformation/motion of a drop suspended in viscous liquids under influence of steady electric fields, Phys. Fluids 20 (2008).

\bibitem{Wang_PRE2003}
L. Wang, K. He, and H. Wang, Phase-field-based lattice Boltzmann model for simulating thermocapillary flows, Phys. Rev. E 108 (2023) 055306.

\bibitem{Chen_JFM2017}
Y. Chen and Z. Deng, Hydrodynamics of a droplet passing through a microfluidic T-junction, J. Fluid Mech. 819 (2017) 401.

\bibitem{Ladd_jfm1994}
A. J. C. Ladd, Numerical simulations of particulate suspensions via a discretized Boltzmann equation. Part 1. Theoretical foundation, J. Fluid Mech. 271 (1994) 285.

\bibitem{Zhang_PRE2012}
T. Zhang, B. Shi, Z. Guo, Z. Chai, and J. Lu, General bounce-back scheme for concentration boundary condition in the lattice-Boltzmann method, Phys. Rev. E 85 (2012) 016701.

\bibitem{Lou_PRE2013}
Q. Lou, Z. Guo, and B. Shi, Evaluation of outflow boundary conditions for two-phase lattice Boltzmann equation, Phys. Rev. E 87 (2013) 063301.

\bibitem{Feng_PRSA1999}
J. Q. Feng, Electrohydrodynamic behaviour of a drop subjected to a steady uniform electric field at finite electric Reynolds number, Proc. R. Soc. Lond. A 455 (1999) 2245.

\bibitem{Luo_pof2020}
K. Luo, J. Wu, H. L. Yi, and H. P. Tan, Numerical analysis of two-phase electrohydrodynamic flows in the presence of surface charge convection, Phys. Fluids 32 (2020).

\bibitem{Xiong_JSC2025}
F. Xiong, L. Wang, J. Huang, and K. Luo, A thermodynamically consistent phase-field lattice Boltzmann method for two-phase electrohydrodynamic flows, J. Sci. Comput. 103 (2025) 1.

\end{thebibliography}

\end{document}